\documentclass[accepted]{article}

\usepackage{microtype}
\usepackage{graphicx}
\usepackage{subfigure}
\usepackage{booktabs} 

\usepackage{hyperref}

\usepackage{icml2025}

\usepackage{amsmath}
\usepackage{amssymb}
\usepackage{mathtools}
\usepackage{amsthm}
\usepackage{amsfonts,oldgerm,mathrsfs,booktabs}
\usepackage[capitalize,noabbrev]{cleveref}

\theoremstyle{plain}
\newtheorem{theorem}{Theorem}[section]

\newtheorem{lemma}[theorem]{Lemma}

\theoremstyle{definition}
\newtheorem{definition}[theorem]{Definition}

\theoremstyle{remark}

\usepackage[textsize=tiny]{todonotes}

\usepackage{graphicx} 
\usepackage{algorithm}
\usepackage{algorithmic}
\newcommand{\setalgospace}{%
    \setlength{\textfloatsep}{5pt} 
}

\usepackage{amsmath}
\usepackage{amsthm}
\usepackage{booktabs}
\usepackage{multirow}
\renewcommand{\arraystretch}{2}
\usepackage{braket}
\usepackage{array}

\newcommand{\statespace}{\mathcal{S}}
\newcommand{\actionspace}{\mathcal{A}}

\newcommand{\hilbertspace}{\mathcal{H}}

\newcommand{\tensorproduct}{\otimes}
\newcommand{\realnumber}{\mathbb{R}}

\newcommand{\ComplexSpace}[1]{\mathbb{C}^{#1}}

\newcommand{\qArgmax}{\textbf{QMS}}
\newcommand{\V}[2]{V^{#1}_{#2}}
\newcommand{\Q}[2]{Q^{#1}_{#2}}
\newcommand{\Expectation}{\mathbb{E}}
\newcommand{\argmax}{\mathop{\mathrm{argmax}}}
\newcommand{\ceil}[1]{\lceil{#1}\rceil}

\newcommand{\MDPQoracle}{O_{\mathcal{QM}}}

\newcommand{\MDP}{\mathcal{M}}

\newcommand{\classicalMDPoracle}{O_{\mathcal{M}}}

\newcommand{\QVIone}{\textbf{QVI-1}(\MDP, \delta)}
\newcommand{\QVItwo}{\textbf{QVI-2}(\MDP, \epsilon, \delta)}
\newcommand{\QVIthree}{\textbf{QVI-3}(\MDP, \epsilon, \delta)}
\newcommand{\QVIfour}{\textbf{QVI-4}(\MDP, \epsilon, \delta)}
\newcommand{\QVIfive}{\textbf{QVI-5}(\MDP, \epsilon, \delta)}
\newcommand{\qAdder}{U_{\text{qAdd}}}
\newcommand{\qMul}{U_{\text{qMul}}}
\newcommand{\QFT}{\text{QFT}}
\newcommand{\IQFT}{\text{IQFT}}
\newcommand{\QMSoracle}{O_{\text{QMS}}}
\newcommand{\QBSC}{U_{\text{QBSC}}}
\newcommand{\MDPQPoracle}[1]{O^{#1}_{\mathcal{QM}}}
\newcommand{\MDPPHoracle}[1]{\Tilde{U}^{#1}_{\mathcal{QM}}}
\newcommand{\ProbabilityOracle}[1]{O_{#1}}
\newcommand{\PhaseOracle}[1]{\Tilde{U}_{#1}}
\newcommand{\BinaryOracle}[1]{B_{#1}}
\newcommand{\fixbinary}[1]{\text{Bi}[#1]}

\newcommand{\supp}[1]{\text{supp}(#1)}
\newcommand{\Phsa}{P_{h|s,a}}
\newcommand{\infiNorm}[1]{\left\|#1\right\|_{\infty}}
\newcommand{\OPNorm}[1]{\left\|#1\right\|_{\text{op}}}

\newcommand{\BTP}{\textbf{BTP}}
\newcommand{\MDPgenerative}{\mathcal{G}}
\newcommand{\policyvalueoperator}[3]{\mathcal{T}_{#3}^{#1}(#2)}
\newcommand{\valueoperator}[2]{\mathcal{T}^{#1}(#2)}
\newcommand{\define}{\coloneqq}
\newcommand{\qEstBO}{\textbf{QMEBO}}
\newcommand{\transpose}{\mathrm{T}}

\newcommand{\qEstone}{\textbf{QME1}}
\newcommand{\qEsttwo}{\textbf{QME2}}
\newcommand{\variance}[1]{\text{Var}[#1 \mid x \sim p]}
\newcommand{\beginproof}{\noindent\textbf{Proof. }}

\newcommand{\positiveintegerset}{\mathbb{Z}^{+}}

\ifx\papertech\undefined
  \ifx\techreport\undefined
    \newcommand{\onlytech}[1]{\ignorespaces}
    \newcommand{\onlypaper}[1]{#1}
  \else
    \newcommand{\onlytech}[1]{#1}
    \newcommand{\onlypaper}[1]{\ignorespaces}
    \makeatletter
    \def\@copyrightspace{\relax}
    \makeatother
  \fi
\else
  \newcommand{\onlytech}[1]{ {\color{blue}[(TECH ONLY) #1]} }
  \newcommand{\onlypaper}[1]{ {\color{olive}[(PAPER ONLY) #1]} }
\fi

\def\done{\hspace*{\fill} \rule{1.8mm}{2.5mm} \\ }

\newtheorem*{rem}{Remark}

\icmltitlerunning{Quantum Algorithms for Finite-horizon Markov Decision Processes}

\begin{document}

\twocolumn[
\icmltitle{Quantum Algorithms for Finite-horizon Markov Decision Processes}


\icmlsetsymbol{equal}{*}

\begin{icmlauthorlist}
\icmlauthor{Bin Luo}{CUHK}
\icmlauthor{Yuwen Huang}{CUHK}
\icmlauthor{Jonathan Allcock}{Tencent}
\icmlauthor{Xiaojun Lin}{CUHK}
\icmlauthor{Shengyu Zhang}{Tencent}
\icmlauthor{John C.S. Lui}{CUHK}
\end{icmlauthorlist}

\icmlaffiliation{CUHK}{The Chinese University of Hong Kong, Hong Kong, China}
\icmlaffiliation{Tencent}{Tencent Quantum Laboratory, Hong Kong, China}

\icmlcorrespondingauthor{Yuwen Huang}{yuwen.huang@link.cuhk.edu.hk}

\icmlkeywords{Machine Learning, ICML}

\vskip 0.3in
]



\printAffiliationsAndNotice{}  

\begin{abstract}\label{sec: abstract}
    In this work, we design quantum algorithms that are more efficient than classical algorithms to solve time-dependent and finite-horizon Markov Decision Processes (MDPs) in two distinct settings: (1) In the exact dynamics setting, where the agent has full knowledge of the environment's dynamics (i.e., transition probabilities), we prove that our \textbf{Quantum Value Iteration (QVI)} algorithm \textbf{QVI-1} achieves a quadratic speedup in the size of the action space $(A)$ compared with the classical value iteration algorithm for computing the optimal policy ($\pi^{*}$) and the optimal V-value function ($V_{0}^{*}$). Furthermore, our algorithm \textbf{QVI-2} provides an additional speedup in the size of the state space $(S)$ when obtaining near-optimal policies and V-value functions. 
    Both \textbf{QVI-1} and \textbf{QVI-2} achieve quantum query complexities that provably improve upon classical lower bounds, particularly in their dependences on $S$ and $A$.
    (2) In the generative model setting, where samples from the environment are accessible in quantum superposition, we prove that our algorithms \textbf{QVI-3} and \textbf{QVI-4} achieve improvements in sample complexity over the state-of-the-art (SOTA) classical algorithm in terms of $A$, estimation error $(\epsilon)$, and time horizon $(H)$. More importantly, we prove quantum lower bounds to show that \textbf{QVI-3} and \textbf{QVI-4} are asymptotically optimal, up to logarithmic factors, assuming a constant time horizon. 
\end{abstract}
This is the full version of \cite{bin2025quanutm}, which was presented at ICML 2025.

\section{Introduction}\label{sec: introduction}
Markov Decision Processes (MDPs) provide a mathematical framework for modeling decision-making problems in uncertain environments. They are an important framework to model discrete-time stochastic control and reinforcement learning (RL) \cite{puterman2014markov, agarwal2019reinforcement}. MDPs have been applied in fields such as networks, robotics, and operations research \cite{alsheikh2015markov, matignon2012coordinated}. Despite their wide applicability, MDPs often face significant computational challenges in practice. A key issue arises when the number of possible states or actions in the system becomes very large. In particular, \onlytech{state or action space may grow exponentially as the number of variables or components increases \cite{rust1997using}. This} \onlypaper{this} "curse of dimensionality" makes solving MDPs computationally infeasible in many practical scenarios \cite{powell2007approximate}.


Quantum computing is a new computing paradigm that harnesses the laws of quantum mechanics. For certain classes of problems, such as unstructured search \cite{grover1996fast}, prime number factoring \cite{shor1994algorithms}, optimization \cite{sidford2023quantum,jordan2005fast,liu2024quantum} and online learning \cite{he2024quantum,he2022quantum,wan2023quantum}, quantum computing demonstrates significant speedups over classical computing.
Recent advancements in quantum hardware \cite{arute2019quantum,aiquantum}
indicate that practical quantum computers can be a reality in the near future.

Given the importance of MDPs and the advancement in quantum computing, researchers have explored various quantum algorithms to reduce the time complexity of solving MDPs. In the stochastic control domain, \cite{naguleswaran2005quantum} suggested two quantum techniques that can potentially be used to accelerate classical algorithms for finite-horizon MDPs \cite{puterman2014markov}. However, this work only focused on problem formulation and did not provide a concrete quantum algorithm for finite-horizon MDPs with performance guarantee. \cite{naguleswaran2006automated} applied quantum walk \cite{magniez2007quantumwalk} to efficiently solve a specific class of MDPs, namely deterministic shortest path problems. 
However, the quantum algorithm and analysis there cannot be applied to general finite-horizon MDPs.
For RL, researchers proposed to replace subroutines of existing RL frameworks by quantum algorithms.
For example, \cite{wiedemann2022quantum} proposed to use a quantum Monte Carlo (MC) \cite{Montanaro_2015} to replace the classical MC method on policy evaluation\onlypaper{.} \onlytech{and use Grover's algorithm \cite{grover1996fast} to search the optimal policy in the whole policy space to achieve policy improvement.} However, their algorithm is  inefficient
as its quantum sample complexity  \onlypaper{is exponential with respect to $S$ in both time-dependent and time-independent settings for obtaining near-optimal policies.} \onlytech{scales as $O(A^{3SH/2}H/(\epsilon\delta))$ for time-dependent setting, and $O(A^{3S/2}H/(\epsilon\delta))$ for time-independent setting to obtain an $\epsilon$-optimal policy with probability $1-\delta$, which are exponential to $S$.}
Improved sample-complexity results that do not increase exponentially in $S$ have been obtained for infinite-horizon MDPs. For example, \cite{Cherrat_2023} utilized a quantum linear system solver \cite{chak2019block} 
to approximate Q-values during the policy evaluation. \onlytech{Policy improvement is achieved by selecting actions based on the highest occurrence frequency in repeated quantum measurements of states encoding these Q-values.} \onlytech{Based on the model-free algorithm for infinite-horizon MDPs \cite{sidford2018near}, \cite{pmlr-v139-wang21w} improved sample complexity by leveraging quantum mean estimation \cite{Montanaro_2015} and quantum maximum searching \cite{durr1999quantumalgorithmfindingminimum}. They also provided quantum lower bounds for obtaining near-optimal policies and values.} \onlypaper{\cite{pmlr-v139-wang21w} proposed nearly minimax optimal quantum algorithms for infinite-horizon MDPs by leveraging quantum mean estimation \cite{Montanaro_2015} and quantum maximum searching \cite{durr1999quantumalgorithmfindingminimum}.} Besides, \cite{cornelissen2018quantum} applied
 quantum gradient estimation \cite{gilyen2019optimizing} in policy improvement. 
 However, these algorithms are only tailored to infinite-horizon problems with a time-invariant value function, thus \textit{preventing their use in finite-horizon and time-dependent scenarios} where the value functions depend on time.
 

Thus, one open question is, \textit{can one design quantum algorithms that are more efficient than classical algorithms in obtaining the optimal or $\epsilon$-optimal policy, V-value and Q-value functions for ``finite-horizon'' and ``time-dependent'' MDPs?} We address this open question in both the exact dynamics setting and the generative model setting. Our contributions are as follows:
\begin{itemize}
\vspace{-5pt}
    \item In the exact dynamics setting (Section \ref{sec: exact dynamics setting}), we propose a \textbf{Quantum Value Iteration (QVI)} algorithm \textbf{QVI-1}, that computes the optimal policy and V-value function with a quadratic speedup in $A$ compared with the classical value iteration algorithm. Additionally, \textbf{QVI-2} achieves a further speedup in $S$ for obtaining near-optimal policies and V-value functions, enabled by our novel quantum subroutine, quantum mean estimation with binary oracles ($\qEstBO$), for mean estimation of arbitrary bounded functions. \onlytech{For MDPs with sparse transition probability matrices, \textbf{QVI-5} offers further improvements in $S$ over \textbf{QVI-2}.} Besides, we also derive new classical lower bounds for computing near-optimal policies and V-value functions. A summary of these results is provided in Table \ref{table: comparisons between classical and quantum query complexity in the exact dynamics setting}.
    \vspace{-5pt}
    \item In the generative model setting (Section \ref{sec: generative model setting}), we propose two quantum algorithms, \textbf{QVI-3} and \textbf{QVI-4}, to efficiently compute $\epsilon$-optimal policies and value functions. Compared with SOTA classical algorithms for time-dependent and finite-horizon MDPs, both \textbf{QVI-3} and \textbf{QVI-4} achieves speedups in $H$, and $\epsilon$, with \textbf{QVI-3} additionally achieving a quadratic speedups in $A$.
    \vspace{-5pt}
    \item Assuming access to a quantum generative oracle for time-dependent and finite-horizon MDPs, we establish quantum lower bounds for obtaining near-optimal policies, V-value functions, and Q-value functions. Our results demonstrate that \textbf{QVI-3} and \textbf{QVI-4} are asymptotically optimal, up to log factors, provided that $H$ is a constant.
    Further, our results also lead to a new lower bound for obtaining Q-values in the classical setting. 
    A summary of the upper and lower bounds in the generative model setting is provided in Table \ref{table: comparisons between classical and quantum sample complexity in the generative model setting}.
    \vspace{-5pt}
\end{itemize}

\section{Preliminaries}\label{sec: preliminaries}
\vspace{-5pt}

\noindent\textbf{Define notations: }For an arbitrary positive integer $n$, we define $[n]$ as the set $\{0, ..., n-1\}$. For any finite set $X$ and any vector $f\in Y^{X}$, we denote the element of $ f $ at entry $x$ by $f(x)$. For any $f\in \realnumber^{X}$, the operations $\sqrt{f}$, $|f|$, and $f^{2}$ are applied component-wise.
Given two vectors $f_{1}, f_{2}\in\realnumber^{X}$, we define  $\max\{f_{1}, f_{2}\}$ as their element-wise maximum, and write  $f_{1} \leq f_{2}$ to indicate component-wise inequality. The bold symbols $\mathbf{0}$ and $\mathbf{1}$ represent vectors of all zeros and ones, respectively, and a scalar $x$ in an equation with vectors should be interpreted as $x \cdot \mathbf{1}$. We usually identify a function $f: X\rightarrow Y$ as a vector $f\in Y^{X}$.

\begin{table*}[t]
\centering
\renewcommand{\arraystretch}{1.5}
\begin{tabular}{|c|c|c|c|}
\hline
 \multirow{2}{*}{\small Goal:} 
  & \multicolumn{2}{c|}{\small Classical query complexity} &  \small Quantum query complexity \\ \cline{2-4} 
  & \small Upper bound & \small Lower bound & \small Upper bound \\\hline
 \small Optimal $\pi^{*}$, $V^{*}_{0}$  & \small $S^{2}AH$ &  \small $S^{2}A$ [Theorem \ref{thm: classical lower bound}]
 & \small $S^{2}\sqrt{A}H$ [Theorem \ref{Thm: Complexity of QVI1}]  \\ \hline
 \multirow{2}{2.6cm}{ \centering\small  $\epsilon$-accurate estimate of $\pi^{*}$ and $\{V^{*}_{h}\}_{h=0}^{H-1}$} &\multirow{2}{*}{\small $S^{2}AH$} & \multirow{2}{*}{\small $S^{2}A$ [Theorem \ref{thm: classical lower bound}]} &\multirow{2}{*}{ $\frac{S^{1.5}\sqrt{A}H^{3}}{\epsilon}$ \small [Theorem \ref{Thm: Complexity of QVI2}]}\\ 
 & & &\\\hline
\end{tabular}
\vspace{-5pt}
\caption{Classical and quantum query complexities for solving time-dependent and finite-horizon MDPs in the exact dynamics setting. 
All quantum upper bounds are $\Tilde{O}(\cdot)$, assuming a constant failure probability $\delta$. The range of error term $\epsilon$ is $(0,H]$. The classical upper bounds \onlytech{for computing the $\epsilon$-accurate estimate of  $\pi^{*}$ and  $\{V^{*}_{h}\}_{h=0}^{H-1}$ and optimal $\pi^{*}$ and  $\{V^{*}_{h}\}_{h=0}^{H-1}$} are $O(\cdot)$, derived from the value iteration algorithm in \cite{puterman2014markov}. The classical lower bounds are $\Omega(\cdot)$, which holds for $\epsilon\in O(H)$.
}
\label{table: comparisons between classical and quantum query complexity in the exact dynamics setting}
\vspace{-5pt}
\end{table*}


\begin{table*}[t]
\centering
\renewcommand{\arraystretch}{1.9}
\begin{tabular}{|c|c|c|c|c|}
\hline
 \multirow{2}{1.7cm}{\centering\small Goal: obtain an $\epsilon$-accurate estimate of} & \multicolumn{2}{c|}{ \small Classical sample complexity} & \multicolumn{2}{c|}{\small Quantum sample complexity} \\ \cline{2-5}
 & \small Upper bound & \small Lower bound & \small Upper bound & \small Lower bound\\ 
 \hline
 $\{Q^{*}_{h}\}_{h=0}^{H-1}$ & $\frac{SAH^{4}}{\epsilon^{2}}$ & $\frac{SAH^{3}}{\epsilon^{2}}$ \small [Theorem \ref{thm: quantum lower bound for finite horizon MDP}]  & $\frac{SAH^{2.5}}{\epsilon}$ \small [Theorem \ref{Thm: Complexity of QVI4}] & $\frac{SAH^{1.5}}{\epsilon}$ \small [Theorem \ref{thm: quantum lower bound for finite horizon MDP}]\\
\multirow{2}{*}{ $\pi^{*}, \{V^{*}_{h}\}_{h=0}^{H-1}$} & \multirow{2}{*}{$\frac{SAH^{4}}{\epsilon^{2}}$} & \multirow{2}{*}{$\frac{SAH^{3}}{\epsilon^{2}}$ \small [Theorem \ref{thm: quantum lower bound for finite horizon MDP}]} & $\frac{SAH^{2.5}}{\epsilon}$ \small [Theorem \ref{Thm: Complexity of QVI4}] & \multirow{2}{*}{$\frac{S\sqrt{A}H^{1.5}}{\epsilon}$ \small [Theorem \ref{thm: quantum lower bound for finite horizon MDP}]}\\
   &   & & $\frac{S\sqrt{A}H^{3}}{\epsilon}$ \small [Theorem \ref{Thm: Complexity of QVI3}] &\\ 
   \hline
\end{tabular}
\vspace{-5pt}
\caption{Classical and quantum sample complexities for solving time-dependent and finite-horizon MDPs in the generative model setting.  All bounds assume a constant maximum failure probability \(\delta\). All upper bounds are $\tilde{O}(\cdot)$, which requires $\epsilon\in O(1/\sqrt{H})$ for [Theorem \ref{Thm: Complexity of QVI4}] and $\epsilon\in (0, H]$ for [Theorem \ref{Thm: Complexity of QVI3}]. All lower bounds are $\tilde{\Omega}(\cdot)$, which holds for $\epsilon\in (0, 1/2)$. The classical upper bounds \onlypaper{for all goals} \onlytech{for computing the $\epsilon$-optimal policy (\(\pi^*\)) and value functions ($\{V_{h}^*\}_{h=0}^{H-1}$ and $\{Q_{h}^{*}\}_{h=0}^{H-1}$)} were shown in \cite{li2020breaking}. The classical lower bound for $\pi^{*}$ and $\{V_{h}^*\}_{h=0}^{H-1}$ was shown in \cite{sidford2018near}. 
}
\label{table: comparisons between classical and quantum sample complexity in the generative model setting}
\vspace{-12pt}
\end{table*}


\noindent\textbf{MDP Preliminaries: } We study time-dependent and finite-horizon MDPs in two settings: (a) the exact dynamics setting (Section \ref{sec: exact dynamics setting}) and (b) the generative model setting (Section \ref{sec: generative model setting}). In both settings, the MDP has a finite and discrete state space $\statespace$ and action space $\actionspace$. In each time step $h\in[H]$, an agent need to decide which action $a\in\actionspace$ to take for each state $s\in\statespace$. 
After taking the action $a$ at the state $s$ in the time step $h\in [H]$, the agent obtains a reward $r_{h}(s, a)\in [0,1]$ and transitions to the next state $s'\in\statespace$ with probability $P_{h}(s'|s, a)$.
We define a finite-horizon and time-dependent MDP as a 5-tuple $\mathcal{M}=(\statespace, \actionspace, \{P_{h}\}_{h=0}^{H-1}, \{r_{h}\}^{H-1}_{h=0}, H)$. \onlytech{For simplicity, we} \onlypaper{We} define $S\define|\statespace|$ and $A\define|\actionspace|$, which are the cardinalities of $\statespace$ and $\actionspace$ respectively. A policy $\pi$ is a mapping from $\statespace\times [H]$ to $\actionspace$, where $\pi(s,h)$ specifies the action that the agent should take in the state $s$ at the time step $h$. The policy space is defined as $\Pi\define\actionspace^{\statespace \times [H]}$. In MDPs, the objective of the agent is to find a policy $\pi$ that maximizes the expected cumulative reward over $H$ time horizon. This can be written as maximizing the V-value function, $V_{h}^{\pi}: \statespace\rightarrow \realnumber$, at each time step $h$.
Specifically, the V-value function at time $h$ for an initial state $s$ under a policy $\pi$ is defined as 
$ \V{\pi}{h}(s)\define\Expectation\bigl[ \sum_{t=h}^{H-1} r_{t}(s_{t}, a_{t})|\pi, s_{h}=s \bigr],$
where $a_{t}=\pi(s_{t}, t)$\onlypaper{.} \onlytech{is the action taken at a state $s_{t}$ at time $t$.} 
Similarly, the Q-value function $\Q{\pi}{h}:\statespace\times\actionspace\rightarrow\realnumber$ is defined as $\Q{\pi}{h}(s, a)\define\Expectation\bigl[ \sum_{t=h}^{H-1}r_{t}(s_{t}, a_{t}) \bigl| \pi, s_{h}=s, a_{h}=a \bigr].$

For a policy $\pi$, we define $P_{h}^{\pi}\in \realnumber^{\statespace\actionspace\times \statespace\actionspace}$ as the matrix with entries $P_{h}^{\pi}\left((s,a), (s',a')\right)= P_{h}(s'|s, a)$ if $a'=\pi(s')$ and $0$ otherwise.
For any $Q\in\realnumber^{\statespace\times \actionspace}$, we define $P_{h}^{\pi}Q\in\realnumber^{\statespace\times\actionspace}$ as $(P_{h}^{\pi}Q)(s,a)=\sum_{s'\in\statespace}P_{h}(s'|s,a)Q\bigl(s',\pi(s') \bigr)$.
With a slight abuse of notation, we define $P_{h}\in \realnumber^{\statespace\actionspace\times \statespace}$ as the matrix satisfying $P_{h}\left((s,a),s'\right)=P_{h}(s'|s,a)$ for any $h\in[H]$. For any fixed $s\in\statespace, a\in\actionspace$ and $h\in [H]$, we define $\Phsa\in\realnumber^{\statespace}$ as the vector satisfying $\Phsa(s')=P_{h}(s'|s,a)$. Therefore, we can express $\Expectation[f(s')|s' \sim \Phsa]=\Phsa^{\transpose}f$ for any $f\in\realnumber^{\statespace}$. 

For any vector $v\in \realnumber^{\statespace}$, we define $\sigma^{2}_{h}(v)\in\realnumber^{\statespace\times\actionspace}$ as a vector satisfying $[\sigma^{2}_{h}(v)](s,a)\define \text{Var}[v(s')|s'\sim P_{h}(\cdot|s,a)]$ for any $h\in[H]$. In the vector notation, it can be written as $\sigma^{2}_{h}(v)=P_{h}v^{2}- (P_{h} v)^{2}$. We also define $\sigma_{h}(v)=\sqrt{\sigma_{h}^{2}(v)}$.

We define $V(Q)\in\realnumber^{\statespace}$ as $[V(Q)]_{s}=\max_{a\in\actionspace}\{Q(s,a)\}$ and $\pi(Q)\in \actionspace^{\statespace}$ as $[\pi(Q)]_{s}=\arg\max_{a\in\actionspace}\{Q(s,a)\}$ for any vector $Q\in\realnumber^{\statespace\times\actionspace}$. \onlytech{With this expression, we have $[V(Q)]_{s}=Q(s, [\pi(Q)]_{s})$.}

Below, we provide formal definitions for some important concepts in the finite-horizon MDP $\MDP$.
\begin{definition}[Value operator associated with a policy]\label{def: value operator associated policy} 
For any policy $\pi\in\Pi$, let $\policyvalueoperator{h}{\cdot}{\pi}$ be the value operator associated with $\pi$ such that, for all $u\in\realnumber^{\statespace}$, $h\in[H]$ and $s\in\statespace$,  $[\policyvalueoperator{h}{u}{\pi}]_{s}\define r\bigl( s,\pi(s,h) \bigr)+P_{h|s,\pi(s,h)}^{\transpose}u$. We let $\{V^{\pi}_{h}\}_{h=0}^{H-1}$ denote the V-value functions of policy $\pi$, which satisfies $\policyvalueoperator{h}{V_{h+1}^{\pi}}{\pi}=V_{h}^{\pi}$ for all $h\in[H]$.
\end{definition}

\begin{definition}[Optimal value and policy]\label{def: optimal value and policy} 
Define the optimal value of an initial state $s\in\statespace$ at each time step $h\in[H]$ of the finite-horizon MDP $\MDP$ as $V^{*}_{h}(s)\define\max_{\pi\in\Pi} V^{\pi}_{h}(s)$. A policy $\pi$ is said to be an optimal policy $\pi^{*}$ if $V^{\pi}_{0}=V^{*}_{0}$. Similarly, we can also define the optimal value of an initial pair of $(s,a)\in\statespace\times\actionspace$ at each time step $h\in[H]$ as $Q^{*}_{h}(s,a)\define \max_{\pi\in\Pi} Q^{\pi}_{h}(s,a)$. 
\end{definition}

\begin{definition}[$\epsilon$-optimal value function and policy]\label{def: epsilon optimal value and policy} 
We say that V-value functions $\{V_{h}\}_{h=0}^{H-1}$ are $\epsilon$-optimal if $\infiNorm{V_{h}^{*}-V_{h}}\leq \epsilon$ for all $h\in[H]$ and a policy $\pi\in\Pi$ is $\epsilon$-optimal if $\infiNorm{V_{h}^{*}-V_{h}^{\pi}}\leq \epsilon$ for all $h\in[H]$, which implies the V-value functions of $\pi$ are $\epsilon$-optimal. Similarly, we say that Q-value functions $\{Q_{h}\}_{h=0}^{H-1}$ are $\epsilon$-optimal if $\infiNorm{Q^{*}_{h}-Q_{h}}\leq \epsilon$ for all $h\in[H]$.
\end{definition}
\vspace{-5pt}
\noindent\textbf{Quantum Preliminaries: }
\onlypaper{Before introducing our quantum algorithms,  a brief overview of Dirac notation \cite{nielsen2010quantum} is given to ensure clarity.}
\onlytech{Before introducing our quantum algorithms, we provide a formal definition of a quantum oracle using the Dirac notation of quantum computing \cite{nielsen2010quantum}. Below, a brief overview of Dirac notation is given to ensure clarity.}
In Dirac notation, vectors $v$ in a complex vector space $\ComplexSpace{n}$ are represented as $\ket{v}$. The symbol $\ket{i}$, where $i\in [n]$, denotes the $i+1$-th standard basis vector, with $\ket{0}$ typically reserved for the first standard basis vector. In this paper, real numbers are encoded in the computational basis using a fixed-point binary representation with precision $2^{-p}$. Specifically, a real number $k$ is encoded as $\ket{\fixbinary{k}}=\ket{k_{1}\ldots k_{q}}\in\ComplexSpace{^{2^{q}}}$, where $k_{1}\ldots k_{q}=k_{1}\ldots k_{q-p}.k_{q-p+1}\ldots k_{q}$ is the binary string of $k$.
\onlytech{Throughout this paper, we} \onlypaper{We} assume that $q$ and $p$ are sufficiently large so that there is no overflow in storing real numbers. 

We now define a quantum oracle for arbitrary functions and vectors, which is often referred to as \textit{binary oracle}.
\begin{definition}[Quantum oracle for functions and vectors]\label{Def: Quantum oracle encoding of functions and vectors}
     Let $\Omega$ be a finite set of size $N$ and  $f\in \realnumber^{\Omega}$. A quantum oracle encoding $f$ is a unitary operator $\BinaryOracle{f}: \ComplexSpace{N}\tensorproduct\ComplexSpace{2^{q}}\rightarrow \ComplexSpace{N}\tensorproduct\ComplexSpace{2^{q}}$ such that 
     $\BinaryOracle{f}: \ket{i}\ket{0}\mapsto\ket{i}\ket{\fixbinary{f(i)}}$
    for all $i\in [N]$, where $\fixbinary{f(i)}$ is the binary representation of $f(i)$ with precision $2^{-p}$.
\end{definition}


\vspace{-10pt}
\section{Exact Dynamics Setting}\label{sec: exact dynamics setting}
In this setting, it is assumed that the environment's dynamics are fully known, i.e., the transition probability matrix $P_{h}$ at each time step $h$ is explicitly provided for the entire state-action space. To formalize this assumption, we introduce the classical oracle for finite-horizon MDPs $\classicalMDPoracle$ in Definition \ref{def: oracle of classical finite horizon MDPs}. Given this classical oracle, 
the classical value iteration algorithm (Algorithm \ref{algo: VI-finite horizon MDP-optimal action and V value}) can obtain an optimal policy $\pi^{*}$ and optimal value $V^{*}_{0}(s)$ for any initial state $s\in\statespace$ with $O(S^{2}AH)$ queries to the oracle $\classicalMDPoracle$ \cite{bellman1958dynamic}.

\begin{definition}[Classical oracle of an MDP]\label{def: oracle of classical finite horizon MDPs}
    We define a classical oracle $\classicalMDPoracle: \statespace\times\actionspace\times [H]\times \statespace \rightarrow [0,1]\times [0,1]$ for a time-dependent and finite-horizon MDP $\MDP$ satisfying 
    $\classicalMDPoracle: (s, a, h, s') \mapsto \big(r_{h}(s, a),\Phsa(s')\big).$
    \vspace{-5pt}
\end{definition}

To understand the limits of classical algorithms under this setting, we establish a lower bound on query complexity for computing near-optimal policies and V-value functions. This result adapts the techniques developed for infinite-horizon MDPs in \cite{chen2017lowerboundcomputationalcomplexity} to the finite-horizon case. The rigorous proof of Theorem \ref{thm: classical lower bound} is presented in Appendix \ref{appendix: classical lower bounds}.
\begin{theorem}[Classical lower bounds]\label{thm: classical lower bound}
     Let $\statespace$ and $\actionspace$ be finite sets of states and actions. Let $H\geq 2$ be a positive integer and $\epsilon\in(0, \frac{H-1}{4})$ be an error parameter. We consider the following time-dependent and finite-horizon MDP $\MDP=(\statespace, \actionspace, \{P_{h}\}_{h=0}^{H-1}, \{r_{h}\}_{h=0}^{H-1}, H)$, where $r_{h}\in[0,1]^{\statespace\times\actionspace}$ for all $h\in[H]$. 
    Given access to the classical oracle $\classicalMDPoracle$, any algorithm $\mathcal{K}$, which takes $\MDP$ as an input and outputs $\epsilon$-approximations of $\{V^{*}_{h}\}_{h=0}^{H-1}$ or $\pi^{*}$ with probability at least $0.9$, must require at least  $\Omega(S^{2}A)$ queries to $\classicalMDPoracle$
     on the worst case of input $\MDP$.
\end{theorem}

\vspace{-10pt}
\subsection{Speedup on $A$}\label{subsec: speedup on A}
Having established the classical baseline, we now turn to investigating whether quantum algorithms can offer improvements, particularly in the dependence on the action space size $A$. To have a fair comparison on the time complexity between a classical algorithm and a quantum algorithm, we first define the quantum analog of the classical oracle $\classicalMDPoracle$. 

\begin{definition}[Quantum oracle of an MDP]\label{def: quantum oracle of MDP}
     A quantum oracle of an MDP $\MDP$ is a unitary operator $\MDPQoracle: \ComplexSpace{S}\tensorproduct\ComplexSpace{A}\tensorproduct \ComplexSpace{H}\tensorproduct\ComplexSpace{S}\tensorproduct\ComplexSpace{2^{q}}\tensorproduct\ComplexSpace{2^{q}}\rightarrow\ComplexSpace{S}\tensorproduct\ComplexSpace{A}\tensorproduct \ComplexSpace{H}\tensorproduct\ComplexSpace{S}\tensorproduct\ComplexSpace{2^{q}}\tensorproduct\ComplexSpace{2^{q}}$ such that
         \vspace{-5pt}
    \begin{equation}
            \begin{aligned}
        &\MDPQoracle: \ket{s}\ket{a}\ket{h}\ket{s'}\ket{0}\ket{0}\\
    &\mapsto\ket{s}\ket{a}\ket{h}\ket{s'}\ket{\fixbinary{r_{h}(s,a)}}\ket{\fixbinary{\Phsa(s')}},
        \end{aligned}    
        \vspace{-5pt}
    \end{equation}
    for all $(s, a, h, s')\in\statespace \times \actionspace\times [H]\times\statespace$, where $\fixbinary{r_{h}(s,a)}$ and $\fixbinary{\Phsa(s')}$ denote the binary representation of $r_{h}(s,a)$ and $\Phsa(s')$ with precision $2^{-p}$.
    \vspace{-5pt}
\end{definition}

We define the number of queries made to the quantum oracle $\MDPQoracle$ or classical oracle $\classicalMDPoracle$ as the quantum or classical query complexity, respectively. Comparing quantum and classical time complexities can be achieved by examining their respective query complexities, because implementing $\MDPQoracle$ has comparable overhead as $\classicalMDPoracle$. Specifically, given a Boolean circuit of $\classicalMDPoracle$ with $N$ logic gates, it can be converted into a quantum circuit of  $\MDPQoracle$ with $O(N)$ quantum gates. This conversion can be efficiently achieved by simple conversion rules at the logic gate level \cite{nielsen2010quantum}. Therefore, $\MDPQoracle$ and $\classicalMDPoracle$ have comparable costs at the elementary gate level. Then, if the classical oracle $\classicalMDPoracle$ can be called in constant time, the quantum oracle $\MDPQoracle$ can be called in constant time as well. Under this assumption, query complexity directly reflects the time complexity for both the classical and quantum algorithms.

With the quantum oracle $\MDPQoracle$, our objective is to design quantum algorithms that can compute $\pi^{*}$ and $V^{*}_{0}(s)$ for all $s\in\statespace$ with probability at least $1-\delta$, while minimizing the total number of queries to $\MDPQoracle$.



\begin{algorithm}[t]
\caption{Quantum Value Iteration $\QVIone$}
\label{algo: quantum VI-finite horizon MDP-optimal action and V value v1}
\begin{algorithmic}[1]
\STATE \textbf{Require:} MDP $\MDP$, quantum oracle $\MDPQoracle$, maximum failure probability $\delta\in(0,1)$.
\STATE \textbf{Initialize:} $\zeta\leftarrow\delta/(SH)$, $\hat{V}_{H}\gets \mathbf{0}$.
\FOR{$h := H-1, \ldots, 0$}
    \STATE create a quantum oracle $\BinaryOracle{\hat{V}_{h+1}}$ for vector $\hat{V}_{h+1}\in \realnumber^{\statespace}$
    \STATE $\forall s\in\statespace$: create a quantum oracle $\BinaryOracle{\hat{Q}_{h,s}}$ encoding  vector $\hat{Q}_{h,s}\in \realnumber^{\actionspace}$ with $\MDPQoracle$ and $\BinaryOracle{\hat{V}_{h+1}}$ satisfying \\ \quad $\hat{Q}_{h,s}(a)\leftarrow r_{h}(s, a) + \Phsa^{\transpose}\hat{V}_{h+1}$
    \STATE $\forall s\in\statespace$: $\hat{\pi}(s,h) \leftarrow \text{\qArgmax}_{\zeta}\{\hat{Q}_{h,s}(a): a\in \actionspace\}$ 
    \STATE $\forall s\in\statespace$: $\hat{V}_{h}(s) \leftarrow \hat{Q}_{h,s}\bigl( \hat{\pi}(s,h) \bigr)$ 
\ENDFOR
\STATE \textbf{Return:} $\hat{\pi}$, $\hat{V}_{0}$
\end{algorithmic}
\end{algorithm}
\setalgospace

We first introduce an existing quantum subroutine, quantum maximum searching algorithm \cite{durr1999quantumalgorithmfindingminimum}, which can efficiently find the maximum of a list of unsorted $N\in\positiveintegerset$ numbers using only $O(\sqrt{N})$ queries to that list. In contrast, the best-possible classical algorithm must examine all $N$ elements in the worst case to find the maximum. 

\begin{theorem}[Quantum maximum searching \cite{durr1999quantumalgorithmfindingminimum}]\label{Thm: quantum maximum finding}
\vspace{-5pt}
Let $\BinaryOracle{f}$ be a quantum oracle encoding a vector $f \in \mathbb{R}^N$, $N\in\positiveintegerset$. There exists a quantum maximum searching algorithm, $\qArgmax$, which, for any $\delta > 0$, can identify an index $i$ such that $f(i)$ is the maximum value in $f$, with a success probability of at least $1 - \delta$. The algorithm requires at most $\Tilde{c} \sqrt{N} \log(1 / \delta)$ queries to $\BinaryOracle{f}$, where $\Tilde{c} > 0$ is a constant.
\vspace{-5pt}
\end{theorem}
We use $\qArgmax_{\delta}\{f(i): i\in[N]\}$ to denote the process of finding the index of the maximum value of a vector $f$ using $\qArgmax$, with a success probability at least $1-\delta$.
Note that the classical value iteration algorithm needs to take the maximum over the whole action space in the Bellman recursion to obtain the estimates of optimal V-value function $V^{*}_{h}$ and optimal action $\pi^{*}(s,h)$ for state $s$ at time stage $h$. We incorporate $\qArgmax$ in this step to reduce the query complexity from $O(A)$ to $O(\sqrt{A})$. Now, we propose our quantum value iteration algorithm $\textbf{QVI-1}$ in Algorithm \ref{algo: quantum VI-finite horizon MDP-optimal action and V value v1}\onlytech{, which takes an MDP $\MDP$ and error term $\epsilon$ as inputs and returns the estimates of optimal policy and V-value}. In order to use $\qArgmax$ correctly, one needs to suitably encode the vector $\hat{V}_{h+1}$ and $\hat{Q}_{h,s}$ with the binary oracles. In summary, $\textbf{QVI-1}$ returns an optimal policy and optimal values (Theorem \ref{Thm: correctness of QVI1}) but only requires $\Tilde{O}(S^{2}\sqrt{A}H)$ queries to the quantum oracle $\MDPQoracle$ (Theorem \ref{Thm: Complexity of QVI1}). 
The proof of Theorems \ref{Thm: correctness of QVI1} and \ref{Thm: Complexity of QVI1} can be found in Appendix \ref{appendix: QVIone}, where we also analyze the cost of the qubit resources of $\textbf{QVI-1}$.

\begin{theorem}[Correctness of \textbf{QVI-1}]\label{Thm: correctness of QVI1}
The outputs $\hat{\pi}$ and $\hat{V}_{0}$ satisfy that $\hat{\pi}=\pi^{*}$ and $\hat{V}_{0}=V_{0}^{*}$ with a success probability at least $1-\delta$.
\end{theorem}

\begin{theorem}[Complexity of \textbf{QVI-1}]\label{Thm: Complexity of QVI1}
The quantum query complexity of \textbf{QVI-1} in terms of the quantum oracle 
$\MDPQoracle$ is 
$O\bigl( S^{2}\sqrt{A}H\log(SH/\delta) \bigr)$.
\vspace{-5pt}
\end{theorem}

\vspace{-7pt}
\subsection{Speedup on $S$}
Since \textbf{QVI-1} achieves a speedup in the action space size $A$,
it is advantageous for problems with a large action space, such as natural language processing, where each text in a large dictionary corresponds to a distinct action \cite{feng2024naturallanguagereinforcementlearning}.
However, in problems modeled by numerous variables, such as Chess or Go, where each position in a vast board is represented as a state,
the state space can be much larger than the action space and time horizon \cite{bellman1962dynamic}.
In such scenarios, $\textbf{QVI-1}$ may not be suitable due to its complexity of $O(S^{2})$. This complexity arises for two reasons: (1) one needs to update $O(S)$ Q-value functions at each time step; (2) computing the ``\textit{precise mean}'' of the V-value function from the last time step needs $O(S)$ queries to the oracle $\MDPQoracle$ when updating each Q-value function. Note that for obtaining an ``\textit{$\epsilon$-estimation of the mean}'' of $n$ Boolean variables, quantum algorithms only need $\Theta(\min\{\epsilon^{-1}, n\})$ queries to a binary oracle \cite{Nayak1999mean,beals1998quantumlowerboundspolynomials}.
This suggests that a quantum speedup in $S$ may be achievable if one is satisfied with a near-optimal policy.
Therefore, next we investigate \textit{whether there exists \onlytech{an error-bounded} \onlypaper{a} quantum algorithm that can obtain $\epsilon$-optimal policies and V-value functions for an MDP $\MDP$ but only requires 
$\Tilde{O}\bigl( S^{c}\text{poly}(\sqrt{A},H,\epsilon^{-1}) \bigr)$ queries to $\MDPQoracle$, where $0<c<2$.}

To achieve this optimization goal, we propose \textbf{QVI-2} in Algorithm \ref{algo: quantum VI-finite horizon MDP-optimal action and V value v2}, where the quantum subroutine \textbf{QMEBO}, as used in the fifth step, is defined in Algorithm \ref{algo: quantum mean estimation with binary oracle}. \onlytech{, which takes an MDP $\MDP$, error term $\epsilon$ and failure probability $\delta$ as inputs and returns $\epsilon$-optimal policy $\hat{\pi}$ and values $\{\hat{V}_{h}\}_{h=0}^{H-1}$} The main difference between \textbf{QVI-1} and \textbf{QVI-2} is that we compute an estimate of the expectation of $\Phsa^{\transpose} \hat{V}_{h+1}$ rather than its precise value in each time step $h$ in \textbf{QVI-2}.
Since the oracle $\MDPQoracle$ that encodes the probability distribution $\Phsa$ is a binary oracle, we cannot directly apply the existing quantum mean estimation algorithms \cite{Montanaro_2015}, which require an oracle that encodes the probability distribution in the amplitude (See Theorem \ref{thm: quantum mean estimation}). 
Hence, we design a new quantum subroutine in Algorithm \ref{algo: quantum mean estimation with binary oracle}, denoted as quantum mean estimation with binary oracles (\textbf{QMEBO}).

\begin{theorem}[Quantum mean estimation with binary oracles]\label{thm: quantum mean estimation with binary oracle}
Let $\Omega$ be a finite set with cardinality $N$, $p=(p_{x})_{x\in\Omega}$ a discrete probability distribution over $\Omega$, and $f: \Omega\rightarrow \realnumber$ a function. Suppose we have access to a binary oracle $\BinaryOracle{p}$ encoding the probability distribution $p$ and a binary oracle $\BinaryOracle{f}$ encoding the function $f$. If the function $f$ satisfies $f(x)\in [0, 1]$ for all $x\in\Omega$, then the algorithm $\textbf{QMEBO}$ requires $O\bigl( (\frac{\sqrt{N}}{\epsilon}+\sqrt{\frac{N}{\epsilon}})\log(1/\delta) \bigr)$ queries to $\BinaryOracle{p}$ and $\BinaryOracle{f}$ to output an estimate $\hat{\mu}$ of $ \mu\define\Expectation[f(x)|x \sim p]=p^{\transpose}f$
such that $Pr(|\Tilde{\mu}-\mu|<\epsilon)>1-\delta$ for any $\delta>0$. 
\end{theorem}

\begin{algorithm}[t]
\caption{Quantum Value Iteration $\QVItwo$}
\label{algo: quantum VI-finite horizon MDP-optimal action and V value v2}
\begin{algorithmic}[1]
\STATE \textbf{Require:} MDP $\MDP$, quantum oracle $\MDPQoracle$, maximum error $\epsilon\in (0, H]$, failure probability $\delta\in(0,1)$.
\STATE \textbf{Initialize:} 
$\zeta\gets \delta/\bigl( 4\Tilde{c}SA^{1.5}H\log(1/\delta) \bigr)$,
$\hat{V}_{H}\gets \mathbf{0}$.
\FOR{$h := H-1, \ldots, 0$}
    \STATE create a quantum oracle $\BinaryOracle{\Tilde{V}_{h+1}}$ encoding $\Tilde{V}_{h+1}\in [0,1]^{\statespace}$ defined by $\Tilde{V}_{h+1}\leftarrow\hat{V}_{h+1}/H$ 
    \STATE $\forall s\in\statespace$: create a quantum oracle $\BinaryOracle{z_{h,s}}$ encoding $z_{h,s}\in\realnumber^{\actionspace}$ defined by\\
     {\tiny $z_{h,s}(a)\gets H\cdot {\scriptsize\qEstBO}_{\zeta}(\Phsa^{\transpose}\Tilde{V}_{h+1}, \MDPQoracle, \BinaryOracle{\Tilde{V}_{h+1}}, \frac{\epsilon}{2H^{2}})-\frac{\epsilon}{2H}$}
    \STATE $\forall s\in\statespace$: create quantum oracle $\BinaryOracle{\hat{Q}_{h,s}}$ encoding $\hat{Q}_{h,s}\in \realnumber^{\actionspace}$ with $\MDPQoracle$ and $\BinaryOracle{z_{h,s}}$ satisfying \\ \quad $\hat{Q}_{h,s}(a)\leftarrow \max\{r_{h}(s, a) + z_{h,s}(a), 0\}$
    \STATE $\forall s\in\statespace$: $\hat{\pi}(s,h) \leftarrow \qArgmax_{\delta}\{\hat{Q}_{h,s}(a): a\in \actionspace\}$ 
    \STATE $\forall s\in\statespace$: $\hat{V}_{h}(s) \leftarrow \hat{Q}_{h,s}\bigl( \hat{\pi}(s,h) \bigr)$ 
\ENDFOR
\STATE \textbf{Return:} $\hat{\pi}$, $\{\hat{V}_{h}\}_{h=0}^{H-1}$
\end{algorithmic}
\end{algorithm}

\vspace{-5pt}
We denote $\qEstBO_{\delta}(p^{\transpose}f, \BinaryOracle{p}, \BinaryOracle{f},\epsilon)$ as an estimate of \onlytech{$\mathbb{E}[f(x)|x\sim p]=p^{\transpose}f$} \onlypaper{$p^{\transpose}f$}, to an error less than $\epsilon$ with probability at least $1-\delta$ obtained by \textbf{QMEBO}. The key step of \textbf{QMEBO} lies in line 4, where
a binary oracle $\BinaryOracle{p}$ is transformed into a unitary oracle $\hat{U}_{p}$. 
Unlike $\BinaryOracle{p}$, $\hat{U}_{p}$ encodes the information of the probability distribution $p$ in amplitude rather than in quantum state.
Using $\hat{U}_{p}$, we prepare the state $\ket{\psi^{(0)}}$ defined as 
\vspace{-5pt}
\begin{equation}
\vspace{-5pt}
\frac{1}{\sqrt{N}}\sum_{i=1}^{N}\sqrt{p_{i}}\ket{i}\ket{0}+\sqrt{\frac{N-1}{N}}\sum_{i=1}^{N}\sqrt{\frac{1-p_{i}}{N-1}}\ket{i}\ket{1}.
\end{equation}
The transformation and the required query complexity are presented in Theorem \ref{thm: convert binary oracle to a sub probability oracle}. 
After encoding the function $f$ in the amplitudes (lines 5-6), the amplitude estimation (Theorem \ref{thm: amplitude estimation}) is applied to compute an estimate $\mu_{k}$ of $p^{\transpose}f/N$ with an error of $\epsilon/N$ in the loop $k$. Finally, guaranteed by the Powering lemma (Lemma \ref{lem: powering lemma}), the output $\hat{\mu}$ is an $\epsilon$-estimate of $p^{\transpose}f$ with probability at least $1-\delta$. The complete version and full analysis of \textbf{QMEBO} is presented in Appendix \ref{appendix: QVItwo}.

\begin{algorithm}[t]
\caption{Quantum Mean Estimation with Binary Oracles $\qEstBO_{\delta}(p^{\transpose}f, \BinaryOracle{p},\BinaryOracle{f}, \epsilon)$}
\label{algo: quantum mean estimation with binary oracle}
\begin{algorithmic}[1]
\STATE \textbf{Require:} $\BinaryOracle{p}$ encoding a probability distribution $p=(p_{i})_{i\in\Omega}$ on a finite set $\Omega$ with cardinality $N$, $\BinaryOracle{f}$ encoding a function $f=(f_{i})_{i\in\Omega}$ where $f_{i}\in [0, 1]$, maximum error $\epsilon$, maximum failure probability $\delta\in(0,1)$.
\STATE \textbf{Initialize:} $K=O(\log1/\delta)$, $T=O(\frac{\sqrt{N}}{\epsilon}+\sqrt{\frac{N}{\epsilon}})$

\FOR{$k\in [K]$}
    \STATE Prepare state $\ket{\psi^{(0)}}=\hat{U}_{p}\ket{0}\ket{0}$ using $\BinaryOracle{p}$
    \STATE Attach $\ket{0}^{\tensorproduct (q+1)}$ qubits on $\ket{\psi^{(0)}}$ and apply $\BinaryOracle{f}$\\ 
    $\small\ket{\psi^{(1)}}=\frac{1}{\sqrt{N}}\sum_{i=1}^{N}\sqrt{p_{i}}\ket{i}\ket{0}\ket{\fixbinary{f_{i}}}\ket{0}+\ket{\Phi^{(1)}}$
    \STATE Perform controlled rotation $R_{f}$ based on $\ket{\fixbinary{f_{i}}}$ and revert $\BinaryOracle{f}$\\
    $\ket{\psi^{(2)}}=\frac{1}{\sqrt{N}}\sum_{i=1}^{N}\sqrt{p_{i}f_{i}}\ket{i}\ket{000}+\ket{\Phi^{(2)}}$
    \STATE Apply $T$ iterations of amplitude estimation with state $\ket{\psi}=\ket{\psi^{(2)}}$, operator $ U=2\ket{\psi}\bra{\psi}-I$, and projector $P=I\tensorproduct \ket{000}\bra{000}$ to obtain $\mu_{k}$
\ENDFOR
\STATE \textbf{Return:} $\hat{\mu}=N\cdot \text{Median}(\{\mu_{k}\}_{k\in[K]})$
\end{algorithmic}
\end{algorithm}

With $\qEstBO$, it only requires $O(\sqrt{S}/\epsilon)$ queries to the oracle $\MDPQoracle$ to obtain an $\epsilon$-estimate of $\Phsa^{\transpose}V$ for any $V\in[0,1]^{\statespace}$. Compared with computing the precise value of $\Phsa^{\transpose}V$ with $O(S)$ queries to $\MDPQoracle$, $\qEstBO$ reduces the query complexity from $O(S)$ to $O(\sqrt{S})$. \onlytech{This is the source of the speedup in $S$.} Finally, \textbf{QVI-2} only requires 
$\Tilde{O}\bigl(S^{1.5}\text{poly}(\sqrt{A},H,1/\epsilon) \bigr) $ queries to $\MDPQoracle$ (Theorem \ref{Thm: Complexity of QVI2}). By suitably controlling the error induced by $\qEstBO$\onlytech{in each time step}, one can ensure that \textbf{QVI-2} can obtain $\epsilon$-optimal policies \onlytech{$\hat{\pi}$} and V-value functions \onlytech{$\{\hat{V}_{h}\}_{h=0}^{H-1}$} (Theorem \ref{Thm: correctness of QVI2}). Note that we subtract the $H$ times error induced by $\qEstBO$ in line~5 of \textbf{QVI-2} which allows the estimates $z_{h,s}(a)$ to have an one-sided error. This is a variant of \textit{the monotonicity technique} which was originally proposed for solving the infinite-horizon MDPs more efficiently \cite{sidford2018near}. This technique ensures that the value function $\hat{V}_{h}$ is bounded by the value function of the policy $V^{\hat{\pi}}_{h}$ at the same time step. 

\begin{theorem}[Correctness of \textbf{QVI-2}]\label{Thm: correctness of QVI2}
The outputs $\hat{\pi}$ and $\{\hat{V}_{h}\}_{h=0}^{H-1}$ satisfy that $V_{h}^{*}-\epsilon\leq \hat{V}_{h}\leq V^{\hat{\pi}}_{h}\leq V_{h}^{*}$
for all $h\in[H]$ with a success probability at least $1-\delta$.
\end{theorem}

\begin{theorem}[Complexity of \textbf{QVI-2}]\label{Thm: Complexity of QVI2}
The quantum query complexity of \textbf{QVI-2} in terms of the quantum oracle of MDPs $\MDPQoracle$ is 
\vspace{-5pt}
\begin{equation}
     O\bigl( \frac{S^{1.5}\sqrt{A}H^{3}\log(SA^{1.5}H/\delta)}{\epsilon} \bigr).
\end{equation}
\end{theorem}

\onlytech{\vspace{-5pt}
Finally, we note that \textbf{QVI-2} is not the only solution. For \textit{a broad class of MDPs whose transition matrices $\{P_{h}\}_{h=0}^{H-1}$ satisfy certain sparse property}, we propose \textbf{QVI-5} in Algorithm \ref{algo: quantum VI-finite horizon MDP-optimal action and V value v5} to achieve a further speedup in $S$ over \textbf{QVI-2}. (See the full discussion in Appendix \ref{appendix: remark on QVIfive})}

\section{Generative Model Setting}\label{sec: generative model setting}
Even though the exact dynamic model allows precise calculation of optimal policies and values, such a model is not always readily available in a complex environment. In this section, we focus on the generative model setting as studied in \cite{li2020breaking}.
Specifically, we assume that the agent lacks full access to the transition probabilities but can query a generative model to sample transitions for specific state-action pairs. 
Note that similar models have often been used in the classical setting, where one is assumed
to have access to a generative model $G$, which can generate $N$ independent samples for each $(s, a, h)\in\statespace\times\actionspace\times[H]$ satisfying
\vspace{-5pt}
\begin{equation}\label{equ: classical generative model}
\vspace{-5pt}
    s_{h}^{i}(s, a)\overset{\text{i.i.d.}}{\sim} \Phsa, \quad i=1,\ldots, N.
\end{equation}
Correspondingly, we define a quantum generative model $\MDPgenerative$ for an MDP $\MDP$ in Definition \ref{def: quantum generative model of an MDP}. It is important to point out that the quantum state output by $\MDPgenerative$ is similar to a sample drawn from the probability distribution $\Phsa$ in Eq. \eqref{equ: classical generative model}.  
\begin{definition}[Quantum generative model of an MDP]\label{def: quantum generative model of an MDP}
    The quantum generative model of a time-dependent and finite-horizon MDP $\MDP$ is a unitary matrix $\MDPgenerative: \ComplexSpace{S}\tensorproduct\ComplexSpace{A}\tensorproduct\ComplexSpace{H}\tensorproduct\ComplexSpace{S}\tensorproduct \ComplexSpace{J}\rightarrow \ComplexSpace{S}\tensorproduct\ComplexSpace{A}\tensorproduct\ComplexSpace{H}\tensorproduct\ComplexSpace{S}\tensorproduct \ComplexSpace{J}$ satisfying
    \begin{equation}
            \begin{aligned}
        \MDPgenerative: &\ket{s}\ket{a}\ket{h}\ket{0}\ket{0}\\
        &\mapsto\ket{s}\ket{a}\ket{h}\Biggl( \sum_{s'}\sqrt{\Phsa(s')}\ket{s'}\ket{j_{s'}} \Biggr),
    \end{aligned}    
    \vspace{-8pt}
    \end{equation}
    where $J\geq 0$ is an arbitrary integer and $\ket{j_{s'}}\in \ComplexSpace{J}$ are arbitrary auxiliary states.
\end{definition}

We define the number of calls that an algorithm makes to the quantum generative model $\MDPgenerative$ or classical generative model $G$ as its quantum or classical sample complexity. 
Note that in Section \ref{subsec: speedup on A}, we have argued that comparing the time complexities of using a quantum oracle and using a classical oracle can be reduced to comparing the quantum and classical query complexities. In this section, although the quantum generative model $\MDPgenerative$ that we use is different from the quantum oracle $\MDPQoracle$, the same reduction from time complexity to sample complexity still holds. The reason is that $\MDPgenerative$ and $G$ have similar costs at the elementary gate level, assuming access to the classical circuit implementing $G$. In addition, it only incurs logarithmic overhead for $\MDPgenerative$ to encode the transition probabilities into quantum amplitudes, provided that quantum random access memory (QRAM) \cite{giovannetti2008quantum} is available.
Finally, the quantum time complexities match the quantum sample complexities up to log factors, provided that these assumptions hold and $G$ can be called in constant time.

We formally state the optimization goals in this setting. For a given time-dependent and finite-horizon MDP $\MDP$, $\epsilon\in(0,H]$ and $\delta\in(0,1)$, we want to obtain $\epsilon$-optimal policies\onlypaper{, V-value functions and Q-value functions} \onlytech{$\hat{\pi}$} with probability at least $1-\delta$. \onlytech{Additionally, we also aim to obtain $\epsilon$-optimal V-value functions \onlytech{$\{\hat{V}_{h}\}_{h=0}^{H-1}$} and Q-value functions \onlytech{$\{\hat{Q}_{h}\}_{h=0}^{H-1}$} with probability at least $1-\delta$.} With these objectives, we aim to design algorithms that require as few queries to the quantum generative model $\MDPgenerative$ as possible.

Before delving into our algorithms, we first introduce another important quantum subroutine, quantum mean estimation, in Theorem \ref{thm: quantum mean estimation}\onlytech{, which will frequently appear in our algorithms}. Quantum mean estimation consists of two similar quantum algorithms, which are $\qEstone$ and $\qEsttwo$. Both of them are referred to as \textbf{QME}. 

\begin{theorem}[Quantum mean estimation \cite{Montanaro_2015}]\label{thm: quantum mean estimation}
There are two quantum algorithms, denoted as $\qEstone$ and $\qEsttwo$, with the following properties. Let $\Omega$ be a finite set, $p = (p_x)_{x \in \Omega}$ a discrete probability distribution over $\Omega$, and $f: \Omega \to \mathbb{R}$ a function. Assume access to a quantum oracle $U_{p}$ for the probability distribution $p$ satisfying $U_{p}: \ket{0}\ket{0}\mapsto \sum_{x\in\Omega}\sqrt{p_{x}}\ket{x}\ket{j_{x}}$ where $\ket{j_{x}}$ are arbitrary auxiliary states, as well as an oracle $\BinaryOracle{f}$ for the function $f$. Then,
\begin{enumerate}
    \item Taking $u, \epsilon>0$ as additional inputs, along with the assumption that $0\leq f(x)\leq u$ for all $x\in\Omega$, $\qEstone$ requires $O\bigl(\frac{u}{\epsilon}+\sqrt{\frac{u}{\epsilon}}\bigr)$ queries to $U_{p}$ and $\BinaryOracle{f}$,
    \item Taking $\sigma>0$ and $\epsilon\in(0,4\sigma)$ as additional inputs, along with the assumption that $\variance{f(x)}\leq \sigma^{2}$, $\qEsttwo$ needs $O\bigl(\frac{\sigma}{\epsilon}\log^{2}(\frac{\sigma}{\epsilon}) \bigr)$ queries to $U_{p}$ and $\BinaryOracle{f}$,
    \vspace{-15pt}
\end{enumerate}
to output an estimate $\Tilde{\mu}$ of $\mu = \mathbb{E}[f(x) \mid x \sim p] = p^T f$ satisfying $\Pr(|\Tilde{\mu}-\mu|>\epsilon)<1/3$. Furthermore, by repeating either $\qEstone$ or $\qEsttwo$ a total 
 of $O(\log(1/\delta))$ times and taking the median of the outputs, one can obtain another estimate $\hat{\mu}$ of $\mu$ such that $\Pr(|\hat{\mu}-\mu|<\epsilon)>1-\delta$.
\end{theorem}

We denote $\textbf{QME}\{i\}_{\delta}(p^{T}f,\epsilon)$ as an estimate of $p^{\transpose}f$ to an error at most $\epsilon$ with probability at least $1-\delta$, obtained via \onlytech{the quantum mean estimation algorithm} $\textbf{QME}\{i\}$ for $i\in\{1,2\}$. Roughly speaking, $\qEstone$ is a quantum version of Hoeffding's inequality, while $\qEsttwo$ corresponds to the Chebyshev's (or Bernstein's) inequality. For example, for a random variable $X\in[0, u]$, Hoeffding's inequality implies that $O(u^{2}/\epsilon^{2})$ samples are required to obtain an $\epsilon$-estimation of $\Expectation[X]$. In comparison, $\qEstone$ only requires $O(u/\epsilon)$ quantum samples when $\epsilon\in (0, u]$.

Next, we will discuss how to apply the quantum subroutines, \textbf{QME} and \textbf{QMS}, into the model-free algorithms for finite-horizon MDPs by \cite{sidford2020variance} and \cite{sidford2018near}, and propose two quantum algorithms \textbf{QVI-3} and \textbf{QVI-4} which have significantly less sample complexity than the SOTA classical algorithms \cite{li2020breaking}.
\vspace{-5pt}
\subsection{Technical Overview of \textbf{QVI-3}}\label{subsec: Technical overview of QVI-3}
We first briefly review the main idea of the classical algorithm \texttt{RandomizedFiniteHorizonVI} proposed in \cite{sidford2020variance}. In the standard value iteration algorithm (See Algorithm \ref{algo: VI-finite horizon MDP-optimal action and V value}), we initialize $V_{H}\in\realnumber^{\statespace}$ with all zero entries and repeatedly apply the Bellman recursion $V_{h}=\valueoperator{h}{V_{h+1}}$ starting from the last time step and moving backward to the first, where the Bellman value operator $\mathcal{T}^{h}: \realnumber^{\statespace}\rightarrow\realnumber^{\statespace}$ is defined as
\begin{equation}\label{equ: bellman recursion}
    [\valueoperator{h}{V_{h+1}}]_{s}\define \max_{a\in\actionspace}\{r_{h}(s,a)+ \Phsa^{\transpose}V_{h+1}\},
\end{equation}
for all $s\in\statespace$. Instead of computing the exact value, they obtain an approximation of $[\valueoperator{h}{V_{h+1}}]_{s}$ by estimating $\Phsa^{\transpose}V_{h+1}$ via sampling from the classical generative model $G$ and taking maximum over the action space $\actionspace$. In order to obtain  $\epsilon$-optimal policies and V-value functions, they control the error of estimating $\Phsa^{\transpose} V_{h+1}$ to be $\epsilon/H$. If it also holds that $\infiNorm{V_{h+1}}\leq H$, then it requires $O\bigl(SAH^{2}/(\epsilon^{2}/H^{2}) \bigr)=O(SAH^{4}/\epsilon^{2})$ queries to $G$ at each time step $h$ to obtain the estimates of $\Phsa^{\transpose}V_{h+1}$ for all state-action pairs, according to the Hoeffding's inequality. Finally, the classical sample complexity of obtaining near optimal policy and values would be $O(SAH^{5}/\epsilon^{2})$. The sample complexity derived from above informal analysis matches the sample complexity of the algorithm in \cite{sidford2020variance} (up to log-factors). 

\begin{algorithm}[t]
\caption{Quantum Value Iteration $\QVIthree$}
\label{algo: quantum VI-finite horizon MDP-optimal action and V value v3}
\begin{algorithmic}[1]
\STATE \textbf{Require:} MDP $\MDP$, quantum generative model $\MDPgenerative$, maximum error $\epsilon\in (0, H]$, maximum failure probability $\delta\in(0,1)$.
\STATE \textbf{Initialize:} 
$\zeta\leftarrow\delta/\bigl( 4\Tilde{c}SA^{1.5}H\log(1/\delta) \bigr)$, $\hat{V}_{H}\gets\mathbf{0}$.
\FOR{$h := H-1, \ldots, 0$}
    \STATE create a quantum oracle $\BinaryOracle{\hat{V}_{h+1}}$ encoding $\hat{V}_{h+1}\in \realnumber^{\statespace}$
    \STATE $\forall s\in\statespace:$ create a quantum oracle $\BinaryOracle{z_{h,s}}$ encoding $z_{h,s}\in\realnumber^{\actionspace}$ with $\MDPgenerative$ and $\BinaryOracle{\hat{V}_{h+1}}$ satisfying\\
    \quad $z_{h,s}(a)\leftarrow \qEstone_{\zeta}\big((\Phsa^{\transpose}\hat{V}_{h+1}), \frac{\epsilon}{2H}\big)-\frac{\epsilon}{2H}$
    \STATE create a quantum oracle $\BinaryOracle{r_{h}}$ encoding $r_{h}\in \realnumber^{\statespace\times\actionspace}$
    \STATE $\forall s\in\statespace:$ create a quantum oracle $\BinaryOracle{\hat{Q}_{h,s}}$ encoding $\hat{Q}_{h,s}\in \realnumber^{\actionspace}$ with $\BinaryOracle{r_{h}}$ and $\BinaryOracle{z_{h,s}}$ satisfying \\ \quad $\hat{Q}_{h,s}(a)\leftarrow \max\{r_{h}(s, a) + z_{h,s}(a), 0\}$
    \STATE $\forall s\in\statespace:$ $\hat{\pi}(s,h) \leftarrow \qArgmax_{\delta}\{\hat{Q}_{h,s}(a): a\in \actionspace\}$ 
    \STATE $\forall s\in\statespace:$ 
    $\hat{V}_{h}(s) \leftarrow \hat{Q}_{h,s}\bigl( \hat{\pi}(s,h) \bigr)$ 
\ENDFOR
\STATE \textbf{Return:} $\hat{\pi}$, $\{\hat{V}_{h}\}_{h=0}^{H-1}$
\end{algorithmic}
\end{algorithm}

Now, we show how to achieve speedup in $A, H$ and $\epsilon$ by using the quantum subroutines \textbf{QME} and \textbf{QMS}. By using quantum mean estimation $\qEstone$, it only requires $O\bigl( SA\sqrt{H^{2}/(\epsilon^{2}/H^{2})} \bigr)=O(SAH^{2}/\epsilon)$ queries to the quantum generative oracle $\MDPgenerative$ to obtain $\epsilon$-approximations of $\Phsa^{\transpose}V_{h+1}$ for all pairs $(s,a)\in\statespace\times\actionspace$ at each time step. Hence, the total quantum query complexity in $H$ iterations becomes $O(SAH^{3}/\epsilon)$. Furthermore, we apply the quantum maximum searching $\qArgmax$ in the Bellman recursion to \onlytech{find the action over the action space $\actionspace$ which can} maximize the value on the RHS of Eq. \eqref{equ: bellman recursion}. Then, the query complexity further reduces to $O(S\sqrt{A}H^{3}/\epsilon)$. These are the fundamental ideas of Algorithm \ref{algo: quantum VI-finite horizon MDP-optimal action and V value v3}, denoted as \textbf{QVI-3}. In order to correctly apply \textbf{QME}, we also apply the monotonicity technique in \textbf{QVI-3} by subtracting the error induced by $\qEstone$ so that the V values $\hat{V}_{h}$ at each time step are bounded in $[0,H]$. Finally, \textbf{QVI-3} can obtain not only an $\epsilon$-optimal policy $\hat{\pi}$ but also $\epsilon$-optimal V-value functions $\{\hat{V}_{h}\}_{h=0}^{H-1}$ (Theorem \ref{Thm: correctness of QVI3}) with probability at least $1-\delta$, which requires only $\Tilde{O}(S\sqrt{A}H^{3}/\epsilon)$ queries to the oracle $\MDPgenerative$ (Theorem \ref{Thm: Complexity of QVI3}). The rigorous proof of the correctness and complexity of \textbf{QVI-3} are provided in Appendix \ref{appendix: QVIthree}.

\begin{theorem}[Correctness of \textbf{QVI-3}]\label{Thm: correctness of QVI3}
The outputs $\hat{\pi}$ and $\{\hat{V}_{h}\}_{h=0}^{H-1}$ satisfy that $V_{h}^{*}-\epsilon\leq \hat{V}_{h}\leq V^{\hat{\pi}}_{h}\leq V_{h}^{*}$
for all $h\in[H]$ with a success probability at least $1-\delta$.
\end{theorem}

\begin{theorem}[Complexity of \textbf{QVI-3}]\label{Thm: Complexity of QVI3}
The quantum query complexity of \textbf{QVI-3} in terms of the quantum generative oracle $\MDPgenerative$ is 
\vspace{-5pt}
\begin{equation}
\vspace{-5pt}
O\bigl(\frac{S\sqrt{A}H^{3}\log(SA^{1.5}H/\delta)}{\epsilon} \bigr).
\end{equation}
\end{theorem}

\begin{algorithm*}[t]
\caption{Quantum Value Iteration $\QVIfour$}
\label{algo: quantum VI-finite horizon MDP-optimal action and V value v4}
\begin{algorithmic}[1]
\STATE \textbf{Require:} MDP $\MDP$, quantum generative model $\MDPgenerative$, maximum error $\epsilon\in (0, \sqrt{H}]$, maximum failure probability $\delta\in(0,1)$.
\STATE \textbf{Initialize:} $K\leftarrow\ceil{\log_{2}(H/\epsilon)}+1$, $\zeta\leftarrow \delta/4KHSA$, $c=0.001$, $b=1$
\STATE \textbf{Initialize:} $\forall h\in[H]: V^{(0)}_{0,h}\gets \mathbf{0}$; $\forall s\in\statespace, h\in[H]: \pi_{0}^{(0)}(s,h)\leftarrow \text{arbitrary action } a\in\actionspace$.
\FOR{$k=0, \ldots, K-1$}
    \STATE $\epsilon_{k}\leftarrow H/2^{k}, V_{k,H}\leftarrow \mathbf{0}, V^{(0)}_{k,H}\leftarrow \mathbf{0}$
    \STATE $\forall (s, a, h)\in \statespace\times \actionspace\times[H]: y_{k,h}(s,a)\leftarrow \max\bigl\{\qEstone_{\zeta}\bigl(\Phsa^{\transpose}(V_{k,h+1}^{(0)})^{2},b\bigr)- \bigl(\qEstone_{\zeta}( \Phsa^{\transpose}V_{k,h+1}^{(0)}, b/H) \bigr)^{2},0 \bigr\}$
    \STATE $\forall (s, a, h)\in \statespace\times \actionspace\times[H]: x_{k,h}(s,a)\leftarrow \qEsttwo_{\zeta}\Bigl(\Phsa^{\transpose}V_{k,h+1}^{(0)}, \frac{c\epsilon}{H^{1.5}}\sqrt{y_{k,h}(s,a)+4b} \Bigr)-\frac{c\epsilon}{H^{1.5}}\sqrt{y_{k,h}(s,a)+4b}$
    \FOR{$h := H-1, \ldots, 0$}
        \STATE $\forall (s,a)\in\statespace\times \actionspace: g_{k,h}(s,a)\leftarrow\qEstone_{\zeta}\bigl(\Phsa^{\transpose}(V_{k,h+1}-V_{k,h+1}^{(0)}), cH^{-1}\epsilon_{k}\bigr)-cH^{-1}\epsilon_{k}$
        \STATE $\forall (s,a)\in\statespace\times\actionspace: Q_{k,h}(s,a)\leftarrow \max\{r_{h}(s,a)+x_{k,h}(s,a)+g_{k,h}(s,a),0\}$
        \STATE  $\forall s\in\statespace: \Tilde{V}_{k,h}(s)\leftarrow V_{k,h}(s)\leftarrow [V(Q_{k,h})]_{s}, \Tilde{\pi}_{k}(s,h)\leftarrow \pi_{k}(s,h)\leftarrow  [\pi(Q_{k,h})]_{s}$
        \STATE $\forall s\in\statespace:$ if $\Tilde{V}_{k,h}(s)\leq V_{k,h}^{(0)}(s)$, then $V_{k,h}(s)\leftarrow V_{k,h}^{(0)}(s)$ and $\pi_{k}(s,h)\leftarrow \pi_{k}^{(0)}(s,h)$
    \ENDFOR
        \STATE $\forall h\in[H]: V_{k+1,h}^{(0)}\leftarrow V_{k,h}$ and $\pi_{k+1}^{(0)}(\cdot,h)\leftarrow \pi_{k}(\cdot,h)$
\ENDFOR
\STATE \textbf{Return:} $\hat{\pi}\define\pi_{K-1}$, $\{\hat{V}_{h}\}_{h=0}^{H-1}\define\{V_{K-1,h}\}_{h=0}^{H-1}$, $\{\hat{Q}_{h}\}_{h=0}^{H-1}\define\{Q_{K-1,h}\}_{h=0}^{H-1}$
\end{algorithmic}
\end{algorithm*}
\setlength{\dbltextfloatsep}{10pt}
\vspace{-10pt}
\subsection{Technical Overview of \textbf{QVI-4}}
Note that \textbf{QVI-3} can only obtain $\epsilon$-optimal policy and V-value functions.
Below, we will introduce another algorithm, \textbf{QVI-4} in Algorithm \ref{algo: quantum VI-finite horizon MDP-optimal action and V value v4}, which can obtain not only $\epsilon$-optimal policies and V-value functions, but also Q-value functions. In this setting, although we can no longer attain a speed-up in $A$, we attain a speed-up in $H$ 
by utilizing two additional techniques in \cite{sidford2018near}: ``\textit{variance reduction}'' and ``\textit{total variance}''. 

We now introduce the essential ideas of the two new techniques and show how to integrate these techniques with the quantum mean estimation \textbf{QME} to reduce the sample complexity. 
First, the main idea of variance reduction technique is that, instead of using the standard value iteration algorithm (Algorithm \ref{algo: VI-finite horizon MDP-optimal action and V value}) directly for a target approximation error $\epsilon$, one repeats the value iteration algorithm for $K=O(\log(H/\epsilon))$ epochs with decreasing $\epsilon_{k}$ satisfying $\epsilon_{k}=\epsilon_{k-1}/2$ and $\epsilon_{K}=\epsilon$. In each epoch $k$, we obtain $\epsilon_{k}$-optimal V-value functions $\{V_{k,h}\}_{h=0}^{H-1}$, Q-value functions $\{Q_{k,h}\}_{h=0}^{H-1}$ and policy $\pi_{k}$.
Note that, at the time step $h$ in epoch $k$, the second term on the RHS of Eq. \eqref{equ: bellman recursion} can be rewritten as follows
\vspace{-5pt}
\begin{equation}\label{equ: new bellman recursion}
  \Phsa^{\transpose}V_{k,h+1}=\Phsa^{\transpose}(V_{k,h+1}-V_{k,h+1}^{(0)})+\Phsa^{\transpose}V_{k,h+1}^{(0)},
  \vspace{-3pt}
\end{equation}
where $V_{k,h+1}^{(0)}\in\realnumber^{\statespace}$ is defined as an initial V-value function for the time step $h+1$ from the previous epoch $k-1$. Note that there are a total $SA$ of these equations, each of which is corresponding to a pair $(s, a)\in\statespace\times\actionspace$. Rather than directly obtaining $\epsilon_{k}/H$-estimation of $\Phsa^{\transpose}V_{k,h+1}$, we instead obtain $\epsilon_{k}/(2H)$ estimations of both  $\Phsa^{\transpose}(V_{k,h+1}-V_{k,h+1}^{(0)})$ and $\Phsa^{\transpose}V_{k,h+1}^{(0)}$. For the first estimation, if we have $\mathbf{0}\leq V_{k,h+1}-V_{k,h+1}^{(0)} \leq \Tilde{c}\epsilon_{k}$ for some constant $\Tilde{c}>0$, it can be done up to error $\epsilon_{k}/(2H)$ using only $O(H^{2})$ classical samples or $O(H)$ quantum samples by the Hoeffding's bound or \textbf{QME1}, respectively. Similarly, for the second estimation, if it holds that $\mathbf{0}\leq V_{k,h+1}^{(0)}\leq H$, it requires 
$O(H^{4}/\epsilon_{k}^{2})$ classical samples or $O(H^{2}/\epsilon_{k})$ quantum samples. The overall classical sample complexity is 
$O\bigl( KHSA(H^{4}/\epsilon_{k}^{2}+H^{2}) \bigr)=\Tilde{O}(SAH^{5}/\epsilon_{k}^{2})$, while the quantum sample complexity is 
$O\bigl( KHSA(H^{2}/\epsilon_{k}+H) \bigr)=\Tilde{O}(SAH^{3}/\epsilon_{k})$\onlytech{, which holds for $\epsilon_{k}\in(0,H]$}. 
Note that unlike Section \ref{subsec: Technical overview of QVI-3}, we do not expect a speedup from $A$ to $\sqrt{A}$ here, since we need to estimate the Q-values for all actions (instead of finding the action with the highest Q-value).
Although the variance reduction technique alone does not achieve a speedup in $H$ compared with \textbf{QVI-3}, we will see the advantage when combined with the subsequent total variance technique.

The total variance technique stems from the observation that the actual error propagation across time steps is much smaller than previously assumed. Previously, the error in estimating 
$\mu_{k,h}^{s,a} \define \Phsa^\transpose V_{k,h+1}^{(0)}$ 
at each time step was set to $\epsilon_k / (2H)$, ensuring that the total error accumulated over $H$ iterations remains bounded by $\epsilon_k / 2$. In fact, the per-step error can be further relaxed to $\epsilon_k \sigma_{k,h}^{s,a} / (2H^{1.5})$, where 
$\sigma_{k,h}^{s,a} \define [\sigma_h(V_{k,h+1}^{(0)})](s,a)$.
This error value can reach up to $\epsilon_k / (2\sqrt{H})$. As the cumulative standard deviation $\sum_{h=0}^{H-1} \sigma_{k,h}^{s,a}$ is associated with an expression that can be non-trivially upper-bounded by $H^{1.5}$ (Lemma \ref{lemma: upper bound on variance}), the total error remains $\epsilon_k / 2$. 
With classical algorithms, $\mu_{k,h}^{s,a}$ can be estimated with an error $\epsilon \sigma_{k,h}^{s,a}$ without explicitly knowing $\sigma_{k,h}^{s,a}$. This requires overall 
$O\bigl( SA(\epsilon / H^{1.5})^{-2} \bigr) = O(SAH^3 / \epsilon^2)$ 
classical samples per time step at each epoch, as guaranteed by Chebyshev's (or Bernstein's) inequality. When combined with the variance reduction technique in estimating the first term on the RHS of Eq. \eqref{equ: new bellman recursion}, this approach achieves an overall classical sample complexity of $\tilde{O}(SAH^4 / \epsilon^2)$, matching the complexity of the algorithm \onlytech{\texttt{FiniteHorizonRandomQVI} presented} in \cite{sidford2018near}\footnote{The result in \cite{sidford2018near} was originally presented for the time-independent case. We adapt it here for the time-dependent case with an additional factor of $H$.}. 

Inspired by \cite{pmlr-v139-wang21w}, we can adapt the total variance technique in the quantum setting. The main challenge is that we cannot directly apply the quantum mean estimation $\qEsttwo$ like its classical counterpart. First, $\qEsttwo$ cannot estimate $\mu_{k,h}^{s,a}$ to an error of $\epsilon \sigma_{k,h}^{s,a}/(2H^{1.5})$ without prior knowledge of $\sigma_{k,h}^{s,a}$. To address this, we can use $\qEstone$ to obtain an estimate $(\hat{\sigma}_{k,h}^{s,a})^{2}$ of $(\sigma_{k,h}^{s,a})^{2}$ with an error $4b > 0$, then use $\qEsttwo$ to estimate $\mu_{k,h}^{s,a}$ with an error $\epsilon\underline{\sigma}_{k,h}^{s,a}/(2H^{1.5})$, where $\underline{\sigma}_{k,h}^{s,a} \define \sqrt{(\hat{\sigma}_{k,h}^{s,a})^{2} - 4b} \leq \sigma_{k,h}^{s,a}$, to maintain the correctness. Second, $\qEsttwo$ also requires upper bounds $C\in\realnumber$ on $\sigma_{k,h}^{s,a}$. Observing that its sample complexity $O(C/\epsilon)$ can be inefficient for large $C$, an ideal way is to use $\overline{\sigma}_{k,h}^{s,a}\define \sqrt{(\hat{\sigma}_{k,h}^{s,a})^{2}+4b}$ as $C$. However, this may lead to an unbounded complexity ratio 
\onlytech{$O(H^{1.5}(\hat{\sigma}_{k,h}^{s,a} + b) / \epsilon(\hat{\sigma}_{k,h}^{s,a} - b))$.}
\onlypaper{$\bigl( (\hat{\sigma}_{k,h}^{s,a})^{2} + 4b \bigr) 
/ \bigl( (\hat{\sigma}_{k,h}^{s,a})^{2}- 4b \bigr)$.}
To resolve this, we estimate $\mu_{k,h}^{s,a}$ with an error proportional to $\overline{\sigma}_{k,h}^{s,a}$, ensuring $C / \overline{\sigma}_{k,h}^{s,a} = 1$. Although the correctness may not hold due to $\overline{\sigma}_{k,h}^{s,a} > \sigma_{k,h}^{s,a}$, we can bound $\overline{\sigma}_{k,h}^{s,a} \leq \sigma_{k,h}^{s,a} + \sqrt{7b}$ and suppress the extra error by setting $b$ and the parameter $c$ in \textbf{QVI-4} as small constants. \onlytech{This adjustment increases the complexity only by a constant factor and keeps the cost of estimating $\sigma_{k,h}$ within budget.} Ultimately, \textbf{QVI-4} can obtain $\epsilon$-optimal policies, V-value functions and Q-value functions (Theorem \ref{Thm: correctness of QVI4}) with $\Tilde{O}(SAH^{2.5}/\epsilon)$ queries to the quantum generative oracle $\MDPgenerative$ (Theorem \ref{Thm: Complexity of QVI4}), which holds for $\epsilon=O(1/\sqrt{H})$. 
\onlytech{Besides, the $H^{2.5}$ scaling of the total sample complexity, compared with the $\Gamma^{1.5}$ scaling in the infinite-horizon case \cite{pmlr-v139-wang21w} (where $\Gamma \define 1/(1-\gamma)$ and $\gamma$ is the discount factor), arises from the time-independence property. Specifically, an additional $H$ factor is introduced when estimating $\mu_{k,h}$, since the transition matrix $P_h$ varies with $h$.} The proof of the correctness and complexity of \textbf{QVI-4} is presented in Appendix \ref{appendix: QVIfour}.

\begin{theorem}[Correctness of \textbf{QVI-4}]\label{Thm: correctness of QVI4}
The outputs $\hat{\pi}$, $\{\hat{V}_{h}\}_{h=0}^{H}$ and $\{\hat{Q}_{h}\}_{h=0}^{H}$ satisfy that 
\begin{align}
        &V_{h}^{*}-\epsilon\leq \hat{V}_{h}\leq V^{\hat{\pi}}_{h}\leq V_{h}^{*},\\
        &Q_{h}^{*}-\epsilon\leq \hat{Q}_{h}\leq Q^{\hat{\pi}}_{h}\leq Q_{h}^{*},
\end{align}
for all $h\in[H]$ with a success probability at least $1-\delta$.
\end{theorem}

\begin{theorem}[Complexity of \textbf{QVI-4}]\label{Thm: Complexity of QVI4}
The quantum query complexity of \textbf{QVI-4} in terms of the quantum generative oracle $\MDPgenerative$ is 
\begin{equation}
\vspace{-5pt}
         O\bigl(SA(\frac{H^{2.5}}{\epsilon}+H^{3})\log^{2}(\frac{H^{1.5}}{\epsilon})\log(\log(\frac{H}{\epsilon})HSA/\delta)\bigr).
\end{equation}
\end{theorem}
\subsection{Quantum Lower Bound for Finite-horizon MDPs}
We now state the quantum lower bound of the sample complexity for obtaining the $\epsilon$-optimal policy, V-value functions and Q-value functions for a finite-horizon and time-dependent MDP $\MDP$. Our proof idea \onlytech{inherits from \cite{sidford2018near} by reducing} \onlypaper{is to reduce} an infinite-horizon MDP problem to a finite-horizon MDP problem. Specifically, we show that, if there is an algorithm that can obtain an $\epsilon$-optimal V-value function for the finite-horizon MDP, it also can give an $2\epsilon$-optimal V-value function to the infinite-horizon MDP. Therefore, the lower bound of solving finite-horizon MDP with a quantum generative oracle inherits from that of the infinite-horizon MDP. The full analysis is presented in Appendix \ref{appendix: lower bounds}. Note that our achievable quantum sample complexities of \textbf{QVI-3} and \textbf{QVI-4} differ from the quantum lower bounds only by a factor of $H$ or $H^{1.5}$, up to logarithmic factors.


\begin{theorem}[Lower bounds for finite-horizon MDPs]\label{thm: quantum lower bound for finite horizon MDP}
    Let $\statespace$ and $\actionspace$ be finite sets of states and actions. Let $H>0$ be a positive integer and $\epsilon\in(0, 1/2)$ be an error parameter. We consider the following time-dependent and finite-horizon MDP $\MDP=(\statespace, \actionspace, \{P_{h}\}_{h=0}^{H-1}, \{r_{h}\}_{h=0}^{H-1}, H)$, where $r_{h}\in[0,1]^{\statespace\times\actionspace}$ for all $h\in[H]$. 
    \begin{itemize}
    \vspace{-5pt}
        \item Given access to a classical generative oracle $G$, any algorithm $\mathcal{K}$, which takes $\MDP$ as an input and outputs $\epsilon$-approximations of $\{Q^{*}_{h}\}_{h=0}^{H-1}$ $\{V^{*}_{h}\}_{h=0}^{H-1}$ or $\pi^{*}$ with probability at least $0.9$, must call $G$ at least  
        $\Omega\bigl( \frac{SAH^{3}}{\epsilon^{2}\log^{3}(\epsilon^{-1})} \bigr)$
        times on the worst case of input $\MDP$.
        \vspace{-10pt}
        \item Given access to a quantum generative oracle $\mathcal{G}$, any algorithm $\mathcal{K}$, which takes $\MDP$ as an input and outputs $\epsilon$-approximations of $\{Q^{*}_{h}\}_{h=0}^{H-1}$ with probability at least $0.9$, must call $\MDPgenerative$ at least 
        $\Omega\bigl( \frac{SAH^{1.5}}{\epsilon\log^{1.5}(\epsilon^{-1})} \bigr)$
        times on the worst case of input $\MDP$. Besides, any algorithm $\mathcal{K}$ that outputs $\epsilon$-approximations of $\{V^{*}_{h}\}_{h=0}^{H-1}$ or $\pi^{*}$ with probability at least $0.9$ must call $\MDPgenerative$ at least $\Omega\bigl( \frac{S\sqrt{A}H^{1.5}}{\epsilon\log^{1.5}(\epsilon^{-1})} \bigr)$ 
        times on the worst case of input $\MDP$. 
    \end{itemize}
\end{theorem}

\section{Conclusion}\label{sec: conclusion}
To the best of our knowledge, this is the first work to rigorously study quantum algorithms for solving ``time-dependent'' and ``finite-horizon'' MDPs. In the exact dynamics setting, our quantum value iteration algorithm \textbf{QVI-1} achieves a quadratic speedup in the size of the action space $(A)$ for computing the optimal policy and V-value function, while \textbf{QVI-2} \onlytech{and \textbf{QVI-5}} achieves an additional speedup in the size of the state space $(S)$ for computing near-optimal policy and V-value functions. Besides, our classical lower bounds show that no classical algorithm can attain comparable query complexities of \textbf{QVI-1} and \textbf{QVI-2} in terms of the dependences on $S$ and $A$. In the generative model setting, our algorithms \textbf{QVI-3} and \textbf{QVI-4} achieve speedups in 
$A$, time horizon $(H)$, and approximation error $(\epsilon)$ over the SOTA classical algorithm and are asymptotically optimal, up to log terms, for computing near-optimal policies, V-value functions, and Q-value functions, provided a constant time horizon.
\begin{rem}
    Shortly after the conference version of this work was presented, the independent work~\cite{ambainis2025bit} appeared, which proposes quantum algorithms for finite-horizon and infinite-horizon time-independent MDPs with discrete and compact state spaces.
\end{rem}
\section*{Acknowledgements}
We especially thank Zongqi Wan for providing insightful guidance on quantum subroutines, including quantum maximum searching \cite{durr1999quantumalgorithmfindingminimum} and quantum mean estimation algorithms \cite{Montanaro_2015}, and for suggesting helpful references, including \cite{cornelissen2018quantum}. 
The work of John C.S. Lui was supported in part by the RGC SRFS2122-4S02.
\section*{Impact Statement}
This paper presents work whose goal is to advance the field of Machine Learning and AI via quantum computing. There are many potential societal consequences of our work, none of which we feel must be specifically highlighted here.


\bibliography{main}
\bibliographystyle{icml2025}

\newpage
\appendix
\onecolumn




\section{Exact Dynamics Setting}\label{appendix: Exact Dynamics Setting}
\subsection{Classical Algorithm for Finite-horizon MDPs}
For the completeness, we restate the classical value iteration (or backward induction) algorithm in \cite{puterman2014markov}.

\begin{algorithm}
\caption{Value Iteration (Backward Induction) Algorithm for Finite Horizon MDPs}
\label{algo: VI-finite horizon MDP-optimal action and V value}
\begin{algorithmic}[1]
\STATE \textbf{Require:} MDP $\MDP$.
\STATE \textbf{Initialize: } $V_{H}\leftarrow \mathbf{0}$
\FOR{$h := H-1, \ldots, 0$}
    \FOR{each $s \in \statespace$}
        \FOR{each $a \in \actionspace$}
            \STATE $Q_{h}(s, a)= r_{h}(s, a) + \sum\limits_{s'\in\statespace}\Phsa(s')V_{h+1}(s')$
        \ENDFOR
        \STATE $\pi(s,h) = \argmax\limits_{a \in \actionspace} Q_{h}(s, a)$ 
        \STATE $V_{h}(s)= Q_{h}\bigl( s, \pi(s,h) \bigr)$ 
    \ENDFOR
\ENDFOR
\STATE \textbf{Return:} $\pi$, $V_{0}$
\end{algorithmic}
\end{algorithm}

\subsection{Classical Lower Bounds}\label{appendix: classical lower bounds}
\beginproof
We define two sets of hard instances of finite-horizon MDP $M_{1}$ and $M_{2}$ which are the same as those in Section 4.1 in \cite{chen2017lowerboundcomputationalcomplexity}. Specifically, suppose that the state space $\statespace$ can be divided into four parts $\statespace=\statespace_U \cup \statespace_G \cup \statespace_{B}  \cup \{s_{N}\}$, where the cardinalities of the sets $\statespace_U, \statespace_G$ and $\statespace_{B}$ satisfy $S_{U}=S_{G}=S_{B}=\frac{S-1}{3}$, and $s_{N}$ is a single action. Let the action space be $\actionspace=\actionspace_U \cup \{a_{N}\}$, where the cardinality of the set $\actionspace_U$ satisfies $A_U=A-1$ and $a_N$ is a single action. We now construct two sets of MDP instances $M_1$ and $M_2$ that are hard to distinguish.
\begin{itemize}
    \item Let $M_1$ be the set of instances satisfying the following conditions. 
    \begin{itemize}
        \item $P_{h}=P\in[0,1]^{\statespace\times\actionspace\times\statespace}$ for all $h\in[H]$ and $r_{h}=r\in [0,1]^{\statespace\times \actionspace}$, where $H\geq 2$;
        \item For any $(s,a)$ satisfies $s\in \statespace_G \cup \statespace_B\cup \{s_{N}\}$ and $a\in\actionspace$, the transition probabilities satisfy $P(s'|s,a)=1$ if $s'=s$ and $P(s'|s,a)=0$ if $s'\neq s$, i.e., the states in $\statespace_G \cup \statespace_B\cup \{s_{N}\}$ are absorbing states. Besides, the reward functions satisfy
        \begin{equation}
            r(s,a) = 
                \begin{cases}
                1, & \text{if } s \in \mathcal{S}_G, \, a \in \mathcal{A} \\
                0, & \text{if } s \in \mathcal{S}_B, \, a \in \mathcal{A} \\
                \frac{1}{2}, & \text{if } s = s_N, \, a \in \mathcal{A}
                \end{cases}.
        \end{equation}
        \item For any $a\in\actionspace_U$ and $s\in\statespace_U$, the transition probability satisfies $P(s'|s,a)=1$ if $s'\in\statespace_{B}$ and $P(s'|s,a)=0$ otherwise, while the reward satisfies $r(s,a)=0$.
        \item For any $a=a_{N}$ and $s\in\statespace_U$, the transition probability satisfies $P(s'|s, a)=1$ if $s'=s_{N}$ and $P(s'|s,a)=0$ otherwise, while the reward satisfies $r(s,a)=0$.
    \end{itemize}
    \item Let $M_{2}$ be the set of instances that are different from those in $M_1$ at one state-action pair, which we denote by $(\overline{s}, \overline{a})\in\statespace_{U}\times \actionspace_{U}$. 
    \begin{itemize}
        \item When $(s,a)=(\overline{s},\overline{a})$, the transition probability satisfies $P(s'|s,a)=1$ for some $s'\in\statespace_{G}$ and $P(s'|s,a)=0$ otherwise, while the reward satisfies $r(s,a)=0$.
    \end{itemize}
\end{itemize}
From the above definitions, we can know that the cardinalities of $M_{1}$ and $M_{2}$ are $|\statespace_{B}|^{|\statespace_{U}\times\actionspace_{U}|}$ and $|\statespace_{U}\times \actionspace_{U}|\times |\statespace_{B}|^{|\statespace_{U}\times \actionspace_{U}|}$, respectively. We now compute the optimal V-value function $V_{0,\MDP_1}^{*}$ for any finite-horizon MDP $\MDP_{1}\in M_{1}$. 
\begin{itemize}
    \item For $s\in\statespace_G$, $V_{H-1, \MDP_1}^{*}(s)=\max_{a\in\actionspace}\{r(s,a)+P_{H|s,a}^{\transpose}V_{H,\MDP_{1}}^{*}\}=1$, because $V_{H,\MDP_{1}}^{*}=\mathbf{0}$. Further, since $s$ is an absorbing state, $V_{h, \MDP_1}^{*}(s)=1+V_{h+1, \MDP_1}^{*}(s)=H-h$. Hence, we can compute $V_{0, \MDP_1}^{*}(s)=H$.
    \item For $s\in\statespace_B$, since $r(s,a)=0$ for all $a\in\actionspace$ and $s$ is an absorbing state, we can compute $V_{h,\MDP_1}^{*}(s)=0$ for all $h\in[H]$.
    \item When $s=s_{N}$, since $r(s,a)=\frac{1}{2}$ for all $a\in\actionspace$ and $s$ is also an absorbing state, we can compute $V_{h,\MDP_1}^{*}(s)=\frac{1}{2}+V_{h+1,\MDP_1}^{*}(s)=\frac{H-h}{2}$ and  $V_{0,\MDP_1}^{*}(s)=\frac{H}{2}$.
    \item For $s\in\statespace_U$, we can compute $V_{H-1,\MDP_{1}}^{*}(s)=\max_{a\in\actionspace}\{r(s,a)\}=0$. Further, by the Bellman optimality equation \cite{bellman1958dynamic}, we can compute 
    \begin{equation}
        \begin{aligned}
            V_{H-2,\MDP_{1}}^{*}(s)&=\max_{a\in\actionspace}\{r(s,a)+\sum_{s'\in\statespace}P(s'|s,a)V_{H-1,\MDP_{1}}^{*}(s')\}\\
            &=\max\{V_{H-1, \MDP_{1}}^{*}(s)1\{s\in\statespace_{B}\}, V_{H-1, \MDP_{1}}^{*}(s_N)\}\\
            &=\max\{0, \frac{1}{2}\}.
        \end{aligned}
    \end{equation}
    The second line comes from the fact that $r(s,a)=0$ for any $a\in\actionspace$ and state $s$ will transition to $s_{N}$ if $a=a_{N}$ or transition to some state $s\in\statespace{B}$ if $a\in\actionspace_{U}$.
    By induction, we know that $V_{h,\MDP_{1}}^{*}(s)=\max\{V_{h+1, \MDP_{1}}^{*}(s)1\{s\in\statespace_{B}\}, V_{h+1, \MDP_{1}}^{*}(s_N)\}=\max\{0, \frac{H-h+1}{2}\}$ and $V_{0,\MDP_{1}}^{*}(s)=\frac{H-1}{2}$. 
\end{itemize}
Similarly, we can compute the optimal V-value function $V_{0,\MDP_{2}}^{*}$ for any finite-horizon MDP $\MDP_{2}\in M_{2}$. Since $\MDP_{2}$ only differs from $\MDP_{1}$ on the state-action pair $(\overline{s},\overline{a})\in \statespace_{U}\times \actionspace_{U}$ and the states $s\in \statespace_{G}\times\statespace_{B}\times \{s_{N}\}$ are absorbing states, $V_{h,\MDP_{2}}^{*}(s)$ only differs from $V_{h,\MDP_{1}}^{*}(s)$ on state $\overline{s}$. Specifically, $V_{H-1, \MDP_{2}}^{*}(\overline{s})=\max_{a\in\actionspace}\{r(\overline{s},a)\}=0$ and 

\begin{equation}
    \begin{aligned}
        V_{h, \MDP_{2}}^{*}(\overline{s})&=\max_{a\in\actionspace}\{r(\overline{s},a)+\sum_{s'\in\statespace}P(s'|\overline{s},a)V_{h+1, \MDP_{2}}^{*}(s')\}\\
        &=\max\{V_{h+1, \MDP_{2}}^{*}(s)1\{s\in\statespace_{B}\}, V_{h+1, \MDP_{2}}^{*}(s)1\{s\in\statespace_{G}\}, V_{h+1, \MDP_{2}}^{*}(s_{N})\}\\
        &=\max\{0, H-h+1, \frac{H-h+1}{2}\}\\
        &=H-h+1.
    \end{aligned}
\end{equation}
The first line comes from the Bellman optimality equation \cite{bellman1958dynamic}. The second line comes from the fact that $r(\overline{s},a)=0$ for all $a\in\actionspace$ and the state $\overline{s}$ will transition to some state $s'\in\statespace_{B}$, $s'\in\statespace_{G}$ or $s'=s_{N}$ under the action $a\in\actionspace_{U}\setminus\{\overline{a}\}$, $a=\overline{a}$ or $a=a_{N}$.
Hence, it implies that $ V_{0, \MDP_{2}}^{*}(\overline{s})=H-1$. However, $V_{0, \MDP_{1}}^{*}(\overline{s})=\frac{H-1}{2}$. Therefore, we can see that $\infiNorm{V_{0, \MDP_{1}}^{*}-V_{0, \MDP_{2}}^{*}}=\frac{H-1}{2}$. 
Using the same proof in Section 5.2 in \cite{chen2017lowerboundcomputationalcomplexity}, we can know that, to achieve $\frac{H-1}{4}$-optimal $V_{0}$ with high probability, any algorithm must distinguish $\MDP_{1}$ from $\MDP_{2}$, requiring to search for two discrepancies in an array of size $|\statespace_{U}\times\actionspace_{U}\times\statespace_{B}|=\Omega(S^{2} A)$ by quering the classical oracle $\classicalMDPoracle$. Therefore, given the classical oracle $\classicalMDPoracle$, the classical lower bound of query complexity for computing an $\epsilon$-optimal $V_{0}$ for the time-independent and finite-horizon MDP is $\Omega(S^{2} A)$ for $\epsilon \in (0,\frac{H-1}{4})$. This implies the classical lower bound of query complexity for obtaining an $\epsilon$-optimal policy or $\epsilon$-optimal V-value functions for the time-dependent and finite-horizon MDP is $\Omega(S^{2} A)$.
\done

\subsection{Correctness, Complexity and Qubit Cost of \textbf{QVI-1} (Algorithm \ref{algo: quantum VI-finite horizon MDP-optimal action and V value v1})}\label{appendix: QVIone}
\subsubsection{Correctness of \textbf{QVI-1} (Proof of Theorem \ref{Thm: correctness of QVI1})}
\vspace{-5pt}
\beginproof
First, we consider the failure probability of the algorithm to achieve above goal. Every $\qArgmax$ is performed with maximum failure probability $\zeta=\delta/(SH)$ and $\qArgmax$ is called $SH$ times when running Algorithm \ref{algo: quantum VI-finite horizon MDP-optimal action and V value v1} one time. By the union bound, the probability that there exists an incorrect output is at most $\delta$.

     Now, we assume the ideal scenario when $\qArgmax$ is always successful to find the action $a^{*}=\arg\max_{a\in\actionspace} \hat{Q}_{h,s}(a)$, i.e., $\hat{\pi}(s,h)=\arg\max_{a\in\actionspace} \hat{Q}_{h,s}(a)$. Note that we assume $\hat{V}_{H}(s)=0$ for all $s\in\statespace$. Then, we have $\hat{Q}_{H-1,s}(a)=r_{H-1}(s,a)$ for any policy $\pi$, indicating that $\hat{Q}_{H-1,s}(a)=Q_{H-1}^{*}(s,a)=Q^{\hat{\pi}}_{H-1}(s,a)$.

     Assume that with our policy $\hat{\pi}(s,h)=\arg\max_{a\in\actionspace} \hat{Q}_{h,s}(a)$ for all $s\in\statespace, h\in [H]$, we have $\hat{Q}_{h,s}(a)= Q^{*}_{h}(s,a)=Q^{\hat{\pi}}_{h}(s,a)$ for all $s\in\statespace, a\in\actionspace, h\in [H]$. 
     Besides, we define $\hat{\pi}_{h}(a|s)$ as the probability that the agent choose action $a$ in the state $s$ at time $h$. Note that $\hat{\pi}_{h}(a|s)=1$ if $a=\hat{\pi}(s,h)$ and $\hat{\pi}_{h}(a|s)=0$ otherwise. By Bellman equations, we have $V_{h}^{\hat{\pi}}(s)=\sum_{a\in\actionspace}Q_{h}^{\hat{\pi}}(s,a)\hat{\pi}_{h}(a|s)$ and $Q_{h}^{\hat{\pi}}(s,a)=r_{h}(s,a)+\Phsa^{\transpose}V_{h+1}^{\hat{\pi}}$. Then, we can know that, for all $s\in\statespace$ and $h\in[H]$, 
     \begin{equation}
     \begin{aligned}
         V^{*}_{h}(s)&=\max_{\hat{\pi}_{h}} \max_{\hat{\pi}_{h+1} \cdots \hat{\pi}_{H-1}}\sum_{a\in\actionspace}  Q^{\hat{\pi}}_{h}(s, a) \hat{\pi}_{h}(a|s)\\
         &=\max_{\hat{\pi}_{h}} \sum_{a\in\actionspace} Q^{*}_{h}(s,a)\hat{\pi}_{h}(a|s)\\
         &= Q_{h}^{*}\bigl(s, \hat{\pi}(s,h) \bigr)\\
         &=\hat{Q}_{h,s}\bigl( \hat{\pi}(s,h) \bigr)\\
         &=\hat{V}_{h}(s).
     \end{aligned}
     \end{equation}
    Besides, since $Q_{h}^{*}(s, \hat{\pi}(s,h))=Q_{h}^{\hat{\pi}}(s, \hat{\pi}(s,h))$ by the assumption, then $V_{h}^{*}(s)=V^{\hat{\pi}}_{h}(s)$ for all $s\in\statespace$ and $h\in[H]$.
     Similarly, assume that with our policy $\hat{\pi}(s,h)=\arg\max_{a\in\actionspace} \hat{Q}_{h,s}(a)$ for all $s\in\statespace, h\in [H]$, we have $\hat{V}_{h}(s)= V^{*}_{h}(s)=V^{\hat{\pi}}_{h}(s)$ for all $s\in\statespace, h\in [H]$. Then, we have
     \begin{equation}
      \begin{aligned}
         Q^{*}_{h}(s, a)&= r_{h}(s,a)+\max_{\pi} \sum_{s'\in\statespace}\Phsa(s')V^{\pi}_{h+1}(s')\\
         &=r_{h}(s,a)+\sum_{s'\in\statespace}\Phsa(s') V^{*}_{h+1}(s')\\
         &= r_{h}(s,a)+\sum_{s'\in\statespace}\Phsa(s') \hat{V}_{h+1}(s')\\
         &=\hat{Q}_{h,s}(a).
     \end{aligned}
     \end{equation}
     Note that we also have $\hat{V}_{h}=V_{h}^{\hat{\pi}}$ for all $h\in[H]$. Then, it also holds that $Q_{h}^{*}=r_{h}+\Phsa^{\transpose}\hat{V}_{h+1}=r_{h}+\Phsa^{\transpose}V^{\hat{\pi}}_{h+1}=Q_{h}^{\hat{\pi}}$
     Since $\hat{Q}_{H-1,s}(a)= Q^{*}_{H-1}(s,a)=Q^{\hat{\pi}}_{H-1}(s,a)$ for all $s\in\statespace, a\in\actionspace$, then we can know that $\hat{V}_{H-1}(s)= V_{H-1}^{*}(s)= V_{H-1}^{\hat{\pi}}(s)$ for all $s\in\statespace$. Furthermore, since $\hat{V}_{H-1}(s)= V_{H-1}^{*}(s)= V_{H-1}^{\hat{\pi}}(s)$ holds for all $s\in\statespace$, then we can deduce that $\hat{Q}_{H-2,s}(a)= Q^{*}_{H-2}(s,a)=Q^{\hat{\pi}}_{H-2}(s,a)$ for all $s\in\statespace$ and $a\in\actionspace$. In the end, we can conclude that $\hat{V}_{0}(s)= V_{0}^{*}(s)= V_{0}^{\hat{\pi}}(s)$ for all $s\in\statespace$ which implies $\hat{\pi}$ is an optimal policy.
\done
\subsubsection{Complexity of \textbf{QVI-1} (Proof of Theorem \ref{Thm: Complexity of QVI1})}
\beginproof
We first assume that all $\qArgmax$ are successful to find the optimal actions, up to the specified error, because the probability that this does not hold is at most $\delta$. Let $C$ be the complexity of $\QVIone$ as if all $\qArgmax$ are carried out with maximum failure probabilities set to constant. Then, since the actual maximum failure probabilities are set to $\zeta= \delta/(SH)$, the actual complexity of $\QVIone$ is \begin{equation}\label{equ: initial complexity of QVIone}
    O\bigl( C\log(SH/\delta) \bigr).
\end{equation}
Now, we check each line of $\QVIone$ to bound $C$. 

In line 4, we encode the vector $\hat{V}_{h+1}$ to an oracle $\BinaryOracle{\hat{V}_{h+1}}$. This process does not need to query $\MDPQoracle$ and only needs to access the classical vector $\hat{V}_{h+1}$. Therefore, the query complexity of $\BinaryOracle{\hat{V}_{h+1}}$ in terms of $\MDPQoracle$ is $O(1)$.

In line 5, we need to construct the quantum oracle $\BinaryOracle{\hat{Q}_{h,s}}$ with $\MDPQoracle$. Since we need to obtain $\ket{\Phsa(s')}$ for all $s'\in\statespace$ and calculate the weighted sum $\ket{\sum_{s'\in\statespace} \Phsa(s')\hat{V}_{h+1}(s')}$, it requires $O(S)$ query cost of the oracle $\MDPQoracle$. Note that the quantum addition and quantum multiplication can be performed by various quantum circuits, such as quantum Fourier transform techniques \cite{ruiz2017quantum, draper2000addition}. Therefore, the query complexity of $\BinaryOracle{\hat{Q}_{h,s}}$ in terms of $\MDPQoracle$ is $O(S)$.

In line 6, we can use quantum maximum searching algorithm $\qArgmax$ in Theorem \ref{Thm: quantum maximum finding}, resulting in a query cost of order $O\bigl(\sqrt{A}\bigr)$ to the oracle $\BinaryOracle{\hat{Q}_{h,s}}$ for all $s\in\statespace$ in each loop $h\in [H]$. 

Therefore, it induces an overall query cost of $C=O\bigl(S^{2}\sqrt{A}H\bigr)$ to the oracle $\MDPQoracle$. Combining with \eqref{equ: initial complexity of QVIone}, the overall quantum query complexity of $\QVIone$ in terms of $\MDPQoracle$ is 
\begin{equation}\label{equ: final complexity of QVIone}
    O\bigl( S^{2}\sqrt{A}H\log(SH/\delta) \bigr).
\end{equation}
\done

\subsubsection{Analysis on the Cost of Qubit Resources}
Considering that qubits are still scarce resources in a quantum computer, it is necessary to minimize the qubits resources required in a quantum algorithm. Note that the line 5 of Algorithm \ref{algo: quantum VI-finite horizon MDP-optimal action and V value v1} is the main source of consuming qubits in the whole algorithm. 
Constructing the oracle $ \BinaryOracle{\hat{Q}_{h,s}}$ for all $ s \in \statespace $ and \( h \in [H] \) requires a large number of auxiliary qubits. This is because the process involves storing information about the vector \( \hat{V}_{h+1} \in \mathbb{R}^{S} \) and the transition probabilities \( \Phsa(s') \) for all \( s' \in \statespace \), which are obtained by querying the quantum oracles \( \BinaryOracle{\hat{V}_{h+1}} \) and \( \MDPQoracle \). After encoding the classical information into qubits, we need to compute the weighted sum $\sum_{s'\in\statespace}\Phsa(s')\hat{V}_{h+1}(s')$. This process requires (non-modular) quantum adder\cite{ruiz2017quantum, draper2000addition, vedral1996quantum} and (non-modular) quantum multiplier \cite{ruiz2017quantum, vedral1996quantum} to compute the additions and multiplications with additional auxiliary qubits. Specifically, with the fixed-point representation, \cite{ruiz2017quantum} constructs quantum circuits of a non-modular quantum adder $\qAdder: \ket{\fixbinary{a}}_{q}\ket{\fixbinary{b}}_{q+1}\mapsto \ket{\fixbinary{a}}_{q}\ket{\fixbinary{a}+\fixbinary{b}}_{q+1}$ and a non-modular quantum multiplier $\qMul: \ket{\fixbinary{a}}_{q}\ket{\fixbinary{b}}_{q}\ket{0}_{2q}\mapsto \ket{\fixbinary{a}}_{q}\ket{\fixbinary{b}}_{q}\ket{\fixbinary{a}\fixbinary{b}}_{2q}$, which can compute the non-modular sum and multiplication of two non-negative real numbers $a$ and $b$.
We refer readers to \cite{wang2024comprehensive} for a comprehensive overview of the existing work on quantum arithmetic circuits.

Inspired by the quantum circuit for computing the controlled weighted sum proposed in Section 8 in
\cite{ruiz2017quantum}, we design a QFT-based circuit to reduce the qubits consumption in constructing the oracle $\BinaryOracle{\hat{Q}_{h,s}}$. We first prepare the following qubits
\begin{equation}
    \ket{a}\ket{s}\ket{h}\ket{0}^{\tensorproduct 4q+q_{s}+1},
\end{equation}
where $q_{s}=\ceil{\log_{2}(S)}$.
Then we apply the rotation matrix $U_{s'}$ to transform the $\ket{0}$ to the target state $\ket{s'}$. Since we encode the state space into orthonormal bases, then $U_{s'}$ is unitary. Hence, it returns the output state 
\begin{equation}
    \ket{a}\ket{s}\ket{h}\ket{s'}\ket{0}^{\tensorproduct 4q+1}.
\end{equation}
By applying the unitary oracle $\MDPQoracle$ and $\BinaryOracle{\hat{V}_{h+1}}$, we can obtain the following state
\begin{equation}
    \ket{a}\ket{s}\ket{h}\ket{s'}\ket{\fixbinary{r_{h}(s,a)}}\ket{\fixbinary{\Phsa(s')}}\ket{\fixbinary{\hat{V}_{h+1}(s')}}\ket{0}^{\tensorproduct q+1}.
\end{equation}
Then we compute the quantum Fourier transform of $\ket{0}^{\tensorproduct q+1}$ where
\begin{equation}
    \QFT\ket{0}^{\tensorproduct q+1}= \frac{1}{\sqrt{2^{q+1}}}\sum_{k=0}^{2^{q+1}-1}e^{i\frac{2\pi 0 k}{2^{q+1}}}\ket{k}=\ket{\phi(0)},
\end{equation}
we obtain the output state
\begin{equation}
    \ket{a}\ket{s}\ket{h}\ket{s'}\ket{\fixbinary{r_{h}(s,a)}}\ket{\fixbinary{\Phsa(s')}}\ket{\fixbinary{\hat{V}_{h+1}(s')}}\ket{\phi(0)}.
\end{equation}
Then we apply the multiplication block $U_{2^{-p}\Phsa(s')\hat{V}_{h+1}(s')}$ defined in the Fig. 4 in  \cite{ruiz2017quantum} and obtain the output state 
\begin{equation}
        \ket{a}\ket{s}\ket{h}\ket{s'}\ket{\fixbinary{r_{h}(s,a)}}\ket{\fixbinary{\Phsa(s')}}\ket{\fixbinary{\hat{V}_{h+1}(s')}}\ket{\phi(0+\fixbinary{\Phsa(s')}\fixbinary{\hat{V}_{h+1}(s')})}.
\end{equation}
By applying the unitary matrix $\BinaryOracle{\hat{V}_{h+1}}^{\dagger}$, $\MDPQoracle^{\dagger}$ and $U_{s'}^{\dagger}$ in sequence, we can undo the operations on auxiliary qubits and obtain the following state
\begin{equation}
        \ket{a}\ket{s}\ket{h}\ket{0}^{\tensorproduct 3q+q_{s}}\ket{\phi(0+\fixbinary{\Phsa(s')}\fixbinary{\hat{V}_{h+1}(s')})}.
\end{equation}
We can repeat the above operations for all $s'\in\statespace$ and obtain
\begin{equation}
\ket{a}\ket{s}\ket{h}\ket{0}^{\tensorproduct 3q+q_{s}}\ket{\phi(0+\sum_{s'\in\statespace}\fixbinary{\Phsa(s')}\fixbinary{\hat{V}_{h+1}(s')})}.
\end{equation}
By applying the inverse quantum Fourier transform on the state $\ket{\phi(0+\sum_{s'\in\statespace}\fixbinary{\Phsa(s')}\fixbinary{\hat{V}_{h+1}(s')})}$, we can obtain
\begin{equation}
\ket{a}\ket{s}\ket{h}\ket{0}^{\tensorproduct 3q+q_{s}}\ket{\sum_{s'\in\statespace}\fixbinary{\Phsa(s')}\fixbinary{\hat{V}_{h+1}(s')}}.
\end{equation}
Since $\sum_{s'\in\statespace}\Phsa(s')=1$ and $\Phsa(s')\in[0,1]$ for all $s'\in\statespace$, there is no overflow when computing the weighted sum. Hence, the weighted sum is non-modular.
Further, we apply the rotation matrix $U_{s'}$ and the oracle $\MDPQoracle$ in sequence to obtain the state 
\begin{equation}\label{equ: inner product of P and V}
    \ket{a}\ket{s}\ket{h}\ket{s'}\ket{\fixbinary{r_{h}(s,a)}}\ket{\fixbinary{\Phsa(s')}}\ket{0}^{\tensorproduct q}\ket{\sum_{s'\in\statespace}\fixbinary{\Phsa(s')}\fixbinary{\hat{V}_{h+1}(s')}}.
\end{equation}
Then we apply the quantum adder $\qAdder$ to obtain the state 
\begin{equation}
    \ket{a}\ket{s}\ket{h}\ket{s'}\ket{\fixbinary{r_{h}(s,a)}}\ket{\fixbinary{\Phsa(s')}}\ket{0}^{\tensorproduct q}\ket{\fixbinary{r_{h}(s,a)}+\sum_{s'\in\statespace}\fixbinary{\Phsa(s')}\fixbinary{\hat{V}_{h+1}(s')} }.
\end{equation}
Since there are $q+1$ qubits in the last register in \eqref{equ: inner product of P and V}, then  the sum of $\fixbinary{r_{h}(s,a)}$ and $\sum_{s'\in\statespace}\fixbinary{\Phsa(s')}\fixbinary{\hat{V}_{h+1}(s')}$ has no overflow and the result is non-modular. Therefore, by applying $\MDPQoracle^{\dagger}$ and $U_{s'}^{\dagger}$ in sequence, we can obtain the state 
\begin{equation}
    \ket{a}\ket{s}\ket{h}\ket{0}^{\tensorproduct 3q+q_{s}}\ket{\fixbinary{r_{h}(s,a)}+\sum_{s'\in\statespace}\fixbinary{\Phsa(s')}\fixbinary{\hat{V}_{h+1}(s')} }.
\end{equation}
\noindent\textbf{Remark:} Since the above operations are unitary, then the oracle $\BinaryOracle{\hat{Q}_{h,s}}$ constructed in this way is unitary. 
Instead of preparing \( \ket{\fixbinary{\Phsa(1)}}\ket{\fixbinary{\hat{V}_{h+1}(1)}}\cdots \ket{\fixbinary{\Phsa(S)}}\ket{\fixbinary{\hat{V}_{h+1}(S)}}\ket{0}^{\tensorproduct q+1} \) and computing the weighted sum on the last register, which requires \( q(2S+1)+1 \) qubits as shown in Section 8 in \cite{ruiz2017quantum}, our method significantly reduces the number of required qubits, needing only \( 3q+1 \) qubits to compute the weighted sum. 
The main idea is to reuse the \( 2q \) auxiliary qubits \( \ket{\fixbinary{\Phsa(s')}}\ket{\fixbinary{\hat{V}_{h+1}(s')}} \) by leveraging the invertible property of the unitary matrices \( \MDPQoracle \) and \( \BinaryOracle{\hat{V}_{h+1}} \) for all \( h \in [H] \). However, this comes at the cost of an additional \( S \) queries to \( \MDPQoracle^{\dagger} \).
We summarize the above results in creating the oracle $\BinaryOracle{\hat{Q}_{h,s}}$ in the following Theorem \ref{Thm: number of qubits of UQHS}.
\begin{theorem}[Number of Qubits Required for the Construction of Oracle $\BinaryOracle{\hat{Q}_{h,s}}$]\label{Thm: number of qubits of UQHS}
The total number of qubits required for creating a quantum oracle $\BinaryOracle{\hat{Q}_{h,s}}$ for each $h\in[H]$ and $s\in\statespace$ in Algorithm \ref{algo: quantum VI-finite horizon MDP-optimal action and V value v1} is $4q+2q_{s}+q_{h}+q_{a}+1$, among which $3q+2q_{s}+q_{h}$ are auxiliary qubits, not counting the auxiliary qubits necessary to implement the oracle $\MDPQoracle$ and $\BinaryOracle{\hat{V}_{h+1}}$ for all $h\in[H]$, where $q_{s}=\ceil{\log_{2}S}$, $q_{a}=\ceil{\log_{2}A}$ and $q_{h}=\ceil{\log_{2}H}$.
\end{theorem}

\begin{theorem}[Number of Qubits Required for \textbf{QVI-1} (Algorithm \ref{algo: quantum VI-finite horizon MDP-optimal action and V value v1})]\label{Thm: number of qubits of algorithm in precise case}
The total number of qubits required for Algorithm \ref{algo: quantum VI-finite horizon MDP-optimal action and V value v1} is $11q+4q_{s}+2q_{h}+4q_{a}+2$, not counting the auxiliary qubits necessary to implement the oracle $\MDPQoracle$ and $\BinaryOracle{\hat{V}_{h+1}}$ for all $h\in[H]$, where $q_{s}=\ceil{\log_{2}S}$, $q_{a}=\ceil{\log_{2}A}$ and $q_{h}=\ceil{\log_{2}H}$.
\end{theorem}
\beginproof
In line 6 of Algorithm \ref{algo: quantum VI-finite horizon MDP-optimal action and V value v1}, we apply $\qArgmax$ algorithm to achieve a quadratic speedup in searching the optimal action which achieves maximum value of the vector $\hat{Q}_{h,s}$ for all $s\in\statespace$ and $h\in[H]$. In this $\qArgmax$ algorithm, it requires an oracle $\QMSoracle$ to mark the indexes $a$ of the vector $\hat{Q}_{h,s}$ which satisfy $\hat{Q}_{h,s}(a)>\hat{Q}_{h,s}(a')$, where $a'$ is a threshold index, by flipping the phase of the $\ket{a}$. Therefore, the oracle $\QMSoracle$ is defined as
\begin{equation}
    \QMSoracle: \ket{a}\ket{a'}\ket{-}\mapsto (-1)^{f_{a'}(a)}\ket{a}\ket{a'}\ket{-},
\end{equation}
where $f_{a'}(a)=1$ if $\hat{Q}_{h,s}(a)>\hat{Q}_{h,s}(a')$ and $f_{a'}(a)=0$ otherwise, and $\ket{-}=\frac{1}{\sqrt{2}}(\ket{0}-\ket{1})$. Figure 8 in \cite{oliveira2007quantum} showed a way to construct the corresponding unitary quantum circuit of the oracle $\QMSoracle$ based on the quantum bit string comparator (QBSC) $\QBSC$. Given two real number $a$ and $b$, $\QBSC$ works as \begin{equation}
    \QBSC: \ket{\fixbinary{a}}\ket{\fixbinary{b}}\ket{0}^{\tensorproduct 3(q-1)}\ket{0}\ket{0}\mapsto\ket{\fixbinary{a}}\ket{\fixbinary{b}}\ket{\psi}\ket{x}\ket{y},
\end{equation}
where $\ket{\psi}$ is a $3(q-1)$-qubit garbage state and the last two qubits store the comparison result. Specifically, we define that if $a=b$ then $x=y=0$, if $a>b$ then $x=1$ and $y=0$, and if $a<b$ then $x=0$ and $y=1$. We restate the construction process under the background of our algorithm here. The first step to construct $\QMSoracle$ is to prepare the following qubits
\begin{equation}
    \ket{a}\ket{0}^{\tensorproduct q}\ket{a'}\ket{0}^{\tensorproduct q}\ket{0}^{\tensorproduct 3(q-1)}\ket{0}\ket{0}\ket{-}.
\end{equation}
 Then we apply $\BinaryOracle{\hat{Q}_{h,s}}$ created in line 5 to the first and third register and obtain the following state
\begin{equation}\label{equ: construction Oqms with Uqhs 1}
    \ket{a}\ket{\fixbinary{\hat{Q}_{h,s}(a)}}\ket{a'}\ket{\fixbinary{\hat{Q}_{h,s}(a')}}\ket{0}^{\tensorproduct 3(q-1)}\ket{0}\ket{0}\ket{-}.
\end{equation}
Then we apply $\QBSC$ to compare $\hat{Q}_{h,s}(a)$ and $\hat{Q}_{h,s}(a')$
\begin{equation}
        \ket{a}\ket{\fixbinary{\hat{Q}_{h,s}(a)}}\ket{a'}\ket{\fixbinary{\hat{Q}_{h,s}(a')}}\ket{\psi}\ket{x}\ket{y}\ket{-}.
\end{equation}
Further, we apply the controlled unitary matrix $U_{c}=(I\tensorproduct \sigma_{1}\tensorproduct I)T(I\tensorproduct \sigma_{1}\tensorproduct I)$ to the last three qubits and obtain the following state
\begin{equation}
        (-1)^{x(1-y)}\ket{a}\ket{\fixbinary{\hat{Q}_{h,s}(a)}}\ket{a'}\ket{\fixbinary{\hat{Q}_{h,s}(a')}}\ket{\psi}\ket{x}\ket{y}\ket{-},
\end{equation}
where $\sigma_{1}$ is the Pauli-X gate and $T$ is the Tofolli gate. By applying $\QBSC^{\dagger}$ and  $\BinaryOracle{\hat{Q}_{h,s}}^{\dagger}$, we can undo the operations on $\ket{\fixbinary{\hat{Q}_{h,s}(a)}}$ and $\ket{\fixbinary{\hat{Q}_{h,s}(a')}}$ and obtain the following state
\begin{equation}\label{equ: construction Oqms with Uqhs 2}
        (-1)^{x(1-y)}\ket{a}\ket{0}^{\tensorproduct q}\ket{a'}\ket{0}^{\tensorproduct q}\ket{0}^{\tensorproduct 3(q-1)}\ket{0}\ket{0}\ket{-}.
\end{equation}

From the above steps, we can see that the construction of the $\QMSoracle$ requires one query to $\BinaryOracle{\hat{Q}_{h,s}}$ and one query to $\BinaryOracle{\hat{Q}_{h,s}}^{\dagger}$ as shown in \eqref{equ: construction Oqms with Uqhs 1} and \eqref{equ: construction Oqms with Uqhs 2}. By Theorem \ref{Thm: number of qubits of UQHS}, we know that it requires $4q+2q_{s}+q_{h}+q_{a}+1$ qubits to construct the oracle $\BinaryOracle{\hat{Q}_{h,s}}$. Then it requires $2(4q+2q_{s}+q_{h}+q_{a}+1)+2q_{a}+3(q-1)+3=11q+4q_{s}+2q_{h}+4q_{a}+2$ qubits to construct the oracle $\QMSoracle$, among which $2(4q+2q_{s}+q_{h}+q_{a}+1)+3(q-1)+2=11q+4q_{s}+2q_{h}+2q_{a}+1$ are auxiliary qubits. 
\done

\subsection{Correctness and Complexity of \textbf{QVI-2} (Algorithm \ref{algo: quantum VI-finite horizon MDP-optimal action and V value v2})}\label{appendix: QVItwo}

\subsubsection{Proof of Theorem \ref{thm: convert binary oracle to a sub probability oracle}}
\begin{theorem}\label{thm: convert binary oracle to a sub probability oracle}
   Let $\Omega$ be a finite set with cardinality $N$, $p=(p_{x})_{x\in\Omega}$ a discrete probability distribution over $\Omega$.  Suppose we have access to a binary oracle $\BinaryOracle{p}: \ket{i}\ket{0}\mapsto \ket{i}\ket{\fixbinary{p_{i}}}$. By using $O(1)$ invocations of the oracle $\BinaryOracle{p}$ and $\BinaryOracle{p}^{\dagger}$, we can implement a unitary oracle $\hat{U}_{p}: \ComplexSpace{N}\tensorproduct \ComplexSpace{2}\rightarrow\ComplexSpace{N}\tensorproduct \ComplexSpace{2}$ satisfying
   \begin{equation}\label{equ: hatUp from Bp}
       \hat{U}_{p}: \ket{i}\ket{0}\mapsto \frac{1}{\sqrt{N}}\sum_{i=1}^{N}\sqrt{p_{i}}\ket{i}\ket{0}+\sqrt{\frac{N-1}{N}}\sum_{i=1}^{N}\sqrt{\frac{1-p_{i}}{N-1}}\ket{i}\ket{1}.
   \end{equation}
\end{theorem}
\beginproof
 First, we need to create the uniform superposition by applying Hadamard gates and query oracle $\BinaryOracle{p}$
    \begin{equation}
        \ket{i}\ket{0}\overset{H^{\tensorproduct n}}{\rightarrow} \frac{1}{\sqrt{N}}\sum_{i=1}^{N}\ket{i}\ket{0}\overset{\BinaryOracle{p}}{\rightarrow}\frac{1}{\sqrt{N}}\sum_{i=1}^{N}\ket{i}\ket{\fixbinary{p_{i}}}.
    \end{equation}
    Second, we add a single auxiliary qubit and perform a controlled rotation $R_{p}$ based on the value stored in $\ket{\fixbinary{p_{i}}}$ defined as $        R_{p}: \ket{\fixbinary{p_{i}}}\ket{0}\mapsto \ket{\fixbinary{p_{i}}} (\sqrt{p_{i}}\ket{0}+\sqrt{1-p_{i}}\ket{1})$:
    \begin{equation}
        \overset{I\tensorproduct R_{p}}{\rightarrow} \frac{1}{\sqrt{N}}\sum_{i=1}^{N}\ket{i}\ket{\fixbinary{p_{i}}}
        \bigl( \sqrt{p_{i}}\ket{0}+\sqrt{1-p_{i}}\ket{1} \bigr).
    \end{equation}
    Third, we undo the oracle $\BinaryOracle{p}$ and drop the auxilliary qubit $\ket{0}$ in Eq. $(a)$ to obtain the desired result.
    \begin{equation}
        \begin{aligned}
        \overset{\BinaryOracle{p}^{\dagger}}{\rightarrow}& \frac{1}{\sqrt{N}}\sum_{i=1}^{N}\ket{i}\ket{0}(\sqrt{p_{i}}\ket{0}+\sqrt{1-p_{i}}\ket{1})\\
        &\overset{(a)}{=}\frac{1}{\sqrt{N}}\sum_{i=1}^{N}\sqrt{p_{i}}\ket{i}\ket{0}+\frac{1}{\sqrt{N}}\sum_{i=1}^{N}\sqrt{1-p_{i}}\ket{i}\ket{1}\\
        &=\frac{1}{\sqrt{N}}\sum_{i=1}^{N}\sqrt{p_{i}}\ket{i}\ket{0}+\sqrt{\frac{N-1}{N}}\sum_{i=1}^{N}\sqrt{\frac{1-p_{i}}{N-1}}\ket{i}\ket{1}.
        \end{aligned}
    \end{equation}
\done

\begin{lemma}[Powering lemma \cite{jerrum1986random}]\label{lem: powering lemma}
    Let $\mathcal{K}$ be a classical or quantum algorithm designed to estimate a quantity $\mu$, where its output $\Tilde{\mu}$ satisfies $|\mu - \Tilde{\mu}| \leq \epsilon$ with probability at least $1 - \gamma$, for some fixed $\gamma < 1/2$. Then, for any $\delta > 0$, by repeating $\mathcal{K}$ $O(\log(1/\delta))$ times and taking the median of the outputs, one can obtain an estimate $\hat{\mu}$ such that $|\hat{\mu} - \mu| < \epsilon$ with probability at least $1 - \delta$.
\end{lemma}
\begin{theorem}[Amplitude estimation \cite{brassard2002quantum}]\label{thm: amplitude estimation}
The amplitude estimation algorithm is designed to estimate the amplitude $a = \bra{\psi}P\ket{\psi} \in [0,1]$ of a quantum state $\ket{\psi}$. It takes the following inputs, a quantum state $\ket{\psi}$, two unitary operators: $U = 2\ket{\psi}\bra{\psi} - I$ and $V = I - 2P$, where $P$ is some suitable projector, and an integer $T$, which determines the number of repetitions.
The algorithm outputs an estimate $\Tilde{a} \in [0,1]$ for the amplitude $a$. The estimate satisfies the error bound:
\begin{equation}
    |\Tilde{a} - a| \leq 2\pi \frac{\sqrt{a(1-a)}}{T} + \frac{\pi^2}{T^2},
\end{equation}
with a success probability of at least $8/\pi^2$. To achieve this, the unitary operators $U$ and $V$ are applied $T$ times each.
\end{theorem}

\subsubsection{Complete Version of Quantum Mean Estimation with Binary Oracle \textbf{QMEBO}}
In Section \ref{sec: exact dynamics setting}, we provide a simplified version of \textbf{QMEBO} by hiding the details of some auxiliary states and operators. For clarity, we provide a complete version in Algorithm \ref{algo: quantum mean estimation with binary oracle-appendix}. Based the Algorithm \ref{algo: quantum mean estimation with binary oracle-appendix}, the auxiliary state in line 5 of \textbf{QMEBO} in Algorithm \ref{algo: quantum mean estimation with binary oracle} should be 
\begin{equation}
    \ket{\Phi^{(1)}}=\frac{1}{\sqrt{N}}\sum_{i=1}^{N}\sqrt{1-p_{i}}\ket{i}\ket{1}\ket{\fixbinary{f_{i}}}\ket{0},
\end{equation}
and the auxiliary state in line 6 satisfies
 \begin{equation}
     \ket{\Phi^{(2)}}=\frac{1}{\sqrt{N}}\sum_{i=1}^{N}\sqrt{p_{i}(1-f_{i})}\ket{i}\ket{001}+\frac{1}{\sqrt{N}}
        \sum_{i=1}^{N}\sqrt{1-p_{i}}\ket{i}\ket{1}\ket{0}
        \bigl( \sqrt{f_{i}}\ket{0}+\sqrt{1-f_{i}}\ket{1} \bigr)
 \end{equation}

\begin{algorithm*}
\caption{Quantum Mean Estimation with Binary Oracles $\qEstBO_{\delta}(p^{\transpose}f, \BinaryOracle{p},\BinaryOracle{f}, \epsilon)$}
\label{algo: quantum mean estimation with binary oracle-appendix}
\begin{algorithmic}[1]
\STATE \textbf{Require:} $\BinaryOracle{p}$ encoding a probability distribution $p=(p_{i})_{i\in\Omega}$ on a finite set $\Omega$ with cardinality $N$, $\BinaryOracle{f}$ encoding a function $f=(f_{i})_{i\in\Omega}$ where $f_{i}\in [0, 1]$, maximum error $\epsilon$, maximum failure probability $\delta\in(0,1)$.
\STATE \textbf{Output: } $\hat{\mu}$ satisfying $|\hat{\mu}-p^{\transpose}f|\leq \epsilon$
\STATE \textbf{Initialize:} $K=O\bigl(\log(1/\delta) \bigr)$, $T=O\Bigl(\frac{\sqrt{N}}{\epsilon}+\sqrt{\frac{N}{\epsilon}}\Bigr)$

\FOR{$k\in [K]$}
    \STATE create a quantum oracle $\hat{U}_{p}$ with $\BinaryOracle{p}$ and obtain the following state $\ket{\psi^{(0)}}=\hat{U}_{p}\ket{0}\ket{0}$:
    \\
    \quad $\ket{\psi^{(0)}}=\hat{U}_{p}\ket{0}\ket{0}=\frac{1}{\sqrt{N}}\sum_{i=1}^{N}\sqrt{p_{i}}\ket{i}\ket{0}+\sqrt{\frac{N-1}{N}}\sum_{i=1}^{N}\sqrt{\frac{1-p_{i}}{N-1}}\ket{i}\ket{1}$.
    \STATE Attach $\ket{0}^{\tensorproduct (q+1)}$ qubits on $\ket{\psi^{(0)}}$ and apply $\BinaryOracle{f}$ on $\ket{0}^{\tensorproduct q}$ to obtain $\ket{\psi^{(1)}}=\BinaryOracle{f}\ket{\psi^{(0)}}\ket{0}^{\tensorproduct (q+1)}$:  \\ 
    \vspace{-10pt}\quad 
    \begin{align*}
\ket{\psi^{(1)}}&=\BinaryOracle{f}\ket{\psi^{(0)}}\ket{0}^{\tensorproduct q+1}\\
&=\frac{1}{\sqrt{N}}\sum_{i=1}^{N}\sqrt{p_{i}}\ket{i}\ket{0}\ket{\fixbinary{f_{i}}}\ket{0}+\underbrace{\frac{1}{\sqrt{N}}\sum_{i=1}^{N}\sqrt{1-p_{i}}\ket{i}\ket{1}\ket{\fixbinary{f_{i}}}\ket{0}}_{=\ket{\Phi^{(1)}}}.
    \end{align*}
    \STATE Apply the controlled rotation $R_{f}$ defined by $        R_{f}: \ket{\fixbinary{f_{i}}}\ket{0}\mapsto \ket{\fixbinary{f_{i}}} (\sqrt{f_{i}}\ket{0}+\sqrt{1-f_{i}}\ket{1})$
    and undo the oracle $\BinaryOracle{f}$:\\
     \vspace{-20pt}
    \quad 
    \begin{align*}
        \ket{\psi^{(2)}}&=(B^{\dagger}_{f}\tensorproduct I)(I\tensorproduct R_{f})\ket{\psi^{(1)}}\\
        &=\frac{1}{\sqrt{N}}\sum_{i=1}^{N}\sqrt{p_{i}}\ket{i}\ket{0}\ket{0}
        \bigl(\sqrt{f_{i}}\ket{0}+\sqrt{1-f_{i}}\ket{1}\bigr)+\frac{1}{\sqrt{N}}
        \sum_{i=1}^{N}\sqrt{1-p_{i}}\ket{i}\ket{1}\ket{0}\bigl(\sqrt{f_{i}}\ket{0}+\sqrt{1-f_{i}}\ket{1}\bigr)\\
        &=\frac{1}{\sqrt{N}}\sum_{i=1}^{N}\sqrt{p_{i}f_{i}}\ket{i}\ket{000}+\underbrace{\frac{1}{\sqrt{N}}\sum_{i=1}^{N}(\sqrt{p_i(1-f_i)}\ket{i}\ket{001}+\sqrt{1-p_i}\ket{i}\ket{10}(\sqrt{f_i}\ket{0}+\sqrt{1-f_i}\ket{1}))}_{=\ket{\Phi^{(2)}}}.
    \end{align*}
    \STATE Apply $T$ iterations of amplitude estimation by setting $\ket{\psi}=\ket{\psi^{(2)}}, U=2\ket{\psi}\bra{\psi}-I$ and $P=I\tensorproduct \ket{000}\bra{000}$ to obtain $\mu_{k}$
\ENDFOR
\STATE \textbf{Return:} $\hat{\mu}=N\cdot \text{Median}(\{\mu_{k}\}_{k\in[K]})$
\end{algorithmic}
\end{algorithm*}

\subsubsection{Proof of Theorem \ref{thm: quantum mean estimation with binary oracle}}\label{appendix: proof of QMEBO}
\beginproof
We first show the correctness of Algorithm \ref{algo: quantum mean estimation with binary oracle}. Note that we obtain 
    \begin{equation}
        \ket{\psi^{(2)}}=\frac{1}{\sqrt{N}}\sum_{i=1}^{N}\sqrt{p_{i}}\ket{i}\ket{0}\ket{0}
        \bigl( \sqrt{f_{i}}\ket{0}+\sqrt{1-f_{i}}\ket{1} \bigr)+\sqrt{\frac{N-1}{N}}\ket{\Phi},
    \end{equation}
    where $\ket{\Phi}=\sum_{i=1}^{N}\sqrt{\frac{1-p_{i}}{N-1}}\ket{i}\ket{1}\ket{0}(\sqrt{f_{i}}\ket{0}+\sqrt{1-f_{i}}\ket{1})$. Besides, we have 
    \begin{equation}
        \bra{\psi^{(2)}}P\ket{\psi^{(2)}}=\frac{1}{N}\sum_{i=1}^{N}p_{i}f_{i}=\frac{1}{N} E[f(x)|x\sim p]=\frac{1}{N}\mu,
    \end{equation} 
    where $P=I\tensorproduct \ket{000}\bra{000}$. Hence, by Theorem \ref{thm: amplitude estimation}, we know that we can obtain $\mu_{k}$ in each loop $k\in[K]$ such that
        \begin{equation}
        \Biggl| \mu_{k}- \frac{1}{N}\sum_{i=1}^{N}p_{i}f_{i} \Biggr|
        \leq 2\pi \frac{\sqrt{\frac{\mu}{N}\Bigl(1-\frac{\mu}{N}\Bigr)}}{T}+\frac{\pi^{2}}{T^{2}},    \end{equation}
        with probability at least $8/\pi^{2}$. Let $\hat{\mu}=N\cdot \Tilde{\mu}$, where $\Tilde{\mu}=\text{Median}(\mu_{0},\ldots,\mu_{K-1})$. By Lemma \ref{lem: powering lemma}, we know that $\Tilde{\mu}=\hat{\mu}/N$ satisfies 
        \begin{equation}\label{equ: amplitude estimation inequality}
        \Biggl| \frac{\hat{\mu}}{N}-\frac{1}{N}\sum_{i=1}^{N}p_{i}f_{i} \Biggr|\leq 2\pi \frac{\sqrt{\frac{\mu}{N}\Bigl(1-\frac{\mu}{N}\Bigr)}}{T}+\frac{\pi^{2}}{T^{2}},    \end{equation}
        with probability at least $1-\delta$ for any $\delta>0$.  
    We proceed to focus on the complexity cost of Algorithm \ref{algo: quantum mean estimation with binary oracle}. Note that $\mu\in[0,1]$ because $f_{i}\in[0,1]$ for all $i=1, \ldots, N$. Hence, we further have that
    \begin{equation}
        2\pi \frac{\sqrt{\frac{\mu}{N}\Bigl(1-\frac{\mu}{N}\Bigr)}}{T}+\frac{\pi^{2}}{T^{2}}
        < \pi^{2}\Biggl( \frac{1}{\sqrt{N}T}+\frac{1}{T^{2}} \Biggr),
    \end{equation}
    In order to let $\hat{\mu}/N$ be an $\epsilon/N$ approximation of $\frac{1}{N}\sum_{i=1}^{N}p_{i}f_{i}$, it suffices to let 
    \begin{equation}
        \pi^{2}\biggl(\frac{1}{\sqrt{N}T}+\frac{1}{T^{2}} \biggr)\leq \frac{\epsilon}{N},
    \end{equation}
    which is equivalent to $\epsilon T^{2}-\pi^{2}\sqrt{N}T-\pi^{2}N\geq 0$.
    Then it suffices to let $T=O\Bigl(\frac{\sqrt{N}}{\epsilon}+\sqrt{\frac{N}{\epsilon}}\Bigr)$ such that 
    $\bigl| \frac{\hat{\mu}}{N}-\frac{1}{N}\sum_{i=1}^{N}p_{i}f_{i} \bigr|\leq \epsilon/N$. This implies that $\bigl|\hat{\mu}-\sum_{i=1}^{N}p_{i}f_{i} \bigr|\leq \epsilon$. 
    
    By Theorem \ref{thm: convert binary oracle to a sub probability oracle}, we know that the query complexity of $\hat{U}_{p}$ in terms of $\BinaryOracle{p}$ is $O(1)$. Therefore, Algorithm \ref{algo: quantum mean estimation with binary oracle} calls $\BinaryOracle{p}$ and $\BinaryOracle{f}$ 
    $O\biggl( \Bigl(\frac{\sqrt{N}}{\epsilon}+\sqrt{\frac{N}{\epsilon}} \Bigr)\log(1/\delta) \biggr)$ times each.
    \done

\subsubsection{Proof of Lemma \ref{lemma: iterative property of the value operator for QVI2}}

\begin{lemma}\label{lemma: iterative property of the value operator for QVI2}
    $\QVItwo$ holds that $ \policyvalueoperator{h}{\hat{V}_{h+1}}{\hat{\pi}}-\frac{\epsilon}{H}\leq \hat{V}_{h}\leq \policyvalueoperator{h}{\hat{V}_{h+1}}{\hat{\pi}}$ for all $h\in[H]$
    with a success probability at least $1-\delta$.
\end{lemma}
\beginproof
The analysis on success probability is the same as Theorem \ref{Thm: correctness of QVI2} and hence we omit it here.
For all $s\in\statespace, a\in\actionspace$ and $h\in[H]$ we have that 
    \begin{equation}
        \Biggl| \frac{z_{h,s}(a)}{H} + \frac{\epsilon}{2H^{2}} - \Phsa^{\transpose}\Tilde{V}_{h+1} \Biggl|\leq \frac{\epsilon}{2H^{2}}.
    \end{equation}
    This implies that 
        \begin{equation}
        \Bigl| z_{h,s}(a) + \frac{\epsilon}{2H} - \Phsa^{\transpose}\hat{V}_{h+1}
        \Bigr|\leq \frac{\epsilon}{2H},
    \end{equation}
    and 
    \begin{equation}\label{equ: gap between zhsa and PV in QVI2}
        \Phsa^{\transpose}\hat{V}_{h+1}-\frac{\epsilon}{H}\leq z_{h,s}(a)\leq \Phsa^{\transpose} \hat{V}_{h+1}.
    \end{equation}
    Now, for all $s\in\statespace, a\in\actionspace$ and $h\in[H]$, we let 
    \begin{equation}
        \Tilde{Q}_{h}(s,a)\define r_{h}(s,a)+\Phsa^{\transpose}\hat{V}_{h+1}.
    \end{equation}
    Note that $\policyvalueoperator{h}{\hat{V}_{h+1}}{\hat{\pi}}=\Tilde{Q}_{h}\bigl( s,\hat{\pi}(s,h) \bigr)= r_{h}(s,\hat{\pi}(s,h))+P_{h|s,\hat{\pi}(s,h)}^{T}\hat{V}_{h+1}$. Therefore,  for all $s\in\statespace, a\in\actionspace$ and $h\in[H]$ we have
    \begin{equation}
        \hat{Q}_{h,s}(a)-\Tilde{Q}_{h}(s,a)=\max\{r_{h}(s,a)+z_{h,s}(a), 0\}-\bigl( r_{h}(s,a)+\Phsa^{\transpose}\hat{V}_{h+1} \bigr).
    \end{equation}
    On one hand, since $z_{h,s}(a)\leq \Phsa^{\transpose}\hat{V}_{h+1}$ and $\hat{V}_{h+1}\geq 0$, then we have 
    \begin{equation}
            \begin{aligned}
        \hat{Q}_{h,s}(a)-\Tilde{Q}_{h}(s,a)
        &\leq
        \max\bigl\{ r_{h}(s,a)+\Phsa^{\transpose}\hat{V}_{h+1}, 0 \bigr\}-\bigl( r_{h}(s,a)+\Phsa^{\transpose}\hat{V}_{h+1} \bigr)\\
        &= r_{h}(s,a)+\Phsa^{\transpose}\hat{V}_{h+1} -\bigl( r_{h}(s,a)+\Phsa^{\transpose}\hat{V}_{h+1} \bigr)\\
        &=0.
    \end{aligned}
    \end{equation}

    On the other hand, we also have
    \begin{equation}
            \begin{aligned}
        \hat{Q}_{h,s}(a)-\Tilde{Q}_{h}(s,a)&=\max\{r_{h}(s,a)+z_{h,s}(a), 0\}-\bigl( r_{h}(s,a)+\Phsa^{\transpose}\hat{V}_{h+1} \bigr)\\
        &\geq r_{h}(s,a)+z_{h,s}(a)- \bigl( r_{h}(s,a)+\Phsa^{\transpose}\hat{V}_{h+1} \bigr)\\
        &= z_{h,s}(a)-\Phsa^{\transpose}\hat{V}_{h+1}\\
        &\geq -\frac{\epsilon}{H}.
    \end{aligned}
    \end{equation}

    The last line comes from Eq. \eqref{equ: gap between zhsa and PV in QVI2}.
    In summary, for all $s\in\statespace, a\in\actionspace$ and $h\in[H]$, we have
    \begin{equation}
        -\frac{\epsilon}{H}\leq \hat{Q}_{h,s}(a)- \Tilde{Q}_{h}(s,a)\leq 0.
    \end{equation} 
        Hence, by letting $a=\hat{\pi}(s,h)$, we will have
    \begin{equation}
        -\frac{\epsilon}{H}\leq \hat{V}_{h}-\policyvalueoperator{h}{\hat{V}_{h+1}}{\hat{\pi}}=\hat{Q}_{h,s}\bigl( \hat{\pi}(s,h) \bigr)-\Tilde{Q}_{h}\bigl( s,\hat{\pi}(s,h) \bigr)\leq 0,
    \end{equation}
        \begin{equation}
        \policyvalueoperator{h}{\hat{V}_{h+1}}{\hat{\pi}}-\frac{\epsilon}{H}\leq \hat{V}_{h}\leq \policyvalueoperator{h}{\hat{V}_{h+1}}{\hat{\pi}}.
    \end{equation}
    \done

\subsubsection{Monotonicity Property of the value operator associated with a policy $\policyvalueoperator{h}{\cdot}{\pi}$ in Definition \ref{def: value operator associated policy}}
Suppose two vectors $u$ and $v$ satisfy $u\leq v \in\realnumber^{\statespace}$, then it implies that $u(s)\leq v(s)$ for all $s\in\statespace$. Consequently, we must have, for any fixed policy $\pi$ and for all $s\in\statespace$ and $h\in[H]$,
\begin{equation}
    \sum_{s'\in\statespace} P_{h|s,\pi(s,h)}(s') u(s') \leq \sum_{s'\in\statespace} P_{h|s,\pi(s,h)}(s') v(s').
\end{equation}
Further, we can know that 
\begin{equation}
    r\bigl( s,\pi(s,h) \bigr)+\sum_{s'\in\statespace} P_{h|s,\pi(s,h)}(s') u(s') \leq 
    r\bigl( s, \pi(s,h) \bigr)+\sum_{s'\in\statespace} P_{h|s,\pi(s,h)}(s') v(s').
\end{equation}
By the definitions of $\policyvalueoperator{h}{u}{\pi}$ and   $\policyvalueoperator{h}{v}{\pi}$, this implies that $[\policyvalueoperator{h}{u}{\pi}]_{s}\leq [\policyvalueoperator{h}{v}{\pi}]_{s}$ for all $s\in\statespace$ and $h\in[H]$. In other words, $\policyvalueoperator{h}{u}{\pi}\leq \policyvalueoperator{h}{v}{\pi}$. This implies that the operator $\mathcal{T}^{h}_{\pi}$ is monotonically increasing for any $\pi$ and $h\in[H]$ in the coordinate-wise order.

\subsubsection{Correctness of \textbf{QVI-2} (Proof of Theorem \ref{Thm: correctness of QVI2})} 

\beginproof
We start by examining the failure probability. The approach is similar to the analysis in Theorem \ref{Thm: correctness of QVI1}, except that we must now account for quantum oracles that can fail. To address this, we use fundamental properties of unitary matrices, particularly a quantum analog of the union bound, which states that the failure probabilities of quantum operators (unitary matrices) combine linearly.

In line 5, since $\BinaryOracle{z_{h,s}}$ is constructed using $\qEstBO$ with a failure probability $\zeta$, it is within $2A\zeta$ of its "ideal version." Specifically, this means that there exists an ideal quantum oracle $\BinaryOracle{z_{h,s}}^{\text{ideal}}$ encoding $H\widetilde{\Phsa^{\transpose}\Tilde{V}_{h+1}} - \epsilon/2H$, where $\widetilde{\Phsa^{\transpose}\Tilde{V}_{h+1}}$ satisfies $\infiNorm{\widetilde{\Phsa^{\transpose} \Tilde{V}_{h+1}} - \Phsa^{\transpose}\Tilde{V}_{h+1}} \leq \epsilon/(2H^{2})$, such that $\OPNorm{\BinaryOracle{z_{h,s}}^{\text{ideal}} - \BinaryOracle{z_{h,s}}} \leq 2A\zeta$. Since $\BinaryOracle{\hat{Q}_{h,s}}$ is formed using one call each to $\BinaryOracle{z_{h,s}}$ and $\BinaryOracle{z_{h,s}}^{\dagger}$, it is within $4A\zeta$ of its ideal counterpart $\BinaryOracle{\hat{Q}_{h,s}}^{\text{ideal}}$. By applying the quantum union bound and substituting the definition of $\zeta$, this shows that the quantum operation executed by $\qArgmax$ is $(\Tilde{c}\sqrt{A}\log(1/\delta)\cdot 4A\zeta = \delta/SH)$-close to its ideal version. Consequently, the output of $\qArgmax$ is incorrect with a probability of at most $\delta/SH$. Given that $\qArgmax$ is invoked a total of $SH$ times, the overall failure probability is bounded by $\delta$, as ensured by the standard union bound.

    
    In line 5, we apply $\textbf{QMEBO}$ to obtain an approximate value $z_{h,s}(a)/H$ of the inner product $\Phsa^{\transpose}\Tilde{V}_{h+1}$. Theorem \ref{thm: quantum mean estimation with binary oracle} guarantees the output in the line 5 satisfying $|z_{h,s}(a)/H + \epsilon/2H^{2} -\Phsa^{\transpose}\Tilde{V}_{h+1}|\leq \epsilon/2H^{2}$ for all $s\in\statespace, a\in\actionspace$ and $h\in[H]$. This implies that $|z_{h,s}(a)-\Phsa^{\transpose}\hat{V}_{h+1}|\leq \epsilon/H$.
    Hence, it holds for all $s\in\statespace$ and $a\in\actionspace$ in every loop $h\in[H]$ that 
    \begin{equation}
            \begin{aligned}
        |\hat{Q}_{h,s}(a)-Q_{h}^{*}(s,a)|
        &= \bigl| r_{h}(s,a)+z_{h,s}(a)- \bigl(r_{h}(s,a)+\Phsa^{\transpose}V_{h+1}^{*} \bigr) \bigr|\\
        &\leq \bigl|z_{h,s}(a)- \Phsa^{\transpose}(V_{h+1}^{*}-\hat{V}_{h+1})-\Phsa^{\transpose}\hat{V}_{h+1} \bigr|\\
        &\leq \bigl|z_{h,s}(a)-\Phsa^{\transpose}\hat{V}_{h+1} \bigr|+\bigl| \Phsa^{\transpose}(V_{h+1}^{*}-\hat{V}_{h+1}) \bigr|\\
        &\leq \frac{\epsilon}{H}+\max_{s\in\statespace}
        \bigl| \hat{V}_{h+1}(s)-V_{h+1}^{*}(s) \bigr|\\
        &= \frac{\epsilon}{H}+\infiNorm{\hat{V}_{h+1}-V_{h+1}^{*}}.
    \end{aligned}
\end{equation}

Furthermore, we have
\begin{equation}
    \begin{aligned}
    \infiNorm{\hat{V}_{h}-V_{h}^{*}}&= \infiNorm{ \hat{Q}_{h,s}\bigl( \hat{\pi}(s,h) \bigr)- \max_{a\in\actionspace} Q^{*}_{h}(s, a)}\\
    &= \infiNorm{ \max_{a\in\actionspace}\hat{Q}_{h,s}(a)- \max_{a\in\actionspace} Q^{*}_{h}(s, a)}\\
    &\leq  \infiNorm{ \max_{a\in\actionspace} |\hat{Q}_{h,s}(a)- Q^{*}_{h}(s, a)|}\\
    &\leq \max_{s\in\statespace}\max_{a\in\actionspace} 
    \bigl| \hat{Q}_{h,s}(a)- Q^{*}_{h}(s, a) \bigr|\\
    &\leq \frac{\epsilon}{H} + \infiNorm{\hat{V}_{h+1}-V_{h+1}^{*}}.
\end{aligned}
\end{equation}

Since it holds that $\hat{V}_{H}(s)= V_{H}^{*}(s)=0$ for all $s\in\statespace$, we can induce that 
\begin{equation}\label{equ: Vh-optimalVh}
    \infiNorm{\hat{V}_{h}-V_{h}^{*}}\leq \frac{(H-h)\epsilon}{H}+ \infiNorm{\hat{V}_{H}-V_{H}^{*}}=\frac{(H-h)\epsilon}{H}.
\end{equation}
Then, we know that $\infiNorm{\hat{V}_{h}-V_{h}^{*}}\leq \epsilon$ for all $h\in[H]$. In particular, it implies that $V_{h}^{*}(s)-\epsilon\leq \hat{V}_{h}(s)$ for all $s\in\statespace$ and $h\in[H]$.

Now, we proceed to prove the $\hat{V}_{h}(s)\leq V_{h}^{\hat{\pi}}(s)$ for all $s\in\statespace$ and $h\in[H]$. By Lemma \ref{lemma: iterative property of the value operator for QVI2}, we know that $\hat{V}_{h}\leq \policyvalueoperator{h}{\hat{V}_{h+1}}{\hat{\pi}}$ for all $h\in[H]$. Therefore, $\hat{V}_{H-1}(s)\leq [\policyvalueoperator{H-1}{\hat{V}_{H}}{\hat{\pi}}]_{s}=[\policyvalueoperator{H-1}{0}{\hat{\pi}}]_{s}=r_{H-1}(s,\hat{\pi}(s,H-1))=V^{\hat{\pi}}_{H-1}(s)$. 
By the monotonicity of the operators $\mathcal{T}^{h}_{\hat{\pi}}$, where $h\in[H]$, we have $\hat{V_{h}}\leq \policyvalueoperator{h}{\hat{V}_{h+1}}{\hat{\pi}}\leq \policyvalueoperator{h}{\policyvalueoperator{h+1}{\hat{V}_{h+2}}{\hat{\pi}}}{\hat{\pi}}\leq \cdots\leq V_{h}^{\hat{\pi}}$ for all $h\in[H]$. Due to the definition of $V^{*}_{h}(s)$, we must have $V^{*}_{h}(s)=\max_{\pi\in\Pi}V^{\pi}_{h}(s)\geq V^{\hat{\pi}}_{h}(s)$ for all $s\in\statespace$ and $h\in[H]$.
\done

\subsubsection{Complexity of \textbf{QVI-2} (Proof of Theorem \ref{Thm: Complexity of QVI2})}
\beginproof We first assume that all $\qArgmax$ and $\qEstBO$ are correct, up to the specified error, because the probability that this does not hold is at most $\delta$. Let $C$ be the complexity of $\QVItwo$ as if all $\qArgmax$ and $\qEstBO$ are carried out with maximum failure probabilities set to constant. Then, since the failure probabilities are set to $\zeta= \delta/(4cHSA^{1.5}\log(1/\delta))$, the actual complexity of $\QVItwo$ is \begin{equation}\label{equ: initial complexity of QVI2}
O\Bigl( C\log\bigl(SA^{1.5}H\log(1/\delta)/\delta\bigr) \Bigr) 
= O\bigl( C\log(SA^{1.5}H/\delta) \bigr).
\end{equation}
Now, we check each line of $\QVItwo$ to bound $C$. 


In line 4, we encode the vector $\Tilde{V}_{h+1}=\hat{V}_{h+1}/H$ to an oracle $\BinaryOracle{\Tilde{V}_{h+1}}$. This process does not need to query $\MDPQoracle$ and only needs to access the classical vector $\hat{V}_{h+1}$. 

In line 5, we implement $\qEstBO$ to compute an estimate of $\Phsa^{\transpose}\Tilde{V}_{h+1}$ with error $\epsilon/2H^{2}$. Besides, the correctness analysis shows that $\hat{V}_{h+1}(s)\leq V^{*}_{h+1}(s)\leq H$ for all $s\in\statespace$ and the definition of $\hat{Q}_{h,s}(a)$ of \textbf{QVI2} implies that $0\leq \hat{V}_{h}(s)$ for all $s\in\statespace$, so it also holds that $0\leq \Tilde{V}_{h+1}(s)=\hat{V}_{h+1}(s)/H\leq 1$ for all $s\in\statespace$ and $h\in[H]$. By Theorem \ref{thm: quantum mean estimation with binary oracle}, we know that $\qEstBO$ needs $O\bigl(\sqrt{S}(H^{2}/\epsilon+\sqrt{H^{2}/\epsilon}) \bigr)=O(\sqrt{S}H^{2}/\epsilon)$ queries to $\MDPQoracle$ for each $s\in\statespace$ at each time step $h\in[H]$, provided $0<\epsilon\leq H^{2}$. Since we have assumed that $\epsilon\leq H$ on the input $\epsilon$, so this holds.

In line 6, $\BinaryOracle{\hat{Q}_{h,s}}$ needs to call $\BinaryOracle{z_{h,s}}$ and $\BinaryOracle{z_{h,s}}^{\dagger}$ once. Then the query complexity of $\BinaryOracle{\hat{Q}_{h,s}}$ in terms of $\BinaryOracle{z_{h,s}}$ is $O(1)$.

In line 7, we use $\qArgmax$ in accelerating the searching for the optimal action $\hat{\pi}(s,h)$ for all $s\in\statespace$. By Theorem \ref{Thm: quantum maximum finding}, $\qArgmax$ requires $O(\sqrt{A})$ queries to the oracle $\BinaryOracle{\hat{Q}_{h,s}}$ for all $s\in\statespace$ and $h\in[H]$. 
Therefore, after summing up $H$ iterations, it induces an overall query cost of \begin{equation}
    C=O\Biggl( S\cdot \sqrt{A}\cdot H\cdot  \frac{\sqrt{S}H^{2}}{\epsilon} \Biggr) = O\Biggl( \frac{S^{\frac{3}{2}}\sqrt{A}H^{3}}{\epsilon} \Biggr).
\end{equation}

Combining the above equation with Eq. \eqref{equ: initial complexity of QVI2}, the overall quantum query complexity of $\QVItwo$ is 
\begin{equation}
    O\Biggl(\frac{S^{\frac{3}{2}}\sqrt{A}H^{3}\log(SA^{1.5}H/\delta)}{\epsilon} \Biggr).
\end{equation}
\done

\onlytech{
\subsection{Correctness and Complexity of \textbf{QVI-5} (Algorithm \ref{algo: quantum VI-finite horizon MDP-optimal action and V value v5})}\label{appendix: QVIfive}

\begin{definition}[Probability Oracle]\label{def: probability oracle}
    For a function $p: X\rightarrow [0,1]$, let $\{\ket{x}: x\in X\}$ be an orthonormal basis of a Hilbert space $\hilbertspace$. We define the probability oracle for the function $p$ to be a unitary matrix $\ProbabilityOracle{p}: \hilbertspace\tensorproduct \hilbertspace'\rightarrow \hilbertspace\tensorproduct\hilbertspace'$
    \begin{equation}
        \ProbabilityOracle{p}: \ket{x}\ket{0}\mapsto \ket{x}\tensorproduct (\sqrt{p(x)}\ket{\psi_{0}(x)}\ket{0}+\sqrt{1-p(x)}\ket{\psi_{1}(x)}\ket{1}),
    \end{equation}
    where $\psi_{0}(x)$ and $\psi_{1}(x)$ are arbitrary quantum states.
\end{definition}

\begin{definition}[Phase Oracle]\label{def: phase oracle}
    For a function $p: X\rightarrow [-1,1]$, let $\{\ket{x}: x\in X\}$ be an orthonormal basis of a Hilbert space $\hilbertspace$. We define the phase oracle for the function $p$ to be a unitary matrix $\PhaseOracle{p}: \hilbertspace\tensorproduct \hilbertspace'\rightarrow \hilbertspace\tensorproduct\hilbertspace'$
    \begin{equation}
        \PhaseOracle{p}: \ket{x}\ket{0}\mapsto e^{ip(x)}\ket{x}\ket{0}.
    \end{equation}
\end{definition}


\begin{algorithm}
\caption{$\QVIfive$}
\label{algo: quantum VI-finite horizon MDP-optimal action and V value v5}
\begin{algorithmic}[1]
\STATE \textbf{Require:} MDP $\MDP$, quantum oracle $\MDPQoracle$, maximum error $\epsilon\in (0, H]$, maximum failure probability $\delta\in(0,1)$.
\STATE \textbf{Initialize:} $\zeta\leftarrow\delta/(4\Tilde{c}SA^{1.5}H\log(1/\delta))$, $\hat{V}_{H}(s)=0, \forall s\in\statespace$.
\STATE create quantum oracle $\MDPQPoracle{P''}$ which is defined by\\
\quad $\MDPQPoracle{P''}\leftarrow\BTP\left(\MDPQoracle, \epsilon/(4SH^{2})\right)$
\FOR{$h := H-1, \ldots, 0$}
    \STATE create a quantum oracle $U_{\hat{V}_{h+1}}$ encoding $\hat{V}_{h+1}\in \realnumber^{\statespace}$
    \STATE $\forall s\in\statespace$: create a quantum oracle $U_{z_{h,s}}$ encoding $z_{h,s}\in\realnumber^{\actionspace}$ with $\MDPQPoracle{P''}$ and $U_{\hat{V}_{h+1}}$ satisfying\\
    \quad $z_{h,s}(a)\leftarrow \qEstone_{\zeta}\big((\Phsa^{''T}\hat{V}_{h+1}), \epsilon/(4H)\big)- \epsilon/(2H)$
    \STATE $\forall s\in\statespace$: create a quantum oracle $U_{\hat{Q}_{h,s}}$ encoding $\hat{Q}_{h,s}\in \realnumber^{\actionspace}$ with $\MDPQoracle$ and $U_{z_{h,s}}$ satisfying \\ \quad $\hat{Q}_{h,s}(a)\leftarrow \max\{r_{h}(s, a) + z_{h,s}(a), 0\}$
    \STATE $\forall s\in\statespace$: $\hat{\pi}(s,h) \leftarrow \qArgmax_{\delta}\{\hat{Q}_{h,s}(a): a\in \actionspace\}$ 
    \STATE $\forall s\in\statespace$: $\hat{V}_{h}(s) \leftarrow \hat{Q}_{h,s}\bigl( \hat{\pi}(s,h) \bigr)$ 
\ENDFOR
\STATE \textbf{Return:} $\hat{\pi}$, $\{\hat{V}_{h}\}_{h=0}^{H-1}$
\end{algorithmic}
\end{algorithm}

In this section, we show how to convert the quantum oracle of an MDP $\MDPQoracle$ (Definition \ref{def: quantum oracle of MDP}) to the probability oracle $\MDPQPoracle{P''}$ in Eq. \eqref{equ: MDPQPoracle} (See Definition \ref{def: probability oracle} for the definition of a probability oracle) following the conversion methods illustrated in \cite{gilyen2019optimizing}. Note that the probability distribution $\Phsa''$ encoded in the oracle $\MDPQPoracle{P''}$ may have some deviation with the original probability distribution $\Phsa$ for all $s\in\statespace, a\in\actionspace$ and $h\in[H]$ (See Lemma \ref{lem: MDP from binary oracle to probability oracle}). After that, we can use the probability oracle to implement $\qEstone$ to obtain an estimate of $\Phsa^{''\transpose}V$ for any vector $V\in \realnumber^{\statespace}$, along with a suitably defined projection operator $P=I\tensorproduct\ket{0}\bra{0}$ and a binary oracle $\BinaryOracle{V}$ encoding the vector $V$. The details of \textbf{QVI-5} are shown in Algorithm \ref{algo: quantum VI-finite horizon MDP-optimal action and V value v5}. The correctness and complexity of \textbf{QVI-5} are illustrated in Theorem \ref{Thm: correctness of QVI5} and Theorem \ref{Thm: Complexity of QVI5}, respectively.


First, we construct the probability oracle $\MDPQPoracle{P''}$ with our original binary oracle $\MDPQoracle$ and analyze the query complexity of the probability oracle $\MDPQPoracle{P''}$ in terms of the oracle $\MDPQoracle$. There are two steps to construct the oracle $\MDPQPoracle{P}$. The first step is to convert the oracle $\MDPQoracle$ to a phase oracle.

\begin{lemma}[Converting binary oracles to phase Oracles]\label{lem: convert binary oracle to phase oracle}
   Let $f$ be an arbitrary function satisfying $f: X\rightarrow [0,1]$ where $X$ is a finite set. 
   By using $O(1)$ invocations of the binary oracle $\BinaryOracle{f}: \ket{x}\ket{0}\mapsto \ket{x}\ket{\fixbinary{f(x)}}$ and $\BinaryOracle{f}^{\dagger}$, we can implement a phase oracle $\PhaseOracle{f'}: \ket{x}\ket{0}\mapsto e^{i f'(x)}\ket{x}\ket{0}$, where $|f'(x)-f(x)|\leq (2\pi+1) 2^{-p}$ for all $x\in X$.
\end{lemma}
\beginproof
    We consider a binary oracle $\BinaryOracle{f}: \ket{x}\ket{0}\mapsto\ket{x}\ket{f(x)}$. We first construct another binary oracle $\BinaryOracle{f}^{c}: \ket{x}\ket{0}\mapsto\ket{x}\ket{\fixbinary{c}\fixbinary{f(x)}}$, where $c$ is an arbitrary non-negative real number. We first prepare the following state 
    \begin{equation}
        \ket{x}\ket{0}^{\tensorproduct q}\ket{\fixbinary{c}}\ket{0}^{\tensorproduct q}.
    \end{equation}
    Then we apply $\BinaryOracle{f}$ on the first two registers and obtain the following state
    \begin{equation}
        \ket{x}\ket{\fixbinary{f(x)}}\ket{\fixbinary{c}}\ket{0}^{\tensorproduct q}.
    \end{equation}
    We apply the quantum multiplier $\qMul$ and obtain the following state 
    \begin{equation}
          \ket{x}\ket{\fixbinary{f(x)}}\ket{\fixbinary{c}}\ket{\fixbinary{c} \fixbinary{f(x)}}.
    \end{equation}
    Since $f(x)\in[0,1]$, then we have $cf(x)\in[0, c]$. Hence, the multiplication will not induce an overflow, i.e., it is a non-modular multiplication.
    By applying $\BinaryOracle{f}^{\dagger}$, we can undo the operations on the second register. Now we proceed to construct the phase oracle $\PhaseOracle{p}$ with the new oracle $\BinaryOracle{cp}$ using the so-called "phase kickback" technique. First, we prepare qubits in state 
    \begin{equation}
        \ket{x}\ket{0}^{\tensorproduct q}.
    \end{equation} 
    Further, we apply $\sigma_{1}^{\tensorproduct 2q}$ to the second register, where $\sigma_{1}$ is the Pauli-X gate, and obtain the following state 
    \begin{equation}
        \ket{x}\ket{1}^{\tensorproduct q}.
    \end{equation} 
    We apply the inverse quantum Fourier transform $\IQFT$ on the second register and obtain the following state
        \begin{equation}
        \ket{x}\frac{1}{\sqrt{N}}\sum_{k=0}^{N-1}e^{i\frac{-2\pi k}{N}}\ket{k},
    \end{equation} 
    where $N=2^{q}$. We further apply $\BinaryOracle{f}^{c}$ where $c=\frac{N}{2\pi}$ and obtain the following state
    \begin{equation}
            \begin{aligned}
         &\ket{x}\frac{1}{\sqrt{N}}\sum_{k=0}^{N-1}e^{i\frac{-2\pi k}{N}}\ket{k+\fixbinary{\frac{N}{2\pi}}\fixbinary{f(x)} \mod{N}}\\
         &=\ket{x}\frac{1}{\sqrt{N}}\sum_{k'=0}^{N-1}e^{i\frac{-2\pi (k'-\fixbinary{\frac{N}{2\pi}}\fixbinary{f(x)}}{N}}\ket{k'}\\
         &=e^{i\frac{2\pi}{N}\fixbinary{\frac{N}{2\pi}}\fixbinary{f(x)}} \ket{x}\frac{1}{\sqrt{N}}\sum_{k'=0}^{N-1}e^{i\frac{-2\pi k'}{N}}\ket{k'}.
    \end{aligned}
    \end{equation}

    By applying $\QFT$ and $\sigma_{1}^{\tensorproduct 2q}$ in sequence, we obtain the following state
    \begin{equation}
        e^{i\frac{2\pi}{N}\fixbinary{\frac{N}{2\pi}}\fixbinary{f(x)}}\ket{x}\ket{0}^{\tensorproduct q}=e^{i f'(x) }\ket{x}\ket{0}^{\tensorproduct q},
    \end{equation}
    where $f'(x)=\frac{2\pi}{N}\fixbinary{\frac{N}{2\pi}}\fixbinary{f(x)}$. Besides, we can deduce  that
    \begin{equation}
            \begin{aligned}
        |f'(x)-f(x)|&=|\frac{2\pi}{N}\fixbinary{\frac{N}{2\pi}}\fixbinary{f(x)}-f(x)|\\
        &=\frac{2\pi}{N}|\fixbinary{\frac{N}{2\pi}}\fixbinary{f(x)}-\frac{N}{2\pi}f(x)|\\
        &=\frac{2\pi}{N}|\fixbinary{\frac{N}{2\pi}}\fixbinary{f(x)}-\frac{N}{2\pi}\fixbinary{f(x)}+\frac{N}{2\pi}\fixbinary{f(x)}-\frac{N}{2\pi}f(x)|\\
        &\leq \frac{2\pi}{N} (|\fixbinary{\frac{N}{2\pi}}-\frac{N}{2\pi}||\fixbinary{f(x)}|+|\fixbinary{f(x)}-f(x)|\frac{N}{2\pi})\\
        &\leq \frac{2\pi}{N} (|\fixbinary{\frac{N}{2\pi}}-\frac{N}{2\pi}||f(x)|+|\fixbinary{f(x)}-f(x)|\frac{N}{2\pi})\\
        &\leq \frac{2\pi}{N} (2^{-p}+2^{-p}\frac{N}{2\pi})\\
        &= 2^{-p}+2\pi2^{-(p+q)}\\
        &\leq (2\pi+1) 2^{-p}.
    \end{aligned}
    \end{equation}

    The fourth line follows from the triangle inequality $|a+b|\leq |a|+|b|$. The fifth line follows from the fact that $|\fixbinary{f(x)}|\leq |f(x)|$. The sixth line follows from the facts that $|\fixbinary{\frac{N}{2\pi}}-\frac{N}{2\pi}|\leq 2^{-p}$, $|\fixbinary{f(x)}-f(x)|\leq 2^{-p}$ and $|f(x)|\leq 1$. The last line comes from $p\geq 0$ and $q\geq 0$.
    In the end, we can implement a phase oracle $\PhaseOracle{f'}:\ket{x}\ket{0}\mapsto e^{if'(x)}\ket{x}\ket{0}$ where $|f'(x)-f(x)|\leq (2\pi+1) 2^{-p}$ for all $x\in X$.
    \done

By Lemma \ref{lem: convert binary oracle to phase oracle}, we know that we can construct the phase oracle $\MDPPHoracle{P'}$ defined as 
\begin{equation}
    \MDPPHoracle{P'}: \ket{s}\ket{a}\ket{h}\ket{s'}\ket{0}\mapsto e^{i\Phsa'(s')}\ket{s}\ket{a}\ket{h}\ket{s'}\ket{0},
\end{equation}
with binary oracle $\MDPQoracle$, where $|\Phsa'(s')-\Phsa(s')|\leq (2\pi+1)2^{-p}$ for all $s, s'\in\statespace$, $a\in\actionspace$ and $h\in[H]$. Note that the error is induced by the number of qubits for encoding a number. For simplicity, we assume qubits resource is sufficient for our problem and hence omit the error. Besides, the query complexity of $\MDPPHoracle{P'}$ in terms of $\MDPQoracle$ is $O(1)$. The second step to construct the probability oracle with the binary oracle $\MDPQoracle$ is to convert the phase oracle $\MDPPHoracle{P'}$ to a probability oracle. 
As shown in Lemma 16 in \cite{gilyen2019optimizing}, there indeed exists an efficient procedure to convert a phase oracle to a probability oracle.

\begin{lemma}[Converting phase oracles to probability oracles\cite{gilyen2019optimizing}]\label{lem: convert phase oracle to probability oracle}
    Let $\epsilon, \eta\in (0, 1/2)$, and suppose $p: X\mapsto [\eta, 1-\eta]$. Suppose we have access to a phase oracle $\PhaseOracle{p}$ then using $O(\log(1/\epsilon)/\eta)$ invocations of the (controlled) $\PhaseOracle{p}$ and $\PhaseOracle{p}^{\dagger}$ oracle, we can implement a probability oracle 
    \begin{equation}
        \ProbabilityOracle{p}: \ket{x}\ket{0}^{\tensorproduct k}\ket{0}\mapsto \ket{x}\tensorproduct (\sqrt{p'(x)}\ket{0}^{\tensorproduct k} \ket{0}+ \sqrt{1-p'(x)}\ket{\Phi^{\perp}}\ket{1}),
    \end{equation}
    where $|\sqrt{p'(x)}-\sqrt{p(x)}|\leq \epsilon$ for every $x\in X$.
\end{lemma}


Therefore, we can have the following lemma illustrating the conversion from binary oracle $\MDPQoracle$ to probability oracle $\MDPQPoracle{P''}$. We define the support set of a function $f: X\rightarrow\realnumber$ as $\supp{f}=\{x\in X: f(x)\neq 0\}$. 

\begin{lemma}[Converting $\MDPQoracle$ to $\MDPQPoracle{P''}$]\label{lem: MDP from binary oracle to probability oracle}
    Let $\epsilon, \eta\in (0, 1/2)$.  Suppose $\Phsa: \supp{\Phsa}\rightarrow [\eta, 1-\eta]$ for all $s\in\statespace, a\in\actionspace$ and $h\in[H]$. By using $O(\log(1/\sqrt{\epsilon})/\eta)$ invocations of the (controlled) $\MDPQoracle$ and $\MDPQoracle^{\dagger}$, we can implement a probability oracle $\MDPQPoracle{P''}$
    \begin{equation}\label{equ: MDPQPoracle}
    \MDPQPoracle{P''}: \ket{s}\ket{a}\ket{h}\ket{s'}\ket{0}\mapsto \ket{s}\ket{a}\ket{h}\ket{s'}\tensorproduct (\sqrt{\Phsa''(s')}\ket{\psi_{0}}\ket{0}+\sqrt{1-\Phsa''(s')}\ket{\psi_{1}}\ket{1}),
\end{equation}
    satisfying $|\Phsa(s')-\Phsa''(s')|\leq \epsilon$ for all $s'\in\supp{\Phsa}$, $s\in\statespace, a\in\actionspace$ and $h\in[H]$. We use $\MDPQPoracle{P''}=\BTP(\MDPQoracle, \epsilon)$ to denote the conversion from $\MDPQoracle$ to $\MDPQPoracle{P''}$ with error $\epsilon$.
\end{lemma}
\beginproof
    By Lemma \ref{lem: convert binary oracle to phase oracle} and omitting the numerical error caused by the qubits, we know that we can construct the phase oracle $\MDPPHoracle{P}$ defined as 
\begin{equation}
    \MDPPHoracle{P}: \ket{s}\ket{a}\ket{h}\ket{s'}\ket{0}\mapsto e^{i\Phsa(s')}\ket{s}\ket{a}\ket{h}\ket{s'}\ket{0},
\end{equation}
with binary oracle $\MDPQoracle$. 
Besides, the query complexity of $\MDPPHoracle{P}$ in terms of $\MDPQoracle$ is $O(1)$. Then, by Lemma \ref{lem: convert phase oracle to probability oracle}, we can implement the probability oracle $\MDPQPoracle{P''}$ defined as
\begin{equation}
    \MDPQPoracle{P''}: \ket{s}\ket{a}\ket{h}\ket{s'}\ket{0}\mapsto \ket{s}\ket{a}\ket{h}\ket{s'}\tensorproduct (\sqrt{\Phsa''(s')}\ket{\psi_{0}}\ket{0}+\sqrt{1-\Phsa''(s')}\ket{\psi_{1}}\ket{1}),
\end{equation}
by 
\begin{equation}
    O(\log(\frac{2}{\sqrt{4+4\epsilon}-2})/\eta=O(\log(1/\sqrt{\epsilon})/\eta)
\end{equation}
invocations of the phase oracle $\MDPPHoracle{P}$, where $|\sqrt{\Phsa''(s')}-\sqrt{\Phsa(s')}|\leq \frac{\sqrt{4+4\epsilon}-2}{2}$ for all $s, s'\in\statespace$, $a\in\actionspace$ and $h\in[H]$. Hence, with $O(\log(1/\sqrt{\epsilon})/\eta)$ queries of the binary oracle $\MDPQoracle$, we can implement the probability oracle $\MDPQPoracle{P''}$ satisfying 
\begin{equation}
    \begin{aligned}
    |\Phsa(s')-\Phsa''(s')|&=|\Phsa(s')-\Phsa''(s')|\\
    &\leq |\sqrt{\Phsa(s')}-\sqrt{\Phsa''(s')}||\sqrt{\Phsa(s')}+\sqrt{\Phsa''(s')}|\\
    &\leq  |\sqrt{\Phsa(s')}-\sqrt{\Phsa''(s')}||2\sqrt{\Phsa(s')}+ \frac{\sqrt{4+4\epsilon}-2}{2}|\\
    &\leq  (\frac{\sqrt{4+4\epsilon}-2}{2})( \frac{\sqrt{4+4\epsilon}+2}{2})\\
    &= \epsilon.
\end{aligned}
\end{equation}

The second line follows from the fact that $|a^{2}-b^{2}|\leq |a-b||a+b|$. The third line follows from $\sqrt{\Phsa''(s')}\leq \frac{\sqrt{4+4\epsilon}-2}{2}+\sqrt{\Phsa(s')}$. The third inequality follows from the fact that $|\sqrt{\Phsa''(s')}-\sqrt{\Phsa(s')}|\leq \frac{\sqrt{4+4\epsilon}-2}{2}$.
\done

\begin{lemma}\label{lemma: iterative property of the value operator QVIfive}
    $\QVIfive$ holds that  $ \policyvalueoperator{h}{\hat{V}_{h+1}}{\hat{\pi}}-\frac{\epsilon}{H}\leq \hat{V}_{h}\leq \policyvalueoperator{h}{\hat{V}_{h+1}}{\hat{\pi}}$ and $\hat{V}_{h}\leq V_{h}^{\hat{\pi}}\leq V_{h}^{*}\leq H$ for all $h\in[H]$ with a success probability at least $1-\delta$.
\end{lemma}
\beginproof
The analysis of success probability is the same as Theorem \ref{Thm: correctness of QVI5}.
We use induction to prove the claim. First, we check the basic case when $h=H-1$.
In fact, for all $s\in\statespace$ and $a\in\actionspace$, we have
\begin{equation}
        \begin{aligned}
        |z_{h,s}(a)+\frac{\epsilon}{2H}- \Phsa^{\transpose}\hat{V}_{h+1}|
        &= |z_{H-1,s}(a)+\frac{\epsilon}{2H}- P_{H-1|s,a}^{\transpose}\hat{V}_{H}|\\
        &=|z_{H-1,s}(a)+\frac{\epsilon}{2H}-P_{H-1|s,a}^{''\transpose}\hat{V}_{H}+P_{H-1|s,a}^{''\transpose}\hat{V}_{H}-P_{H-1|s,a}^{\transpose}\hat{V}_{H}|\\
        &\leq |z_{h,s}(a)+\frac{\epsilon}{2H}-P_{H-1|s,a}^{''\transpose}\hat{V}_{H}|+|(P_{H-1|s,a}^{''}-P_{H-1|s,a})^{\transpose}\hat{V}_{H}|\\
        &\leq |z_{h,s}(a)+\frac{\epsilon}{2H}-P_{H-1|s,a}^{''\transpose}\hat{V}_{H}|+S\max_{s'\in\statespace}|\hat{V}_{H}(s')|\max_{s'\in\statespace}|P_{H-1|s,a}^{''}(s')-\Phsa(s')|\\
        &\leq \frac{\epsilon}{4H}\\
        &<\frac{\epsilon}{2H}.
    \end{aligned}
\end{equation}

The fifth line comes from the facts that $\qEstone$ induces an error $\epsilon/4H$ in estimating $P_{H-1|s,a}^{''\transpose}\hat{V}_{H}$ and $\hat{V}_{H}(s)=0$ for all $s\in\statespace$. Therefore, we have
\begin{equation}
    -\frac{\epsilon}{H}<z_{H-1,s}(a)-P_{H-1|s,a}^{\transpose}\hat{V}_{H}<0.
\end{equation}

 Now, for all $s\in\statespace, a\in\actionspace$ and $h\in[H]$, we let 
    \begin{equation}
        \Tilde{Q}_{h}(s,a)\define r_{h}(s,a)+\Phsa^{T}\hat{V}_{h+1} \text{ and } \Tilde{V}_{h}(s)\define \max_{a\in\actionspace} \Tilde{Q}_{h}(s,a).
    \end{equation}
    Note that $\policyvalueoperator{h}{\hat{V}_{h+1}}{\hat{\pi}}=\Tilde{Q}_{h}\bigl( s,\hat{\pi}(s,h) \bigr)= r_{h}(s,\hat{\pi}(s,h))+P_{h|s,\hat{\pi}(s,h)}^{T}\hat{V}_{h+1}$.

Hence, for all $s\in\statespace, a\in\actionspace$ and $h=H-1$, we have 
    \begin{equation}
        \hat{Q}_{H-1}(s,a)-\Tilde{Q}_{H-1}(s,a)=\max\{r_{H-1}(s,a)+z_{H-1,s}(a), 0\}-(r_{H-1}(s,a)+P_{H-1|s,a}^{\transpose}\hat{V}_{H}).
    \end{equation}
On the one hand, since $z_{H-1,s}(a)<P_{H-1|s,a}^{\transpose}\hat{V}_{H}$ and $\hat{V}_{H}(s)=0$ for all $s\in\statespace$, it holds that
\begin{equation}
    \begin{aligned}
    \hat{Q}_{H-1}(s,a)-\Tilde{Q}_{H-1}(s,a)&\leq \max\{r_{H-1}(s,a)+P_{H-1|s,a}^{\transpose}\hat{V}_{H}, 0\}-(r_{H-1}(s,a)+P_{H-1|s,a}^{\transpose}\hat{V}_{H})\\
    &=r_{H-1}(s,a)+P_{H-1|s,a}^{\transpose}\hat{V}_{H}-(r_{H-1}(s,a)+P_{H-1|s,a}^{\transpose}\hat{V}_{H})\\
    &=0.
\end{aligned}
\end{equation}

On the other hand, we also have 
\begin{equation}
        \begin{aligned}
        \hat{Q}_{H-1}(s,a)-\Tilde{Q}_{H-1}(s,a)&=\max\{r_{H-1}(s,a)+z_{H-1,s}(a), 0\}-(r_{H-1}(s,a)+ P_{H-1|s,a}^{\transpose}\hat{V}_{H})\\
        &\geq r_{H-1}(s,a)+z_{H-1,s}(a)- (r_{H-1}(s,a)+ P_{H-1|s,a}^{\transpose}\hat{V}_{H})\\
        &= z_{h,s}(a)- P_{H-1|s,a}^{\transpose}\hat{V}_{h+1}\\
        &> -\frac{\epsilon}{H}.
    \end{aligned}
\end{equation}

 In summary, for all $s\in\statespace, a\in\actionspace$, we have
    \begin{equation}
        -\frac{\epsilon}{H}\leq \hat{Q}_{H-1,s}(a)- \Tilde{Q}_{H-1}(s,a)\leq 0,
    \end{equation} 
        Hence, by letting $a=\hat{\pi}(s,H-1)$, we will have
    \begin{equation}
        -\frac{\epsilon}{H}\leq \hat{V}_{H-1}-\policyvalueoperator{H-1}{\hat{V}_{H}}{\hat{\pi}}=\hat{Q}_{H-1,s}(\hat{\pi}(s,H-1))-\Tilde{Q}_{H-1}(s,\hat{\pi}(s,H-1))\leq 0,
    \end{equation}
        \begin{equation}
        \policyvalueoperator{H-1}{\hat{V}_{H}}{\hat{\pi}}-\frac{\epsilon}{H}\leq \hat{V}_{H-1}\leq \policyvalueoperator{H-1}{\hat{V}_{H}}{\hat{\pi}}= V_{H-1}^{\hat{\pi}}.
    \end{equation}
    By the definition of $V_{H-1}^{*}=\max_{\pi\in\Pi}V_{H-1}^{\pi}$, we know that $V_{H-1}^{\hat{\pi}}\leq V^{*}_{H-1}$. Since it holds that $V_{H-1}^{*}\leq H$, then it also holds that $\hat{V}_{H-1}\leq H$.

    We assume that for the case of $h'=h+1, \ldots H-1$, it still holds that 
    \begin{align}
        &\policyvalueoperator{h'}{\hat{V}_{h'+1}}{\hat{\pi}}-\frac{\epsilon}{H}\leq \hat{V}_{h'}\leq \policyvalueoperator{h'}{\hat{V}_{h'+1}}{\hat{\pi}},\\
        &\hat{V}_{h'}\leq V_{h'}^{\hat{\pi}}\leq V_{h'}^{*}\leq H.
    \end{align}

    We now check the case of $h'=h$. For all $s\in\statespace, a\in\actionspace$, we have that 
    \begin{equation}
            \begin{aligned}
        |z_{h,s}(a)+\frac{\epsilon}{2H}- \Phsa^{T}\hat{V}_{h+1}|&=|z_{h,s}(a)+\frac{\epsilon}{2H}-\Phsa^{''T}\hat{V}_{h+1}+\Phsa^{''T}\hat{V}_{h+1}-\Phsa^{T}\hat{V}_{h+1}|\\
        &\leq |z_{h,s}(a)+\frac{\epsilon}{2H}-\Phsa^{''T}\hat{V}_{h+1}|+|(\Phsa^{''}-\Phsa)^{T}\hat{V}_{h+1}|\\
        &\leq |z_{h,s}(a)+\frac{\epsilon}{2H}-\Phsa^{''T}\hat{V}_{h+1}|+S\max_{s'\in\statespace}|\hat{V}_{h+1}(s')|\max_{s'\in\statespace}|\Phsa^{''}(s')-\Phsa(s')|\\
        &\leq \frac{\epsilon}{4H}+SH \cdot \frac{\epsilon}{4SH^{2}}\\
        &\leq \frac{\epsilon}{2H}.
    \end{aligned}
    \end{equation}

    The second line follows from the triangle inequality. The fourth line follows from the induction assumption that $\hat{V}_{h+1}(s')\leq H$ for all $s'\in\statespace$ and the fact that $|\Phsa^{''}(s')-\Phsa(s')|\leq \epsilon/(4SH^{2})$ as required in line 3 of Algorithm \ref{algo: quantum VI-finite horizon MDP-optimal action and V value v5}.
    Besides, it implies that 
    \begin{equation}
         \Phsa^{\transpose}\hat{V}_{h+1}-\frac{\epsilon}{H}\leq z_{h,s}(a)\leq \Phsa^{\transpose} \hat{V}_{h+1}.
    \end{equation}
    Therefore,  for all $s\in\statespace, a\in\actionspace$ and $h\in[H]$ we have
    \begin{equation}
        \hat{Q}_{h,s}(a)-\Tilde{Q}_{h}(s,a)=\max\{r_{h}(s,a)+z_{h,s}(a), 0\}-\bigl( r_{h}(s,a)+\Phsa^{\transpose}\hat{V}_{h+1} \bigr).
    \end{equation}
    On one hand, since $z_{h,s}(a)\leq \Phsa^{\transpose}\hat{V}_{h+1}$ and $\hat{V}_{h+1}\geq 0$, then we have 
    \begin{equation}
            \begin{aligned}
        \hat{Q}_{h,s}(a)-\Tilde{Q}_{h}(s,a)&\leq\max\bigl\{r_{h}(s,a)+\Phsa^{\transpose}\hat{V}_{h+1}, 0\bigr\}-\bigl( r_{h}(s,a)+\Phsa^{\transpose}\hat{V}_{h+1} \bigr)\\
        &= r_{h}(s,a)+\Phsa^{\transpose}\hat{V}_{h+1} -\bigl( r_{h}(s,a)+\Phsa^{\transpose}\hat{V}_{h+1} \bigr)\\
        &=0.
    \end{aligned}
    \end{equation}

    On the other hand, we also have
    \begin{equation}
    \begin{aligned}
        \hat{Q}_{h,s}(a)-\Tilde{Q}_{h}(s,a)&=\max\{r_{h}(s,a)+z_{h,s}(a), 0\}-\bigl( r_{h}(s,a)+\Phsa^{\transpose}\hat{V}_{h+1} \bigr)\\
        &\geq r_{h}(s,a)+z_{h,s}(a)- \bigl( r_{h}(s,a)+\Phsa^{\transpose}\hat{V}_{h+1} \bigr)\\
        &= z_{h,s}(a)-\Phsa^{\transpose}\hat{V}_{h+1}\\
        &\geq -\frac{\epsilon}{H}.
    \end{aligned}
    \end{equation}

    The last line comes from Eq. \eqref{equ: gap between zhsa and PV in QVI2}.
    In summary, for all $s\in\statespace, a\in\actionspace$ and $h\in[H]$, we have
    \begin{equation}
        -\frac{\epsilon}{H}\leq \hat{Q}_{h,s}(a)- \Tilde{Q}_{h}(s,a)\leq 0,
    \end{equation} 

    Hence, by letting $a=\hat{\pi}(s,h)$, we will have
    \begin{equation}
        -\frac{\epsilon}{H}\leq \hat{V}_{h}-\policyvalueoperator{h}{\hat{V}_{h+1}}{\hat{\pi}}=\hat{Q}_{h,s}\bigl( \hat{\pi}(s,h) \bigr)-\Tilde{Q}_{h}\bigl( s,\hat{\pi}(s,h) \bigr)\leq 0,
    \end{equation}
        \begin{equation}
        \policyvalueoperator{h}{\hat{V}_{h+1}}{\hat{\pi}}-\frac{\epsilon}{H}\leq \hat{V}_{h}\leq \policyvalueoperator{h}{\hat{V}_{h+1}}{\hat{\pi}}.
    \end{equation}
    Since it holds that $\hat{V}_{h'}\leq \policyvalueoperator{h'}{\hat{V}_{h'+1}}{\hat{\pi}}$ for all $h'=h+1,\ldots, H-1$, then we have $\hat{V_{h}}\leq \policyvalueoperator{h}{\hat{V}_{h+1}}{\hat{\pi}}\leq \policyvalueoperator{h}{\policyvalueoperator{h+1}{\hat{V}_{h+2}}{\hat{\pi}}}{\hat{\pi}}\leq \cdots\leq V_{h}^{\hat{\pi}}$ by the monotonicity property of the operator $\mathcal{T}^{h}_{\hat{\pi}}$. Further, we have $V_{h}^{\hat{\pi}}\leq V_{h}^{*}$ by the definition of $V_{h}^{*}$. Since $V_{h}^{*}\leq H$, we must have $\hat{V}_{h}\leq V_{h}^{\hat{\pi}}\leq V_{h}^{*}\leq H$.
\done

\begin{theorem}[Correctness of \textbf{QVI-5} (Algorithm \ref{algo: quantum VI-finite horizon MDP-optimal action and V value v5})]\label{Thm: correctness of QVI5}
The outputs $\hat{\pi}$ and $\{\hat{V}_{h}\}_{h=0}^{H-1}$ satisfy that 
\begin{equation}\label{equ: correction of approximate case}
   V_{h}^{*}-\epsilon\leq \hat{V}_{h}\leq V^{\hat{\pi}}_{h}\leq V_{h}^{*}
\end{equation} for all  $h\in[H]$ with a success probability at least $1-\delta$.
\end{theorem}
\beginproof
We start by examining the failure probability. The analysis is similar to Theorem \ref{Thm: correctness of QVI2} where we consider quantum oracles that can fail. Again, we use the quantum union bound for quantum operators here. 

In Line 6, since $\BinaryOracle{z_{h,s}}$ is constructed using $\qEstone$ with a failure probability $\zeta$, it is within $2A\zeta$ of its "ideal version." Specifically, this means that there exists an ideal quantum oracle $\BinaryOracle{z_{h,s}}^{\text{ideal}}$ encoding $\widetilde{\Phsa^{''\transpose}\hat{V}_{h+1}} - \epsilon/2H$, where $\widetilde{\Phsa^{''\transpose}\hat{V}_{h+1}}$ satisfies $\infiNorm{\widetilde{\Phsa^{''\transpose}\hat{V}_{h+1}} - \Phsa^{''\transpose}\hat{V}_{h+1}} \leq \epsilon/(4H)$, such that $\OPNorm{\BinaryOracle{z_{h,s}}^{\text{ideal}} - \BinaryOracle{z_{h,s}}} \leq 2A\zeta$. Since $\BinaryOracle{\hat{Q}_{h,s}}$ is formed using one call each to $\BinaryOracle{z_{h,s}}$ and $\BinaryOracle{z_{h,s}}^{\dagger}$, it is within $4A\zeta$ of its ideal counterpart $\BinaryOracle{\hat{Q}_{h,s}}^{\text{ideal}}$. By applying the quantum union bound and substituting the definition of $\zeta$, this shows that the quantum operation executed by $\qArgmax$ is $(\Tilde{c}\sqrt{A}\log(1/\delta)\cdot 4A\zeta = \delta/SH)$-close to its ideal version. Consequently, the output of $\qArgmax$ is incorrect with a probability of at most $\delta/SH$. Given that $\qArgmax$ is invoked a total of $SH$ times, the overall failure probability is bounded by $\delta$, as ensured by the standard union bound.

    
    In line 5, the quantum oracle $\MDPQPoracle{P''}$ is constructed by querying $\MDPQoracle$ satisfying $|\Phsa''(s')-\Phsa(s')|\leq \epsilon/(4SH^{2})$. Besides, Theorem \ref{thm: quantum mean estimation} guarantees the output in the line 6 satisfying $|z_{h,s}(a)-\Phsa^{''T}\hat{V}_{h+1}|\leq \epsilon/(4H)$. From the discussion in Lemma \ref{lemma: iterative property of the value operator QVIfive},  for every $s\in\statespace$ and $a\in\actionspace$, we have 
    \begin{equation}
         \Phsa^{\transpose}\hat{V}_{h+1}-\frac{\epsilon}{H}\leq z_{h,s}(a)\leq \Phsa^{\transpose} \hat{V}_{h+1}.
    \end{equation}
    Hence, it holds for all $s\in\statespace$ and $a\in\actionspace$ in every loop $h\in[H]$ that 
    \begin{equation}
            \begin{aligned}
        |\hat{Q}_{h,s}(a)-Q_{h,s}^{*}(a)|
        &= |r_{h}(s,a)+z_{h,s}(a)- (r_{h}(s,a)+\Phsa^{T}V_{h+1}^{*})|\\
        &\leq |z_{h,s}(a)- \Phsa^{T}(V_{h+1}^{*}-\hat{V}_{h+1})-\Phsa^{T}\hat{V}_{h+1}|\\
        &\leq |z_{h,s}(a)-\Phsa^{T}\hat{V}_{h+1}|+|\Phsa^{T}(V_{h+1}^{*}-\hat{V}_{h+1})|\\
        &\leq \frac{\epsilon}{H}+\max_{s\in\statespace}|\hat{V}_{h+1}(s)-V_{h+1}^{*}(s)|\\
        &= \frac{\epsilon}{H}+\infiNorm{\hat{V}_{h+1}-V_{h+1}^{*}}.
    \end{aligned}
    \end{equation}

Further, we have
\begin{equation}
    \begin{aligned}
    \infiNorm{\hat{V}_{h}-V_{h}^{*}}&= \infiNorm{ \hat{Q}_{h,s}\bigl( \hat{\pi}(s,h) \bigr)- \max_{a\in\actionspace} Q^{*}_{h,s}(a)}\\
    &= \infiNorm{ \max_{a\in\actionspace}\hat{Q}_{h,s}(a)- \max_{a\in\actionspace} Q^{*}_{h,s}(a)}\\
    &\leq  \infiNorm{ \max_{a\in\actionspace} |\hat{Q}_{h,s}(a)- Q^{*}_{h,s}(a)|}\\
    &\leq \max_{s\in\statespace}\max_{a\in\actionspace} |\hat{Q}_{h,s}(a)- Q^{*}_{h,s}(a)|\\
    &\leq \frac{\epsilon}{H} + \infiNorm{\hat{V}_{h+1}-V_{h+1}^{*}}.
\end{aligned}
\end{equation}

Since it holds that $\hat{V}_{h}(s)= V_{H}^{*}(s)=0$ for all $s\in\statespace$, we can induce that
\begin{equation}\label{equ: Vh-optimalVh2}
    \infiNorm{\hat{V}_{h}-V_{h}^{*}}\leq \frac{(H-h)\epsilon}{H} + \infiNorm{\hat{V}_{H}-V_{H}^{*}}=\frac{(H-h)\epsilon}{H}.
\end{equation}
Therefore, it holds that $\infiNorm{\hat{V}_{h}-V_{h}^{*}}\leq \epsilon$ for all $h\in[H]$. Specifically, $\hat{V}_{h}(s)\geq V_{h}^{*}(s)-\epsilon$ for all $s\in\statespace$ and $h\in[H]$.
Together with Lemma \ref{lemma: iterative property of the value operator QVIfive}, we obtain the desired claim.
\done

\begin{theorem}[Complexity of \textbf{QVI-5} (Algorithm \ref{algo: quantum VI-finite horizon MDP-optimal action and V value v5})]\label{Thm: Complexity of QVI5}
The quantum query complexity of Algorithm \ref{algo: quantum VI-finite horizon MDP-optimal action and V value v5} in terms of the quantum oracle of MDPs $\MDPQoracle$ is $$  O\biggl(\frac{S\sqrt{A}H^{3}\log(\sqrt{SH^{2}/\epsilon})\log(SA^{1.5}H/\delta)}{\epsilon\eta}\biggr),$$
where $\eta\in (0, 1/2)$ is the lower bound of $\Phsa(\supp{\Phsa})$ for all $s\in\statespace, a\in \actionspace$ and $h\in[H]$. 
\end{theorem}
\beginproof
We first assume that all $\qArgmax$ and $\qEstone$ are correct, up to the specified error, because the probability that this does not hold is at most $\delta$. Let $C$ be the complexity of $\QVIfive$ as if all $\qArgmax$ and $\qEstone$ are carried out with maximum failure probabilities set to constant. Then, since the failure probabilities are set to $\zeta= \delta/(4cSA^{1.5}H\log(1/\delta))$, the actual complexity of $\QVIfive$ is \begin{equation}\label{equ: initial complexity of QVIthree}
    O(C\log(SA^{1.5}H\log(1/\delta)/\delta))= O(C\log(SA^{1.5}H/\delta)).
\end{equation}
Now, we check each line of $\QVIfive$ to bound $C$. 

In line 3, we create quantum oracle $\MDPQPoracle{P''}$ with oracle $\MDPQoracle$ where $|\Phsa''(s')-\Phsa(s')|\leq \epsilon/(4SH^{2})$ for all $s,s'\in\statespace, a\in\actionspace$ and $h\in[H]$. By Lemma \ref{lem: MDP from binary oracle to probability oracle}, we can know that the query complexity of $\MDPQPoracle{P''}$ in terms of $\MDPQoracle$ is $O(\log(\sqrt{SH^{2}/\epsilon})/\eta)$.

In line 5, we encode the vector $\hat{V}_{h+1}$ to an oracle $U_{\hat{V}_{h+1}}$. This process does not need to query $\MDPQoracle$ and only needs to access the classical vector $\hat{V}_{h+1}$.

In line 6, we implement $\qEstone$ to compute the approximate inner product of $\Phsa^{''T}\hat{V}_{h+1}$ with error $\epsilon/(4H)$. Besides, the correctness analysis shows that $\hat{V}_{h+1}(s)\leq V^{*}_{h+1}(s) \leq H$ for all $s\in\statespace$ and $h\in[H]$. Besides, we also know that $\hat{V}_{h+1}(s)\geq 0$ by the definition of $\hat{Q}_{h,s}(a)$ in Line 7 of \textbf{QVI-5}. By Theorem \ref{thm: quantum mean estimation}, we know that $\qEstone$ needs $O(H^{2}/\epsilon+\sqrt{H^{2}/\epsilon})=O(H^{2}/\epsilon)$ queries to $\MDPQPoracle{P''}$, provided $0<\epsilon\leq H^{2}$. Since we have assumed that $\epsilon\leq H$ on the input $\epsilon$, so this holds.

In Line 7, $U_{\hat{Q}_{h,s}}$ needs to call $U_{z_{h,s}}$ and $U_{z_{h,s}}^{\dagger}$ once. Then the query complexity of $U_{\hat{Q}_{h,s}}$ in terms of $U_{z_{h,s}}$ is $O(1)$.

In Line 8, we use $\qArgmax$ in accelerating the searching for the optimal action $\hat{\pi}(s,h)$ for all $s\in\statespace$. By Theorem \ref{Thm: quantum maximum finding}, $\qArgmax$ requires $O(\sqrt{A})$ queries to the oracle $U_{\hat{Q}_{h,s}}$ for all $s\in\statespace$ and $h\in[H]$. 
Therefore, it induces an overall query cost of \begin{equation}
    C=O\biggl(\frac{S\sqrt{A}H^{3}\log(\sqrt{SH^{2}/\epsilon})}{\epsilon\eta}\biggr).
\end{equation}

Combining the above equation with Equation \eqref{equ: initial complexity of QVIthree}, the overall quantum query complexity of $\QVIfive$ is 
\begin{equation}
O\biggl(\frac{S\sqrt{A}H^{3}\log(\sqrt{SH^{2}/\epsilon})\log(SA^{1.5}H/\delta)}{\epsilon\eta}\biggr).
\end{equation}
\done
\subsubsection{Remark on \textbf{QVI-5}}\label{appendix: remark on QVIfive}
From Theorem \ref{Thm: Complexity of QVI5}, we know that the query complexity of \textbf{QVI-5} is $\Tilde{O}(S\sqrt{A}H^{3}/(\epsilon\eta))$, where $\eta\in(0, 1/2)$ is the smallest non-zero value in the conditional probability distributions $\Phsa=\left(P_{h}(s'|s,a)\right)_{s'\in\statespace}$ for all $(s,a,h)\in\statespace\times\actionspace\times[H]$. 
In general, it must hold that
\begin{equation}
    S_{\Phsa}\eta\leq \sum_{s'\in\statespace}\Phsa(s')=1\leq S_{\Phsa}(1-\eta),
\end{equation}
where $S_{\Phsa}\define|\supp{\Phsa}|$ represents the cardinality of the support set of the function $\Phsa$. Then, it immediately implies $\max\{S_{\Phsa}, S_{\Phsa}/(S_{\Phsa}-1)\}\leq 1/\eta$ holds. When $S_{\Phsa}\geq 2$, it always holds that $ S_{\Phsa}/(S_{\Phsa}-1)\leq S_{\Phsa}\leq 1/\eta$. Besides, for the special case when $S_{\Phsa}=1$, it also holds that $S_{\Phsa}=1\leq 1/\eta$. Therefore, it always holds that $S_{\Phsa}\leq 1/\eta$ in any case for all triples $(s,a,h)\in\statespace\times\actionspace\times[H]$.
\begin{itemize}
    \item Suppose that there exists a triple $(s,a,h)$ such that $\Phsa(s')>0$ holds for all $s'\in\statespace$, i.e., $\supp{\Phsa}=\statespace$ and $S_{\Phsa}=S$, then it must hold that $S\leq 1/\eta$. Together with the results obtained in Theorem \ref{Thm: Complexity of QVI5}, we can know that the query complexity of \textbf{QVI-5} is no greater than $\Tilde{O}(S^{2}\sqrt{A}H^{3}/\epsilon)$. Hence, \textbf{QVI-5} can not achieve any speedup in terms of $S$, compared with \textbf{QVI-1}.
    \item Suppose that it holds that the probability distributions $\Phsa$ are sparse, i.e., $S_{\Phsa}<S$ for all triples $(s, a, h)\in\statespace\times\actionspace\times[H]$, then it is possible for $1/\eta$ satisfying $S_{\Phsa}\leq 1/\eta<S$. Then, \textbf{QVI-5} indeed achieves speedup in $S$, compared with \textbf{QVI-1}. In particular, when the probability distributions $\Phsa$ of an MDP $\MDP$ only have a few non-zero entries simultaneously, i.e., $S_{\Phsa}=O(1)$ for all $(s,a,h)\in \statespace\times\actionspace\times[H]$, then the query complexity of \textbf{QVI-5} for such an MDP $\MDP$ is $\Tilde{O}(S\sqrt{A}H^{3}/\epsilon)$.
\end{itemize}
}

\section{Generative Model Setting}\label{appendix: generative model setting}

\subsection{Correctness and Complexity of \textbf{QVI-3} (Algorithm \ref{algo: quantum VI-finite horizon MDP-optimal action and V value v3})}\label{appendix: QVIthree}

\subsubsection{Proof of Lemma \ref{lemma: iterative property of the value operator for QVI3}}
\begin{lemma}\label{lemma: iterative property of the value operator for QVI3}
    $\QVIthree$ holds that $\policyvalueoperator{h}{\hat{V}_{h+1}}{\hat{\pi}}-\frac{\epsilon}{H}\leq \hat{V}_{h}\leq \policyvalueoperator{h}{\hat{V}_{h+1}}{\hat{\pi}}$ for all $h\in[H]$ with a success probability at least $1-\delta$.
\end{lemma}
\beginproof
The analysis of success probability is the same as Theorem \ref{Thm: correctness of QVI3} and hence is omitted here. We proceed to show the correctness of the claim.
For all $s\in\statespace, a\in\actionspace$ and $h\in[H]$ we have that 
    \begin{equation}
        \Bigl| z_{h,s}(a)+\frac{\epsilon}{2H}- \Phsa^{\transpose}\hat{V}_{h+1} 
        \Bigr|\leq \frac{\epsilon}{2H}.
    \end{equation}
    This implies, for all $s\in\statespace, a\in\actionspace$ and $h\in[H]$,
    \begin{equation}\label{equ: gap between zhsa and PV in QVI3}
        \Phsa^{\transpose}\hat{V}_{h+1}- \frac{\epsilon}{H} \leq z_{h,s}(a) \leq   \Phsa^{\transpose}\hat{V}_{h+1}.
    \end{equation}
    Now, for all $s\in\statespace, a\in\actionspace$ and $h\in[H]$, we let 
    \begin{equation}
        \Tilde{Q}_{h}(s,a)\define r_{h}(s,a)+\Phsa^{\transpose}\hat{V}_{h+1} \text{ and } \Tilde{V}_{h}(s)\define \max_{a\in\actionspace} \Tilde{Q}_{h}(s,a).
    \end{equation}
    Note that $\policyvalueoperator{h}{\hat{V}_{h+1}}{\hat{\pi}}=\Tilde{Q}_{h}\bigl( s,\hat{\pi}(s,h) \bigr)= r_{h}(s,\hat{\pi}(s,h))+P_{h|s,\hat{\pi}(s,h)}^{T}\hat{V}_{h+1}$. Therefore,  for all $s\in\statespace, a\in\actionspace$ and $h\in[H]$, 
    we have
    \begin{equation}
        \hat{Q}_{h,s}(a)-\Tilde{Q}_{h}(s,a)=\max\{r_{h}(s,a)+z_{h,s}(a), 0\}
        -\bigl( r_{h}(s,a)+\Phsa^{\transpose}\hat{V}_{h+1} \bigr).
    \end{equation}
    On one hand, since $z_{h,s}(a)\leq \Phsa^{\transpose}\hat{V}_{h+1}$ and $\hat{V}_{h+1}\geq 0$, then we have 
    \begin{equation}
            \begin{aligned}
        \hat{Q}_{h,s}(a)-\Tilde{Q}_{h}(s,a)&\leq\max\bigl\{r_{h}(s,a)+\Phsa^{\transpose}\hat{V}_{h+1}, 0\bigr\}
        -\bigl( r_{h}(s,a)+\Phsa^{\transpose}\hat{V}_{h+1} \bigr)\\
        &= r_{h}(s,a)+\Phsa^{\transpose}\hat{V}_{h+1} -\bigl( r_{h}(s,a)+\Phsa^{\transpose}\hat{V}_{h+1} \bigr)\\
        &=0.
    \end{aligned}
    \end{equation}

    On the other hand, we also have
    \begin{equation}
            \begin{aligned}
        \hat{Q}_{h,s}(a)-\Tilde{Q}_{h}(s,a)&=\max\{r_{h}(s,a)+z_{h,s}(a), 0\}-\bigl( r_{h}(s,a)+\Phsa^{\transpose}\hat{V}_{h+1} \bigr),\\
        &\geq r_{h}(s,a)+z_{h,s}(a)- \bigl( r_{h}(s,a)+\Phsa^{\transpose}\hat{V}_{h+1} \bigr),\\
        &= z_{h,s}(a)-\Phsa^{\transpose}\hat{V}_{h+1},\\
        &\geq -\frac{\epsilon}{H}.
    \end{aligned}
    \end{equation}

    The last line comes from Eq. \eqref{equ: gap between zhsa and PV in QVI3}.
    In summary, for all $s\in\statespace, a\in\actionspace$ and $h\in[H]$, we have
    \begin{equation}
        -\frac{\epsilon}{H}\leq \hat{Q}_{h,s}(a)- \Tilde{Q}_{h}(s,a)\leq 0,
    \end{equation} 
    Hence, by letting $a=\hat{\pi}(s,h)$, we will have
    \begin{equation}
        -\frac{\epsilon}{H}\leq \hat{V}_{h}-\policyvalueoperator{h}{\hat{V}_{h+1}}{\hat{\pi}}=
        \hat{Q}_{h,s}\bigl( \hat{\pi}(s,h) \bigr)
        -\Tilde{Q}_{h}\bigl( s,\hat{\pi}(s,h) \bigr)\leq 0,
    \end{equation}
        \begin{equation}
        \policyvalueoperator{h}{\hat{V}_{h+1}}{\hat{\pi}}-\frac{\epsilon}{H}\leq \hat{V}_{h}\leq \policyvalueoperator{h}{\hat{V}_{h+1}}{\hat{\pi}}.
    \end{equation}
    \done
\subsubsection{Correctness of \textbf{QVI-3} (Proof of Theorem \ref{Thm: correctness of QVI3})}
\beginproof We start by examining the failure probability. The analysis is similar to Theorem \ref{Thm: correctness of QVI2} where we need to consider quantum oracles that can fail. Again, we use the quantum union bound for quantum operators here. 

In line 5, since $\BinaryOracle{z_{h,s}}$ is constructed using $\qEstone$ with a failure probability $\zeta$, it is within $2A\zeta$ of its "ideal version." Specifically, this means that there exists an ideal quantum oracle $\BinaryOracle{z_{h,s}}^{\text{ideal}}$ encoding $\widetilde{\Phsa^{\transpose}\hat{V}_{h+1}} - \epsilon/2H$, where $\widetilde{\Phsa^{\transpose}\hat{V}_{h+1}}$ satisfies $\infiNorm{\widetilde{\Phsa^{\transpose}\hat{V}_{h+1}} - \Phsa^{\transpose}\hat{V}_{h+1}} \leq \epsilon/(2H)$, such that $\OPNorm{\BinaryOracle{z_{h,s}}^{\text{ideal}} - \BinaryOracle{z_{h,s}}} \leq 2A\zeta$. Since $\BinaryOracle{\hat{Q}_{h,s}}$ is formed using one call each to $\BinaryOracle{z_{h,s}}$ and $\BinaryOracle{z_{h,s}}^{\dagger}$, it is within $4A\zeta$ of its ideal counterpart $\BinaryOracle{\hat{Q}_{h,s}}^{\text{ideal}}$. By applying the quantum union bound and substituting the definition of $\zeta$, this shows that the quantum operation executed by $\qArgmax$ is $(\Tilde{c}\sqrt{A}\log(1/\delta)\cdot 4A\zeta = \delta/SH)$-close to its ideal version. Consequently, the output of $\qArgmax$ is incorrect with a probability of at most $\delta/SH$. Given that $\qArgmax$ is invoked a total of $SH$ times, the overall failure probability is bounded by $\delta$, as ensured by the standard union bound.

    
    In line 5, we apply $\qEstone$ to obtain an approximate value $z_{h,s}(a)$ of the inner product $\Phsa^{\transpose}\hat{V}_{h+1}$. Theorem \ref{thm: quantum mean estimation} guarantees the output in the line 5 satisfying $|z_{h,s}(a)-\Phsa^{\transpose}\hat{V}_{h+1}|\leq \epsilon/H$ for all $s\in\statespace, a\in\actionspace$ and $h\in[H]$. 
    Hence, it holds for all $s\in\statespace$ and $a\in\actionspace$ in every loop $h\in[H]$ that 
    \begin{equation}
            \begin{aligned}
        \bigl| \hat{Q}_{h,s}(a)-Q_{h}^{*}(s,a) \bigr|
        &= \Bigl| r_{h}(s,a)+z_{h,s}(a) 
        - \bigl( r_{h}(s,a)+\Phsa^{\transpose}V_{h+1}^{*} \bigr) \Bigl|\\
        &\leq \bigl| z_{h,s}(a)- \Phsa^{\transpose}(V_{h+1}^{*}-\hat{V}_{h+1})-\Phsa^{\transpose}\hat{V}_{h+1} \bigr|\\
        &\leq \bigl| z_{h,s}(a)-\Phsa^{\transpose}\hat{V}_{h+1} \bigr|
        + \bigl| \Phsa^{\transpose}(V_{h+1}^{*}-\hat{V}_{h+1}) \bigr|\\
        &\leq \frac{\epsilon}{H}
        + \max_{s\in\statespace} \bigl| \hat{V}_{h+1}(s)-V_{h+1}^{*}(s) \bigr|\\
        &= \frac{\epsilon}{H}+\infiNorm{\hat{V}_{h+1}-V_{h+1}^{*}}.
    \end{aligned}
    \end{equation}

Further, we have
\begin{equation}
    \begin{aligned}
    \infiNorm{\hat{V}_{h}-V_{h}^{*}}&= \infiNorm{ \hat{Q}_{h,s}\bigl( \hat{\pi}(s,h) \bigr)- \max_{a\in\actionspace} Q^{*}_{h}(s, a)}\\
    &= \infiNorm{ \max_{a\in\actionspace}\hat{Q}_{h,s}(a)- \max_{a\in\actionspace} Q^{*}_{h}(s, a)}\\
    &\leq  \infiNorm{ \max_{a\in\actionspace} 
    \bigl| \hat{Q}_{h,s}(a)- Q^{*}_{h}(s, a) \bigr| }\\
    &\leq \max_{s\in\statespace}\max_{a\in\actionspace} 
    \bigl| \hat{Q}_{h,s}(a)- Q^{*}_{h}(s, a) \bigr|\\
    &\leq \frac{\epsilon}{H} + \infiNorm{\hat{V}_{h+1}-V_{h+1}^{*}}.
\end{aligned}
\end{equation}

Since it holds that $\hat{V}_{h}(s)= V_{h}^{*}(s)=0$ for all $s\in\statespace$, we can induce that 
\begin{equation}\label{equ: Vh-optimalVh}
    \infiNorm{\hat{V}_{h}-V_{h}^{*}}\leq \frac{(H-h)\epsilon}{H}+ \infiNorm{\hat{V}_{H}-V_{H}^{*}}=\frac{(H-h)\epsilon}{H}.
\end{equation}
Then, we know that $\infiNorm{\hat{V}_{h}-V_{h}^{*}}\leq \epsilon$ for all $h\in[H]$. In particular, it implies that $V_{h}^{*}(s)-\epsilon\leq \hat{V}_{h}(s)$ for all $s\in\statespace$ and $h\in[H]$.
Now, we proceed to prove the $\hat{V}_{h}(s)\leq V_{h}^{\hat{\pi}}(s)$ for all $s\in\statespace$ and $h\in[H]$. By Lemma \ref{lemma: iterative property of the value operator for QVI3}, we know that $\hat{V}_{h}\leq \policyvalueoperator{h}{\hat{V}_{h+1}}{\hat{\pi}}$ for all $h\in[H]$. Therefore, $\hat{V}_{H-1}(s)\leq [\policyvalueoperator{H-1}{\hat{V}_{H}}{\hat{\pi}}]_{s}=[\policyvalueoperator{H-1}{0}{\hat{\pi}}]_{s}=r_{H-1}\bigl( s,\hat{\pi}(s,H-1) \bigr)=V^{\hat{\pi}}_{H-1}(s)$. By the monotonicity of the operators $\mathcal{T}^{h}_{\hat{\pi}}$, where $h\in[H]$, we have $\hat{V_{h}}\leq \policyvalueoperator{h}{\hat{V}_{h+1}}{\hat{\pi}}\leq \policyvalueoperator{h}{\policyvalueoperator{h+1}{\hat{V}_{h+2}}{\hat{\pi}}}{\hat{\pi}}\leq \cdots\leq V_{h}^{\hat{\pi}}$ for all $h\in[H]$. Due to the definition of $V^{*}_{h}(s)$, we must have $V^{*}_{h}(s)=\max_{\pi\in\Pi}V^{\pi}_{h}(s)\geq V^{\hat{\pi}}_{h}(s)$ for all $s\in\statespace$ and $h\in[H]$.
\done

\subsubsection{Complexity of \textbf{QVI-3} (Proof of Theorem \ref{Thm: Complexity of QVI3})}
\beginproof We first assume that all $\qArgmax$ and $\qEstone$ are correct, up to the specified error, because the probability that this does not hold is at most $\delta$. Let $C$ be the complexity of $\QVIthree$ as if all $\qArgmax$ and $\qEstone$ are carried out with maximum failure probabilities set to constant. Then, since the actual failure probabilities are set to $\zeta= \delta/(4cSA^{1.5}H\log(1/\delta))$, the actual complexity of $\QVIthree$ is \begin{equation}\label{equ: initial complexity of QVIthree}
    O\Bigl( C\log\bigl(SA^{1.5}H\log(1/\delta)/\delta\bigr) \Bigr)
    = O\bigl( C\log(SA^{1.5}H/\delta) \bigr).
\end{equation}
Now, we check each line of $\QVIthree$ to bound $C$.


In line 4, we encode the vector $\hat{V}_{h+1}$ to an oracle $\BinaryOracle{\hat{V}_{h+1}}$. This process does not need to query $\MDPgenerative$ and only needs to access the classical vector $\hat{V}_{h+1}$. 

In line 5, we implement $\qEstone$ to compute the approximate inner product of $\Phsa^{\transpose}\hat{V}_{h+1}$ with error $\epsilon/H$. Besides, the correctness analysis shows that it holds that $ \hat{V}_{h+1}(s)\leq V^{*}_{h+1}(s)\leq H$ for all $s\in\statespace$ and $h\in[H]$ by Theorem \ref{Thm: correctness of QVI3} and $\hat{V}_{h+1}(s)\geq 0$ for all $s\in\statespace$ by the definition of itself, $\hat{Q}_{h+1,s}(a)$ and $z_{h+1,s}(a)$ in \textbf{QVI-3}. By Theorem \ref{thm: quantum mean estimation}, we know that $\qEstone$ needs 
$O\bigl( H^{2}/\epsilon+\sqrt{H^{2}/\epsilon} \bigr)=O(H^{2}/\epsilon)$ queries to $\MDPgenerative$ for each $s\in\statespace$, provided $0<\epsilon\leq H^{2}$. Since we have assumed that $\epsilon\leq H$ on the input $\epsilon$, so this holds.

In line 7, $\BinaryOracle{\hat{Q}_{h,s}}$ needs to call $\BinaryOracle{z_{h,s}}$ and $\BinaryOracle{z_{h,s}}^{\dagger}$ once. Then the query complexity of $\BinaryOracle{\hat{Q}_{h,s}}$ in terms of $\BinaryOracle{z_{h,s}}$ is $O(1)$.

In line 8, we use $\qArgmax$ in accelerating the searching for the optimal action $\hat{\pi}(s,h)$ for all $s\in\statespace$. By Theorem \ref{Thm: quantum maximum finding}, $\qArgmax$ requires $O(\sqrt{A})$ queries to the oracle $\BinaryOracle{\hat{Q}_{h,s}}$ for all $s\in\statespace$ and $h\in[H]$. 
Therefore, it induces an overall query cost of \begin{equation}
    C=O\biggl(S\cdot \sqrt{A} \cdot H \cdot \frac{H^{2}}{\epsilon} \biggr)=O\biggl( \frac{S\sqrt{A}H^{3}}{\epsilon} \biggr),
\end{equation}
in $H$ iterations.
Combining the above equation with Equation \eqref{equ: initial complexity of QVIthree}, the overall quantum query complexity of $\QVIthree$ is 
\begin{equation}
    O\Biggl( \frac{S\sqrt{A}H^{3}
    \log(SA^{1.5}H/\delta)}{\epsilon} \Biggr).
\end{equation}
\done

\subsection{Correctness and Complexity of \textbf{QVI-4}}\label{appendix: QVIfour}
\begin{lemma}[Upper Bound on Variance \cite{sidford2018near}]\label{lemma: upper bound on variance}
    For any policy $\pi: \statespace\times[H]\rightarrow\actionspace$, it must hold that 
    \begin{equation}\label{equ: upper bound on variance}
        \infiNorm{\sum_{h'=h}^{H-1}
        \Biggl(\prod_{i=h+1}^{h'}P_{i}^{\pi} \Biggr)\sigma_{h'}(V_{h'+1}^{\pi})}\leq H^{3/2},
    \end{equation}
    where $\sigma_{h'}(V_{h'+1}^{\pi})=\sqrt{P_{h'}(V_{h'+1}^{\pi})^{2}-(P_{h'}V_{h'+1}^{\pi})^{2}}$.
\end{lemma}

\subsubsection{Proof of Lemma \ref{lemma: monotone property of V value}}
\begin{lemma}\label{lemma: monotone property of V value}
For all $k\in[K]$ and $h\in[H]$, Algorithm \ref{algo: quantum VI-finite horizon MDP-optimal action and V value v4} holds that 
\begin{align}
    V_{k,h}&\leq V_{h}^{\pi_{k}}\leq V_{h}^{*},\\
    Q_{k,h}&\leq Q_{h}^{\pi_{k}}\leq Q_{h}^{*},
\end{align}
with probability at least $1-\delta$.
\end{lemma}
\beginproof
    We first consider the success probability. Note that all the quantum subroutines $\qEstone$ and $\qEsttwo$ are implemented with maximum failure probability $\zeta=\delta/4KHSA$. In total, $\qEstone$ and $\qEsttwo$ are implemented $4KHSA$ times in line 6, 7 and 9. By the union bound, the probability that there exists an incorrect estimate is at most $\delta$.

    Now, we proceed to prove the inequalities
    \begin{equation}
    V_{k,h}\leq V_{h}^{\pi_{k}}\leq V_{h}^{*}.
\end{equation}
    Note that the second inequality is trivial due to the definition of $V_{h}^{*}=\max_{\pi\in\Pi}V_{h}^{\pi}$. Therefore, we only need to prove the inequality $ V_{k,h}\leq V_{h}^{\pi_{k}}$ for all $h\in[H]$ and $k\in[K]$. In fact, it suffices to show that for all $k\in[K]$, we have
    \begin{equation}\label{equ: monotone property of value operator in QVI4}
        V_{k,h}\leq \policyvalueoperator{h}{V_{k,h+1}}{\pi_{k}}.
    \end{equation}
    First, by the definition of $x_{k,h}$ and $g_{k,h}$ in line 7 and line 9 respectively, we have, for all $(s,a)\in\statespace\times \actionspace$,
    \begin{equation}\label{equ: upper bound on xkh}
         x_{k,h}(s,a)\leq \Phsa^{\transpose}V_{k,h+1}^{(0)},
    \end{equation}
    \begin{equation}\label{equ: upper bound on gkh}
        g_{k,h}(s,a)\leq \Phsa^{\transpose}(V_{k,h+1}-V_{k,h+1}^{(0)}).
    \end{equation}
    We continue to prove Eq. \eqref{equ: monotone property of value operator in QVI4} by induction on $k$. We first consider the base case where $k=0$. For any $h\in[H]$, if there exists some $s\in\statespace$ such that $\pi_{k}(s,h)\neq\pi_{k}^{(0)}(s,h)$, then we have
    \begin{equation}
            \begin{aligned}
        V_{k,h}(s)&=\Tilde{V}_{k,h}(s)\\
                  &= Q_{k,h}\bigl( s,\pi_{k}(s,h) \bigr)\\
                  &=\max\Bigl\{
                   r_{h}\bigl( s, \pi_{k}(s, h) \bigr) 
                   + x_{k,h}\bigl( s,\pi_{k}(s,h) \bigr)
                   + g_{k,h}\bigl( s,\pi_{k}(s,h) \bigr), 0 \Bigr\}\\
                  &\leq 
                  \max\Bigl\{r_{h}\bigl( s, \pi_{k}(s, h) \bigr)+P_{h|s,\pi_{k}(s,h)}^{\transpose}V_{k,h+1}^{(0)}+P_{h|s,\pi_{k}(s,h)}^{\transpose}(V_{k,h+1}-V_{k,h+1}^{(0)}),0 \Bigr\}\\
                &= \max\Bigl\{r_{h}\bigl( s, \pi_{k}(s, h) \bigr)+P_{h|s,\pi_{k}(s,h)}^{\transpose}V_{k,h+1},0 \Bigr\}\\
                    &=r_{h}\bigl( s, \pi_{k}(s, h) \bigr)+P_{h|s,\pi_{k}(s,h)}^{\transpose}V_{k,h+1}\\
                    &=\bigl[ \policyvalueoperator{h}{V_{k,h+1}}{\pi_{k}} \bigr]_{s}.
    \end{aligned}
    \end{equation}

    If there exists some $s\in\statespace$ such that $\pi_{k}(s,h)=\pi_{k}^{(0)}(s,h)$, then we have $V_{k,h}(s)=V_{k,h}^{(0)}(s)=V_{0,h}^{(0)}(s)=0$. Since $V_{k,h+1}(s)\geq V_{k,h+1}^{(0)}(s)=V_{0,h+1}^{(0)}(s)=0$ for all $s\in\statespace$, then we must have $V_{k,h}(s)=0\leq [\policyvalueoperator{h}{V_{k,h+1}}{\pi_{k}}]_{s}$. Therefore, when $k=0$, it holds that $V_{k,h}\leq \policyvalueoperator{h}{V_{k,h+1}}{\pi_{k}}$ for all $h\in[H]$. We assume that for any $k'=0, 1, \ldots, k-1$, it also holds that $V_{k',h}\leq \policyvalueoperator{h}{V_{k',h+1}}{\pi_{k}}$ for all $h\in[H]$. Next, we show the above statement holds for $k'=k$. In fact, if there exists some $s\in\statespace$ such that $\pi_{k}(s,h)\neq \pi_{k}^{(0)}(s,h)$, then we also have $V_{k,h}(s)\leq [\policyvalueoperator{h}{V_{k,h+1}}{\pi}]_{s}$ by following the same analysis in the case of $k=0$. If there exists some $s\in\statespace$ such that $\pi_{k}(s,h)=\pi_{k}^{(0)}(s,h)$, then we have 
    \begin{equation}
        V_{k,h}(s)=V_{k,h}^{(0)}(s)
        =V_{k-1,h}(s)
    \leq \bigl[ \policyvalueoperator{h}{V_{k-1,h+1}}{\pi_{k-1}} \bigr]_{s}
        =\bigl[ \policyvalueoperator{h}{V^{(0)}_{k,h+1}}{\pi_{k-1}} \bigr]_{s}
        \leq \bigl[ \policyvalueoperator{h}{V_{k,h+1}}{\pi_{k-1}} \bigr]_{s}
         = \bigl[ \policyvalueoperator{h}{V_{k,h+1}}{\pi_{k}} \bigr]_{s}.
    \end{equation}
    The first inequality comes from the induction hypothesis. The second inequality comes from the fact that $V_{k,h+1}^{(0)}\leq V_{k,h+1}$. The last equation comes from the fact that $\pi_{k}(s,h)=\pi_{k}^{(0)}(s,h)=\pi_{k-1}(s,h)$. Therefore, we already showed that $V_{k,h}\leq \policyvalueoperator{h}{V_{k,h+1}}{\pi_{k}}$ for the case $k'=k$ and finish the induction. Since we have $V_{k,h}\leq \policyvalueoperator{h}{V_{k,h+1}}{\pi_{k}}$ for all $k\in[K]$ and $h\in [H]$ and $V_{k,H}(s)=0, \forall s\in\statespace$, then for any fixed $k\in[K]$, $V_{k,h}\leq \policyvalueoperator{h}{\cdots\policyvalueoperator{H-1}{V_{k,H}}{\pi_{k}}}{\pi_{k}}=V_{h}^{\pi_{k}}$ for all $h\in[H]$.

    Furthermore, since we already proved $V_{k,h}\leq V_{h}^{\pi_{k}}$ for all $h\in[H]$ and $k\in[K]$, we also have, for all $(s,a)\in\statespace\times\actionspace$,
    \begin{equation}
        Q_{k,h}(s,a)\leq r_{h}(s,a)+\Phsa^{\transpose}V_{k,h+1}\leq r_{h}(s,a)+\Phsa^{\transpose}V_{h+1}^{\pi_{k}}=Q_{h}^{\pi_{k}}(s,a)\leq Q_{h}^{*}(s,a).
    \end{equation}
    The first inequality follows from Eq. \eqref{equ: upper bound on xkh} and \eqref{equ: upper bound on gkh}.
    \done

\subsubsection{Proof of Lemma \ref{lemma: gap between Vkh and optimal}}
\begin{lemma}\label{lemma: gap between Vkh and optimal}
    For all $k\in[K]$ and $h\in[H]$, Algorithm \ref{algo: quantum VI-finite horizon MDP-optimal action and V value v4} holds that
    \begin{equation}\label{equ: gap between Vkh and optimal}
        V_{h}^{*}-\epsilon_{k}\leq V_{k,h},
    \end{equation}
        \begin{equation}\label{equ: gap between Qkh and optimal}
        Q_{h}^{*}-\epsilon_{k}\leq Q_{k,h},
    \end{equation}
    with the probability at least $1-\delta$.
\end{lemma}
\beginproof
The success probability analysis is the same as Lemma \ref{lemma: monotone property of V value}, so we omit it here. We continue to use induction on $k$ to prove Eq. \eqref{equ: gap between Vkh and optimal}. First, we consider the base case where $k=0$ and show \eqref{equ: gap between Vkh and optimal} holds for all $h\in[H]$.
    By the definition of $x_{k,h}$ and $g_{k,h}$ in line 7 and line 9 of \textbf{QVI-4}, we know that, for all $(s,a)\in\statespace\times \actionspace$,
    \begin{align}
        x_{k,h}(s,a)&\geq \Phsa^{\transpose}V_{k,h+1}^{(0)}-2cH^{-1.5}\epsilon\sqrt{y_{k,h}(s,a)+4b},\\
        g_{k,h}(s,a)&\geq \Phsa^{\transpose}(V_{k,h+1}-V_{k,h+1}^{(0)})-2cH^{-1}\epsilon_{k}.
    \end{align}
    We define $\xi_{k,h}(s,a)\define 2cH^{-1}\epsilon_{k}+2cH^{-1.5}\epsilon\sqrt{y_{k,h}(s,a)+4b}$. Then, we can show that 
    \begin{equation}
            \begin{aligned}
    Q_{h}^{*}-Q_{k,h}&=r_{h}+P_{h}V_{h+1}^{*}-\max\{r_{h}+x_{k,h}+g_{k,h},0\}\\
        &\leq P_{h}V_{h+1}^{*}-x_{k,h}-g_{k,h}\\
        &\leq P_{h}V_{h+1}^{*}-P_{h}V_{k,h+1}+2cH^{-1}\epsilon_{k}+2cH^{-1.5}\epsilon\sqrt{y_{k,h}+4b}\\
        &=P_{h}(V_{h+1}^{*}-V_{k,h+1})+\xi_{k,h}\\
        &=P_{h}V(Q_{h+1}^{*})-P_{h}V_{k,h+1}+\xi_{k,h}.
    \end{aligned}
    \end{equation}

    Since we have $V(Q_{k,h+1})\leq V_{k,h+1}$, then it holds that 
    \begin{equation}
            \begin{aligned}
        Q_{h}^{*}-Q_{k,h}&\leq P_{h}V(Q_{h+1}^{*})-P_{h}V(Q_{k,h+1})+\xi_{k,h}\\
        &= P_{h}^{\pi^{*}}Q^{*}_{h+1}-P_{h}V(Q_{k,h+1})+\xi_{k,h}\\
        &\leq P_{h}^{\pi^{*}}Q^{*}_{h+1}-P_{h}^{\pi^{*}}Q_{k,h+1}+\xi_{k,h}.
    \end{aligned}
    \end{equation}

    The second line comes from the fact that $V_{h}^{*}(s)=Q^{*}_{h+1}(s,\pi^{*}(s,h))$ for all $s\in\statespace$ and $h\in[H]$. The last line comes from the fact that $\pi^{*}(s,h)$ may not be the same as $\arg\max_{a\in\actionspace}Q_{k,h+1}(s,a)$ for some $s\in\statespace$. Since it must hold that $V_{H}^{*}(s)=0, \forall s\in\statespace
    $ and we require that $V_{k,H}(s)=0, \forall s\in\statespace$, then we have $V_{H}^{*}(s)-V_{k,H}(s)=0, \forall s\in\statespace$. By solving the recursion on $Q_{h}^{*}-Q_{k,h}$, we can obtain
    \begin{equation}
        Q_{h}^{*}-Q_{k,h}\leq 
        \sum_{h'=h}^{H-1} \Biggl( 
        \prod_{i=h+1}^{h'}P_{i}^{\pi^{*}} \Biggr)\xi_{k,h'},
    \end{equation}
    where $\xi_{k,h'}(s,a)=2cH^{-1}\epsilon_{k}+2cH^{-1.5}\epsilon\sqrt{y_{k,h'}(s,a)+4b}$ for all $(s,a)\in\statespace\times\actionspace$.
    Note that
    a product over an empty index set evaluates to 1.
    Now, we try to bound $\sqrt{y_{k,h'}(s,a)+4b}$ for all $(s,a)\in\statespace\times\actionspace$.
    By the definition of $y_{k,h}(s,a)$ in line 6 of \textbf{QVI-4}, we know that there exists a $b'$ satisfying $|b'|\leq b$ such that
    \begin{equation}
            \begin{aligned}
        \sqrt{y_{k,h'}(s,a)+4b}&\leq 
        \max\Bigl\{ \bigl( P_{h'|s,a}^{\transpose}(V_{k,h'+1}^{(0)})^{2}+b-(P_{h'|s,a}^{\transpose}V_{k,h'+1}^{(0)}-b'/H)^{2}+4b \bigr)^{1/2}, \sqrt{4b}\Bigr\}\\
        &\leq \bigl( 
        \sigma_{h'}^{2}(V_{k,h'+1}^{(0)})+5b+2bH^{-1}P_{h'|s,a}^{\transpose}V_{k,h'+1}^{(0)} \bigr)^{1/2}\\
        &\leq \bigl( \sigma_{h'}^{2}(V_{k,h'+1}^{(0)}) + 7b \bigr)^{1/2}.
    \end{aligned}
    \end{equation}

    Since it holds that $V_{k,h'+1}^{(0)}(s)=0$ for all $s\in\statespace$ and $h'\in[H]$ when $k=0$, then $\sigma_{h'}^{2}(V_{k,h'+1}^{(0)})=0$. This implies that $ \sqrt{y_{k,h'}(s,a)+4b}\leq \sqrt{7b}$.  Then we can show that 
    \begin{equation}
            \begin{aligned}
        Q_{h}^{*}-Q_{k,h}&\leq 
        \sum_{h'=h}^{H-1}\Biggl( \prod_{i=h+1}^{h'}P_{i}^{\pi^{*}} \Biggr)
        \Biggl( 2cH^{-1}\epsilon_{k}+2cH^{-1.5}\epsilon\sqrt{y_{k,h'}+4b} \Biggr)
        \\
        &\leq 2c\epsilon_{k}+2cH^{-0.5}\epsilon\sqrt{7b}\\
        &\leq 2c\epsilon_{k}+2c\epsilon\sqrt{7b}\\
        &\leq \Bigl( 2c+4c\sqrt{7b} \Bigr)\epsilon_{k}\\
        &\leq \epsilon_{k}.
    \end{aligned}
    \end{equation}
    
    The second line comes from the fact that $\infiNorm{\sum_{h'=h}^{H-1}\Biggl( \prod_{i=h+1}^{h'}P_{i}^{\pi^{*}} \Biggr) \mathbf{1}}\leq H-h\leq H$ for all $h\in[H]$. The third line comes from the fact that $H\geq 1$. The fourth line comes from the fact that $\epsilon\leq 2\epsilon_{k}=2\epsilon_{0}=2H$. The last line comes from the fact that $c=0.001$ and $b=1$.
    Therefore, we have $V_{k,h}(s)\geq V(Q_{k,h})(s)=\max_{a\in\actionspace}Q_{k,h}(s,a)\geq \max_{a\in \actionspace}\{Q_{h}^{*}(s,a)-\epsilon_{k}\}=V_{h}^{*}(s)-\epsilon_{k}$ for the base case $k=0$. 
    
    Now, we assume that for any $k'=1,\ldots,k-1$, it also holds that $V_{k,h}(s)\geq V_{h}^{*}(s)-\epsilon_{k}$ for all $h\in{H}$. Then, we proceed to prove the claim for the case of $k'=k$. In fact, the analysis for the case of $k'=k$ is quite similar to the base case, except for the part of the upper bound for $\sqrt{y_{k,h'}(s,a)+4b}$. We can show that there exists a $b'$ satisfying $|b'|\leq b$
    \begin{equation}
        \begin{aligned}
            \sqrt{y_{k,h'}(s,a)+4b} &\leq 
            \max\Bigl\{ \bigl( P_{h'|s,a}^{\transpose}(V_{k,h+1}^{(0)})^{2}+b-(P_{h'|s,a}^{\transpose}V_{k,h+1}^{(0)}-b'/H)^{2}+4b \bigr)^{1/2}, \sqrt{4b}
            \Bigr\}\\
            &\leq \bigl( \sigma_{h'}^{2}(V_{k,h'+1}^{(0)})+5b+2bH^{-1}P_{h'|s,a}^{\transpose}V_{k,h'+1}^{(0)} \bigr)^{1/2}\\
            &\leq \bigl( \sigma_{h'}^{2}(V_{k,h'+1}^{(0)})+7b \bigr)^{1/2}\\
            &\leq \sigma_{h'}(V_{k,h'+1}^{(0)})+\sqrt{7b}\\
            &\leq \sigma_{h'}(V_{h'+1}^{*})+\sigma(V_{k,h'+1}^{(0)}-V_{h'+1}^{*})+\sqrt{7b}.
        \end{aligned}
    \end{equation}
    The third line comes from the fact that $V_{k,h'+1}^{(0)}(s)\leq H$ for all $s\in\statespace$. The fourth line comes from the fact that $\sqrt{a+b}\leq \sqrt{a}+\sqrt{b}$ when $a,b\geq 0$. The last line comes from the fact that, for any random variables $X$ and $Y$, we must have $\sigma^{2}(X+Y)=\text{Var}[X+Y]=\text{Var}[X]+\text{Var}[Y]+2\text{Cov}[X,Y]\leq (\sqrt{\text{Var}[X]}+\sqrt{\text{Var}[Y]})^{2}=(\sigma(X)+\sigma(Y))^{2}$. Note that $\sigma(V_{k,h'+1}^{(0)}-V_{h'+1}^{*})\leq \infiNorm{V_{k,h'+1}^{(0)}-V_{h'+1}^{*}}=\infiNorm{V_{k-1,h'+1}-V_{h'+1}^{*}}\leq \epsilon_{k-1}=2\epsilon_{k}$ for all $h'\in[H]$.
    Therefore, we can show that 
    \begin{equation}
        \begin{aligned}
        Q_{h}^{*}-Q_{k,h}&\leq 
        \sum_{h'=h}^{H-1}\Biggl( \prod_{i=h+1}^{h'}P_{i}^{\pi^{*}} \Biggr)\biggl(2cH^{-1}\epsilon_{k}+2cH^{-1.5}\epsilon\sqrt{y_{k,h'}+4b} \biggr)\\
        &\leq 2c\epsilon_{k}+2cH^{-1.5}\epsilon\sum_{h'=h}^{H-1}
        \Biggl( \prod_{i=h+1}^{h'}P_{i}^{\pi^{*}} \Biggr)
        \Bigl( \sigma(V_{h'+1}^{*})
        +\sigma(V_{k,h'+1}^{(0)}-V_{h'+1}^{*})+\sqrt{7b} \Bigr)\\
        &\leq 2c\epsilon_{k}+2cH^{-1.5}\epsilon\sum_{h'=h}^{H-1}\Biggl( \prod_{i=h+1}^{h'}P_{i}^{\pi^{*}} \Biggr)
        \Bigl( \sigma(V_{h'+1}^{*})+2\epsilon_{k}+\sqrt{7b} \Bigr)\\
        &\leq 2c\epsilon_{k}+2c\epsilon+2cH^{-0.5}
        \epsilon \Bigl( 2\epsilon_{k}+\sqrt{7b} \Bigr)\\
        &\leq 2c \Bigl(1+2+2+\sqrt{7} \Bigr)\epsilon_{k}\\
        &\leq \epsilon_{k}.
    \end{aligned}
    \end{equation}
    The fourth line comes from the Lemma \ref{lemma: upper bound on variance} and the fact that $\infiNorm{\sum_{h'=h}^{H-1}\Biggl( \prod_{i=h+1}^{h'}P_{i}^{\pi^{*}} \Biggr) \mathbf{1}}\leq H-h\leq H$ for all $h\in[H]$. The fifth line comes from the fact that  we require the input $\epsilon\in (0,\sqrt{H}]$. The last line comes from the fact that $c=0.001$.
 Therefore, we have $V_{k,h}(s)\geq V(Q_{k,h})(s)=\max_{a\in\actionspace}Q_{k,h}(s,a)\geq \max_{a\in \actionspace}\{Q_{h}^{*}(s,a)-\epsilon_{k}\}=V_{h}^{*}(s)-\epsilon_{k}$ for the case of $k'=k$. 
 \done
 
\subsubsection{Correctness of \textbf{QVI-4} (Proof of Theorem \ref{Thm: correctness of QVI4})}
    By combining Lemma \ref{lemma: gap between Vkh and optimal} and Lemma \ref{lemma: monotone property of V value}, we can obtain that, for all $k\in[K]$,
    \begin{align}
        V_{h}^{*}-\epsilon_{k}&\leq V_{k,h}\leq V_{h}^{\pi_{k}}\leq V_{h}^{*},\\
        Q_{h}^{*}-\epsilon_{k}&\leq Q_{k,h}\leq Q_{h}^{\pi_{k}}\leq Q_{h}^{*},
    \end{align}
    with probability at least $1-\delta$.
    When $k=K-1=\ceil{\log_{2}(H/\epsilon)}\geq \log_{2}(H/\epsilon)$, $\epsilon_{k}=H/2^{k}\leq \epsilon$. Therefore, it implies that
        \begin{align}
        V_{h}^{*}-\epsilon\leq V_{h}^{*}-\epsilon_{K-1}&\leq V_{K-1,h}=\hat{V}_{h}\leq V_{h}^{\pi_{K-1}}=V_{h}^{\hat{\pi}}\leq V_{h}^{*},\\
        Q_{h}^{*}-\epsilon\leq Q_{h}^{*}-\epsilon_{K-1}&\leq Q_{K-1,h}=\hat{Q}_{h}\leq Q_{h}^{\pi_{K-1}}=Q_{h}^{\hat{\pi}}\leq Q_{h}^{*},
    \end{align}
        with probability at least $1-\delta$.
\done

\subsubsection{Complexity of \textbf{QVI-4} (Proof of Theorem \ref{Thm: Complexity of QVI4})}
\beginproof The success probability analysis is analogous to Lemma \ref{lemma: monotone property of V value}. Hence, we omit it here. We first assume that all estimations are correct, up to the specified error, because the probability that this does not hold is at most $\delta$. Let $C$ be the complexity of $\QVIfour$ as if all estimations are carried out with maximum failure probabilities set to constant. Then, since the actual maximum failure probabilities are set to $\zeta= \delta/(4KHSA)$, the actual complexity of $\QVIfour$ is \begin{equation}\label{equ: initial complexity of QVIfour}
    O\bigl( C\log(KHSA/\delta) \bigr).
\end{equation}
Now, we check each line of $\QVIfour$ to bound $C$. 

In line 6, since we have $0\leq V_{k,h+1}^{(0)}(s)=V_{k-1,h+1}(s)\leq V^{*}_{h+1}(s)\leq H$ for all $k>0$ and $0=V_{k,h+1}^{(0)}(s)=V_{k-1,h+1}(s)\leq V^{*}_{h+1}(s)\leq H$ for all $s\in\statespace$ when $k=0$, therefore, we can use quantum mean estimation algorithm $\qEstone$, which induces a total query complexity in the order
\begin{equation}
    KHSA\biggl( H^{2}/b+\sqrt{H^{2}/b}+H^{2}/b+\sqrt{H^{2}/b} \biggr)=O(KSAH^{3}).
\end{equation}

Now, we focus on line 7. By the definition of $y_{k,h}(s,a)$ in line 6, we know that there exists a $b'$ satisfying $|b'|\leq b$ such that
\begin{equation}
    \begin{aligned}
    y_{k,h}(s,a)&\geq 
    \max\Bigl\{
    \Phsa^{\transpose}(V_{k,h+1}^{(0)})^{2}-b-
    \bigl( \Phsa^{\transpose}V_{k,h+1}^{(0)}+b'/H \bigr)^{2},0 \Bigr\}\\
    &\geq \Phsa^{\transpose}(V_{k,h+1}^{(0)})^{2}-b
    -\bigl( \Phsa^{\transpose}V_{k,h+1}^{(0)}+b'/H \bigr)^{2}\\
    &=\bigl[ \sigma^{2}(V_{k,h+1}^{(0)}) \bigr]_{(s,a)}-b-(2b'/H)\Phsa^{\transpose} V_{k,h+1}^{(0)}-(b')^{2}/H^{2}.
\end{aligned}
\end{equation}
This implies that 
\begin{equation}
        \bigl[ \sigma^{2}(V_{k,h+1}^{(0)}) \bigr]_{(s,a)}\leq y_{k,h}(s,a)+b+(2b/H)\Phsa^{\transpose} V_{k,h+1}^{(0)}+b^{2}/H^{2}\leq y_{k,h}(s,a)+4b.
\end{equation}
The last inequality follows from $b=1$ and $V_{k,h+1}^{(0)}(s)=V_{k-1,h+1}(s)\leq V_{h+1}^{*}(s)\leq H$ for all $s\in\statespace$ when $k\geq 1$ and $V_{0,h+1}^{(0)}(s)=0$ for all $s\in\statespace$.
We also note that, since we have $y_{k,h}(s,a) \geq 0$ (by the definition in line 6), then it holds that $0<cH^{-1.5}\epsilon\sqrt{y_{k,h}(s,a)+4b}<4\sqrt{y_{k,h}(s,a)+4b}$. Therefore, we can use quantum mean estimation algorithm $\qEsttwo$ with error $cH^{-1.5}\epsilon\sqrt{y_{k,h}(s,a)+4b}$ and variance upper bound set to $y_{k,h}(s,a)+4b$, which induces a total query complexity of order
\begin{equation}
    KH\sum_{(s,a)\in\statespace\times\actionspace}w(s,a)
    \log^{2}\bigl( w(s,a) \bigr)
    =O\bigl( KSAH^{2.5}\epsilon^{-1}\log^{2}(H^{1.5}/\epsilon) \bigr),
\end{equation}
where $w(s,a)=\bigl( \sqrt{y_{k,h}(s,a)+4b} \bigr)
\bigl( cH^{-1.5}\epsilon\sqrt{y_{k,h}(s,a)+4b} \bigr)^{-1}=O(H^{1.5}/\epsilon)$.

In line 9, we can bound $0\leq V_{k,h+1}(s)-V_{k,h+1}^{(0)}(s)\leq V_{h+1}^{*}(s)-V_{k,h+1}^{(0)}(s)= V_{h+1}^{*}(s)-V_{k-1,h+1}(s)\leq \epsilon_{k-1}=2\epsilon_{k}$ for all $s\in\statespace$ and $k\geq 1$. When $k=0$, since $V_{0,h+1}(s)\geq V_{0,h+1}^{(0)}(s)=0$, then we also have $0\leq V_{0,h+1}(s)-V_{0,h+1}^{(0)}(s)=V_{0,h+1}(s)\leq V^{*}_{h+1}(s)\leq H=\epsilon_{0}$ for all $s\in\statespace$. Therefore, we can use quantum mean estimation algorithm $\qEstone$ which induces a total query complexity of order
\begin{equation}
    KHSA\left(\frac{2\epsilon_{k}}{cH^{-1}\epsilon_{k}}+\sqrt{\frac{2\epsilon_{k}}{cH^{-1}\epsilon_{k}}} \right)=O(KSAH^{2}).
\end{equation}
Therefore, we can show that
\begin{equation}
    C=O\bigl( KSA(H^{2.5}/\epsilon+H^{3}+H^{2})\log^{2}(H^{1.5}/\epsilon) \bigr)=O\bigl( SA(H^{2.5}/\epsilon+H^{3})\log^{2}(H^{1.5}/\epsilon) \bigr).
\end{equation}
Then the total query complexity is 
\begin{equation}
    O\Bigl( SA(H^{2.5}/\epsilon+H^{3})\log^{2}(H^{1.5}/\epsilon)\log\bigl( \log(H/\epsilon)HSA/\delta \bigr) \Bigr).
\end{equation}
\done 
\subsection{Lower Bounds}\label{appendix: lower bounds}
\subsubsection{Infinite-horizon MDPs}
\noindent\textbf{Preliminaries of infinite-horizon MDPs: }An infinite-horizon MDP is formally defined as a tuple $\Tilde{\MDP} \define (\statespace, \actionspace, P, r, \gamma)$, where $\statespace$ is a finite set of states representing the possible configurations of the environment, and $\actionspace$ is a finite set of actions available to the agent at each state. The transition probability $P(s'|s,a)$ specifies the likelihood of transitioning to state $s'$ after taking action $a$ in state $s$, ensuring that $\sum_{s' \in \statespace} P(s'|s,a) = 1$ for all $s \in \statespace$ and $a \in \actionspace$. The reward function $r(s,a)$, bounded within $[0,1]$, assigns a scalar reward for executing action $a$ in state $s$. Finally, the discount factor $\gamma \in [0,1)$ determines the relative importance of future rewards compared to immediate ones, with $\Gamma \define \frac{1}{1-\gamma}$.
Given such an MDP, the agent's objective is to select actions that maximize the expected sum of discounted rewards over an infinite time horizon. The primary goal is to compute a policy $\pi: \statespace \rightarrow \actionspace$ that specifies the action $a = \pi(s)$ the agent should take in each state $s \in \statespace$ to optimize its performance with high probability. For a given policy $\pi$, the state-value function (or V-value) $V^{\pi}:\statespace \rightarrow [0, \Gamma]$ and the state-action-value function (or Q-value) $Q^{\pi}: \statespace \times \actionspace \rightarrow [0, \Gamma]$ are defined as follows:
\begin{equation}
    V^{\pi}(s)=\Expectation\left[\sum_{t=0}^{\infty} \gamma^{t}r(s_{t}, a_{t})  \middle| \pi,  s_{0}=s\right],
\end{equation}
\begin{equation}
    Q^{\pi}(s,a)=\Expectation\left[\sum_{t=0}^{\infty} \gamma^{t}r(s_{t}, a_{t})  
\middle| \pi,  s_{0}=s, a_{0}=a\right].
\end{equation}
A policy $\pi$ is an optimal policy $\pi^{*}$ if $V^{\pi}=\max_{\pi\in\Pi} V^{\pi}=V^{\pi^{*}}$ where $\Pi$ is the space of all policies. For simplicity, we denote $V^{*}\define V^{\pi^{*}}$ and $Q^{*}\define Q^{\pi^{*}}$.

\noindent\textbf{Optimization goals in infinite-horizon MDPs:} The primary computational objectives in infinite-horizon MDPs are as follows: given an infinite-horizon MDP $\Tilde{\MDP}$, an approximation error $\epsilon$, and a failure probability $\delta$, the goal is to compute $\epsilon$-estimates $\hat{\pi}$, $\hat{V}$, and $\hat{Q}$ such that $\|V^{\hat{\pi}} - V^{*}\|_\infty \leq \epsilon$, $\|\hat{V} - V^{*}\|_\infty \leq \epsilon$, and $\|\hat{Q} - Q^{*}\|_\infty \leq \epsilon$ with a probability of at least $1 - \delta$.

\noindent\textbf{Classical generative model for infinite-horizon MDPs: }We denote the classical generative model for infinite-horizon MDPs as $\Tilde{G}$. Assuming access to $\Tilde{G}$, one can collect $N$ independent samples  
\[
    s_{s,a}^{i} \overset{\text{i.i.d.}}{\sim} P(\cdot| s,a), \quad i=1,\ldots, N,
\]
for each state-action pair $(s, a) \in \statespace \times \actionspace$.

\begin{theorem}[Classical and quantum lower bounds for infinite-horizon MDP~\cite{pmlr-v139-wang21w}]\label{thm: quantum lower bound for infinite horizon MDP}
    Fix any integers $S, A\geq 2$ and $\gamma\in [0.9, 1)$. Let $\Gamma=(1-\gamma)^{-1}\geq 10$ and fix any $\epsilon\in(0, \Gamma/4)$. There exists an infinite-horizon MDP $\Tilde{\MDP}=(\statespace, \actionspace, P, r, \gamma)$ with $S$ states, $A$ actions, and discount parameter $\gamma$ such that the following lower bound hold:
    \begin{itemize}
        \item Given access to a classical generative oracle $\Tilde{G}$, any algorithm that computes an $\epsilon$-approximation to $Q^{*}$, $V^{*}$, or $\pi^{*}$ must make $\Omega(\frac{SA\Gamma^{3}}{\epsilon^{2}})$ queries.
        \item 
    Given access to a quantum generative oracle $\Tilde{\mathcal{G}}$ defined as
    \begin{equation}
        \Tilde{\mathcal{G}}: \ket{s}\tensorproduct\ket{a}\tensorproduct\ket{0}\tensorproduct\ket{0}\mapsto\ket{s}\tensorproduct\ket{a}
        \tensorproduct \Biggl(\sum_{s'\in\statespace}\sqrt{P(s'|s,a)}\ket{s'}\tensorproduct\ket{v_{s'}} \Biggr),
    \end{equation}
    where $\ket{v_{s'}}$ are arbitrary auxiliary states,
    any algorithm that computes an $\epsilon$-approximation to $Q^{*}$ must take $\Omega(\frac{SA\Gamma^{1.5}}{\epsilon})$ queries and  any algorithm that computes an $\epsilon$-approximation to $V^{*}$ or $\pi^{*}$ must take $\Omega(\frac{S\sqrt{A}\Gamma^{1.5}}{\epsilon})$ queries.
    \end{itemize}
\end{theorem}

\subsubsection{Finite-horizon MDPs}
\begin{lemma}\label{lem: quantum lower bound for epsilon optimal V value}
    Let $\statespace$ and $\actionspace$ be finite sets of states and actions. Let $H>0$ be a positive integer and $\epsilon\in (0, 1/2)$ be an error parameter. We consider the following finite-horizon MDP $\MDP\define(\statespace, \actionspace, \{P_{h}\}_{h=0}^{H-1}, \{r_{h}\}_{h=0}^{H-1},H)$ where $P_{h}=P\in\realnumber^{\statespace\times \actionspace\times\statespace}$ and $r_{h}=r\in [0,1]^{\statespace\times\actionspace}$ for all $h\in[H]$. 
    \begin{itemize}
        \item Given access to a classical generative model,
     any algorithm $\mathcal{K}$, which takes $\MDP$ as an input and outputs a value function $\hat{V}_{0}$ such that $\infiNorm{\hat{V}_{0}-V_{0}^{*}}\leq \epsilon$ with probability at least $0.9$, needs to call the classical generative oracle at least 
    \begin{equation}
        \Omega\Biggl(\frac{SAH^{3}}{\epsilon^{2}\log^{3}(\epsilon^{-1})}\Biggr)
    \end{equation}
    times on the worst case of input $\MDP$.
        \item Given access to a quantum generative oracle $\mathcal{G}$ defined in Definition \ref{def: quantum generative model of an MDP}
     any algorithm $\mathcal{K}$, which takes $\MDP$ as an input and outputs a value function $\hat{V}_{0}$ such that $\infiNorm{\hat{V}_{0}-V_{0}^{*}}\leq \epsilon$ with probability at least $0.9$, needs to call the quantum generative oracle at least 
    \begin{equation}
        \Omega\Biggl(\frac{S\sqrt{A}H^{1.5}}{\epsilon\log^{1.5}(\epsilon^{-1})}\Biggr)
    \end{equation}
    times on the worst case of input $\MDP$.
    \end{itemize}
    
\end{lemma}
\beginproof
    We first introduce some definitions about infinite horizon MDPs. Let $s_{0}\in\statespace$ to be a state.
    Suppose we have an infinite-horizon MDP $\Tilde{\MDP}=(\Tilde{\statespace}, \Tilde{\actionspace}, \Tilde{P}, \Tilde{r},\gamma)$ with a quantum generative oracle, where $\Tilde{\statespace}=\statespace\setminus\{s_{0}\}$ to be a subset of $\statespace$ and $\gamma\in [0,1)$. For a better differentiation on the notations between finite-horizon and infinite-horizon MDPs, we let $\Tilde{V}^{*}\in\realnumber^{\statespace}$ represent the optimal V-value function of $\Tilde{\MDP}$. First, we define a Bellman operator $\mathcal{T}$ for the infinite-horizon MDP $\Tilde{\MDP}$ satisfying, for any $u\in\realnumber^{\Tilde{\statespace}}$ and $s\in\Tilde{\statespace}$,
    \begin{equation}
        \mathcal{T}(u)_{s}=
        \max_{a\in\Tilde{\actionspace}}\Biggl[\Tilde{r}(s,a)+\gamma \sum_{s'\in\Tilde{\statespace}}\Tilde{P}(s'|s,a)u(s') \Biggr].
    \end{equation}
    Note that for any $u,v\in\realnumber^{\Tilde{\statespace}}$ satisfying $u(s)\leq v(s)$ for all $s\in\Tilde{\statespace}$, we have $\mathcal{T}(u)_{s}\leq \mathcal{T}(v)_{s}$ for all $s\in\Tilde{\statespace}$. This is the so-called monotonicity property of $\mathcal{T}$. Besides, it also holds that $\mathcal{T}(\Tilde{V}^{*})_{s}=\Tilde{V}^{*}(s)$ for all $s\in\Tilde{\statespace}$. 
    
    Now, we proceed to prove that obtaining an $2\epsilon$-approximation value of $\Tilde{V}^{*}$ for any infinite horizon MDP $\Tilde{\MDP}$ can be reduced to obtaining an $\epsilon$-approximation value of $V_{0}^{*}$ for a finite horizon MDP. We consider the following finite-horizon MDP $\MDP=(\statespace,\actionspace, \{P_{h}\}_{h=0}^{H-1}, \{r_{h}\}_{h=0}^{H-1}, H)$ where $P_{h}=P\in\realnumber^{\statespace\times \actionspace\times \statespace}$ and $r_{h}=r\in[0,1]^{\statespace\times \actionspace}$. Besides, the time horizon $H$ satisfies $H=\ceil{2(1-\gamma)^{-1}\log(2\epsilon^{-1})}=\Theta((1-\gamma)^{-1}\log(\epsilon^{-1}))$. Besides, under any action $a\in\actionspace=\Tilde{\actionspace}$, there is a $(1-\gamma)$ probability for each state $s\in\Tilde{\statespace}$ to transition to $s_{0}$ and $\gamma$ probability to follow the original transitions in $\Tilde{\MDP}$. However, when the agent is in $s_{0}$, it can only transition to itself with probability $1$, no matter which action $a\in\actionspace$ it takes. Hence, $s_{0}$ is an absorbing state in $\MDP$. Overall, we have the following definitions for the transition probability kernel $P$ in $\MDP$.
    \begin{align}
        \forall s,s'\in\Tilde{\statespace}, a\in\actionspace, & P(s'|s,a)=\gamma\Tilde{P}(s'|s,a),P(s_{0}|s,a)=(1-\gamma),\\
        & P(s'|s_{0}, a)=0, P(s_{0}|s_{0},a)=1.
    \end{align}
    Besides, we define $r(s_{0},a)=0, r(s,a)=\Tilde{r}(s,a)\in[0,1]$ for all $s\in\Tilde{\statespace}$ and $a\in\actionspace$.

    Now, we proceed to prove that $\infiNorm{V_{0}^{*}|_{\Tilde{\statespace}}-\Tilde{V}^{*}}\leq \epsilon$, i.e., $|V_{0}^{*}(s)-\Tilde{V}^{*}(s)|\leq \epsilon$ for all $s\in\Tilde{\statespace}$. First, we note that $V_{H-1}^{*}=\max_{a\in\actionspace}r(s,a)\leq \Tilde{V}^{*}$. Then, by the monotonicity of the $\mathcal{T}$ operator, we have $\mathcal{T}(V_{H-1}^{*})_{s}\leq \mathcal{T}(\Tilde{V}^{*})_{s}=\Tilde{V}^{*}(s)$ for all $s\in\Tilde{\statespace}$. In fact, by the definition of $P$ in $\MDP$, we have
    \begin{equation}
            \begin{aligned}
        \forall s\in\Tilde{\statespace}, \mathcal{T}(V^{*}_{H-1})_{s}&=\max_{a\in\actionspace}\Biggl[\Tilde{r}(s,a)+\gamma\sum_{s'\in\Tilde{\statespace}}\Tilde{P}(s'|s,a)V_{H-1}^{*}(s')
        \Biggr]\\
        &=\max_{a\in\actionspace}\Biggl[r(s,a)+\sum_{s'\in\Tilde{\statespace}}P(s'|s,a)V_{H-1}^{*}(s')+P(s_{0}|s,a)V_{H-1}^{*}(s_{0}) \Biggr]\\
        &=\max_{a\in\actionspace}\Biggl[
        r(s,a)+\sum_{s'\in\statespace}P(s'|s,a)V_{H-1}^{*}(s') \Biggr]\\
        &=V_{H-2}^{*}(s).
    \end{aligned}
    \end{equation}
    The second line above comes from the fact that $V_{H-1}^{*}(s_{0})=\max_{a\in\actionspace}r(s_{0},a)=0$. By induction, we have $V_{h}^{*}(s_{0})=\max_{a\in\actionspace}[r(s_{0},a)+P(s_{0}|s_{0},a)V_{h+1}^{*}(s_{0})]=0$ for all $h\in[H]$. Hence, we have $V_{H-2}^{*}(s)\leq \Tilde{V}^{*}(s)$ for all $s\in\Tilde{\statespace}$. By induction, we have $V_{h}^{*}(s)\leq \Tilde{V}^{*}(s)$ for all $h\in[H]$ and $s\in\Tilde{\statespace}$. In particular, we have $V_{0}^{*}(s)\leq \Tilde{V}^{*}(s)$ for all $s\in\Tilde{\statespace}$. Let $\Tilde{\pi}^{*}\in\actionspace^{\statespace}$ be an optimal policy for the infinite-horizon MDP $\Tilde{\MDP}$. However, $\Tilde{\pi}\in \actionspace^{\statespace\times [H]}$, where $\Tilde{\pi}(\cdot, h)=\Tilde{\pi}^{*}$ for all $h\in[H]$, may not be an optimal policy for finite-horizon MDP $\MDP$. Then we have $V_{0}^{\Tilde{\pi}}(s)\leq V_{0}^{*}(s)$ for all $s\in\statespace$. In fact, for any $s\in\Tilde{\statespace}$, we have 
    \begin{equation}
            \begin{aligned}
        V_{0}^{\Tilde{\pi}}(s)&=
        r\bigl( s, \Tilde{\pi}^{*}(s) \bigr)+ 
        \sum_{s'\in\statespace}P\bigl( s'|s,\Tilde{\pi}^{*}(s) \bigr)
        r\bigl( s, \Tilde{\pi}^{*}(s) \bigr)+\cdots+
        \sum_{s'\in\statespace}P^{H}\bigl( s'|s,\Tilde{\pi}^{*}(s) \bigr)r\bigl( s, \Tilde{\pi}^{*}(s) \bigr)\\
        &=\Tilde{r}\bigl(s, \Tilde{\pi}^{*}(s) \bigr)+ \gamma\sum_{s'\in\Tilde{\statespace}}\Tilde{P}(s'|s,\Tilde{\pi}^{*}(s))\Tilde{r}\bigl(s, \Tilde{\pi}^{*}(s) \bigr)+\cdots+\gamma^{H}\sum_{s'\in\Tilde{\statespace}}
        \Tilde{P}^{H}\bigl( s'|s,\Tilde{\pi}^{*}(s) \bigr)
        \Tilde{r}\bigl(s, \Tilde{\pi}^{*}(s) \bigr)\\
        &= \Tilde{V}_{H}^{\Tilde{\pi}^{*}},
    \end{aligned}
    \end{equation}
    where $\Tilde{V}_{H}^{\Tilde{\pi}^{*}}$ is the V-value induced by the policy $\Tilde{\pi}^{*}$ over $H$ iterations. Note that for any policy $\Tilde{\pi}$ for the infinite horizon MDP $\Tilde{\MDP}$, $\infiNorm{\Tilde{V}_{k}^{\pi}- \Tilde{V}^{\pi}}\leq \gamma^{k}\infiNorm{\Tilde{V}_{0}^{\pi}-\Tilde{V}^{\pi}}\leq \gamma^{k}\bigl(\infiNorm{\Tilde{V}_{0}^{\pi}}+\infiNorm{\Tilde{V}^{\pi}} \bigr)\leq 2\exp\bigl(-(1-\gamma)k \bigr)/(1-\gamma)$. 
    The last inequality follows from $\infiNorm{\Tilde{V}_{0}^{\pi}}\leq 1/(1-\gamma)$ and $\infiNorm{\Tilde{V}^{\pi}}\leq 1/(1-\gamma)$. Besides, combining the fact that $\log(\gamma)\leq \gamma-1$ for all $\gamma\in(0,1)$ and $\exp(x)$ is monotonically increasing, we can induce the inequalities $k\log(\gamma)\leq -k(1-\gamma)$ and $\gamma^{k}=\exp(k\log(\gamma))\leq \exp(-k(1-\gamma))$. 
    Then, $\forall \epsilon>0$, it suffices to let $k\geq \log(2/((1-\gamma)\epsilon))/(1-\gamma)$ so that $\infiNorm{\Tilde{V}_{k}^{\pi}- \Tilde{V}^{\pi}}\leq \epsilon$. In fact, 
    \begin{equation}
            \begin{aligned}
        \frac{1}{1-\gamma}\log\left(\frac{2}{(1-\gamma)\epsilon}\right)&
        =\frac{1}{1-\gamma}\Biggl(
        \log\left(\frac{1}{1-\gamma}\right)+\log\left(\frac{2}{\epsilon} \right) \Biggr)\\
        &=\frac{1}{1-\gamma}\biggl(-\log(1-\gamma)+\log\left(\frac{2}{\epsilon} \right) \biggr)\\
        &\leq \frac{1}{1-\gamma}\biggl(\gamma+\log\left(\frac{2}{\epsilon} \right)
        \biggr)\\
        &\leq \frac{2}{1-\gamma}\log\left(\frac{2}{\epsilon} \right).
    \end{aligned}
    \end{equation}
    The third line comes from the fact that  $\log(1-\gamma)\leq -\gamma, \forall \gamma\in [0,1)$. The last line comes from the fact that $\log(2/\epsilon)>1, \forall \epsilon\in(0, 1/2)$.
    Since we have $H=\ceil{\frac{2}{1-\gamma}\log\left(\frac{2}{\epsilon} \right)}$ and $\Tilde{\pi}^{*}$ is an optimal policy for $\Tilde{\MDP}$, then we must have $V_{0}^{\Tilde{\pi}}(s)=\Tilde{V}_{H}^{\Tilde{\pi}^{*}}(s)\geq \Tilde{V}^{\Tilde{\pi}^{*}}(s)-\epsilon=\Tilde{V}^{*}(s)-\epsilon$.
    Therefore, we have $\Tilde{V}^{*}(s)-\epsilon\leq V_{0}^{*}(s)\leq \Tilde{V}^{*}(s)$ for all $s\in\Tilde{\statespace}$, which implies $\infiNorm{\hat{V}_{0}^{*}|_{\Tilde{\statespace}}-\Tilde{V}^{*}}\leq \epsilon$. 

    Therefore, an $\epsilon$-approximation of $V_{0}^{*}$ will give an $2\epsilon$-approximation to $\Tilde{V}^{*}$. Specifically, if we let $\hat{V}_{0}$ be an $\epsilon$-approximation of $V_{0}^{*}$, then
    \begin{equation}
            \begin{aligned}
        \infiNorm{\hat{V}_{0}|_{\Tilde{\statespace}}-\Tilde{V}^{*}}&\leq \infiNorm{\hat{V}_{0}|_{\Tilde{\statespace}}-V_{0}^{*}|_{\Tilde{\statespace}}}+\infiNorm{V_{0}^{*}|_{\Tilde{\statespace}}-\Tilde{V}^{*}}\\
        &\leq \infiNorm{\hat{V}_{0}-V_{0}^{*}}+\infiNorm{V_{0}^{*}|_{\Tilde{\statespace}}-\Tilde{V}^{*}}\\
        &\leq 2\epsilon.
    \end{aligned}
    \end{equation}
    Therefore, obtaining $2\epsilon$-approximation $\Tilde{V}^{*}$ for $\Tilde{\MDP}$ with a quantum generative oracle reduced to obtaining $\epsilon$-approximation value $\hat{V}_{0}^{*}$ for $\MDP$ with a quantum generative oracle. Then, it implies that the algorithm $\mathcal{K}$ inherits the lower bound for obtaining $2\epsilon$-approximation $\Tilde{V}^{*}$ for $\Tilde{\MDP}$ with a quantum generative oracle. Note that $\MDP$ is a time-independent MDP. Then the quantum generative oracle $\mathcal{G}$ is the same as $\Tilde{\mathcal{G}}$ defined in Theorem \ref{thm: quantum lower bound for infinite horizon MDP}. By Theorem \ref{thm: quantum lower bound for infinite horizon MDP}, we know that the lower bound for obtaining $2\epsilon$-approximation $\Tilde{V}^{*}$ for $\Tilde{\MDP}$ with a quantum generative oracle is $\Omega(S\sqrt{A}\Gamma^{1.5}/\epsilon)$. This implies the quantum lower bound for finite horizon MDP $\MDP$ to obtain an $\epsilon$-optimal value function $\hat{V}_{0}$ is $\Omega(S\sqrt{A}H^{1.5}/(\epsilon\log^{1.5}(\epsilon^{-1})))$. 

    Note that the above content also shows that obtaining $2\epsilon$-approximation $\Tilde{V}^{*}$ for $\Tilde{\MDP}$ with a classical generative oracle reduced to obtaining $\epsilon$-approximation value $\hat{V}_{0}^{*}$ for $\MDP$ with a classical generative oracle. By Theorem \ref{thm: quantum lower bound for infinite horizon MDP}, we know that the lower bound for obtaining $2\epsilon$-approximation $\Tilde{V}^{*}$ for $\Tilde{\MDP}$ with a classical generative oracle is $\Omega(SA\Gamma^{3}/\epsilon)$. Therefore, the classical lower bound for finite horizon MDP $\MDP$ to obtain an $\epsilon$-optimal value function $\hat{V}_{0}$ is $\Omega(SAH^{3}/(\epsilon^{2}\log^{3}(\epsilon^{-1})))$. 
\done

\begin{lemma}\label{lem: quantum lower bound for epsilon optimal Q value}
    Let $\statespace$ and $\actionspace$ be finite sets of states and actions. Let $H>0$ be a positive integer and $\epsilon\in (0, 1/2)$ be an error parameter. We consider the following finite-horizon MDP $\MDP\define(\statespace, \actionspace, \{P_{h}\}_{h=0}^{H-1}, \{r_{h}\}_{h=0}^{H-1},H)$ where $P_{h}=P\in\realnumber^{\statespace\times \actionspace\times\statespace}$ and $r_{h}=r\in [0,1]^{\statespace\times\actionspace}$ for all $h\in[H]$. 
    \begin{itemize}
    \item Given access to a classical generative oracle,
     any algorithm $\mathcal{K}$, which takes $\MDP$ as an input and outputs a value function $\hat{Q}_{0}$ such that $\infiNorm{\hat{Q}_{0}-Q_{0}^{*}}\leq \epsilon$ with probability at least $0.9$, needs to call the classical generative oracle at least 
    \begin{equation}
        \Omega\Biggl( \frac{SAH^{3}}{\epsilon^{2}\log^{3}(\epsilon^{-1})} \Biggr)
    \end{equation}
    times on the worst case of input $\MDP$.
        \item  Given access to a quantum generative oracle $\mathcal{G}$ defined in Definition \ref{def: quantum generative model of an MDP}
     any algorithm $\mathcal{K}$, which takes $\MDP$ as an input and outputs a value function $\hat{Q}_{0}$ such that $\infiNorm{\hat{Q}_{0}-Q_{0}^{*}}\leq \epsilon$ with probability at least $0.9$, needs to call the quantum generative oracle at least 
    \begin{equation}
        \Omega\Biggl( \frac{SAH^{1.5}}{\epsilon\log^{1.5}(\epsilon^{-1})} \Biggr)
    \end{equation}
    times on the worst case of input $\MDP$.
    \end{itemize}
   
\end{lemma}
\beginproof
Following the same idea in Lemma \ref{lem: quantum lower bound for epsilon optimal V value}, we consider an infinite-horizon MDP $\Tilde{\MDP}=(\Tilde{\statespace}, \Tilde{\actionspace}, \Tilde{P}, \Tilde{r},\gamma)$ with a quantum generative oracle, where $\Tilde{\statespace}=\statespace\setminus\{s_{0}\}$ to be a subset of $\statespace$ and $\gamma\in [0,1)$. 
With a slight abuse of the notations for the infinite-horizon MDPs, we let $\Tilde{V}^{*}\in\realnumber^{\statespace}$  and $\Tilde{Q}^{*}\in\realnumber^{\statespace\times\actionspace}$ be the optimal V-value and Q-value functions of $\Tilde{\MDP}$. 
    Now, we proceed to prove that obtaining an $2\epsilon$-approximation value of $\Tilde{Q}^{*}$ for any infinite horizon MDP $\Tilde{\MDP}$ can be reduced to obtaining an $\epsilon$-approximation value of $Q_{0}^{*}$ for a finite horizon MDP. We consider following finite-horizon MDP $\MDP=(\statespace,\actionspace, \{P_{h}\}_{h=0}^{H-1}, \{r_{h}\}_{h=0}^{H-1}, H)$ where $P_{h}=P\in\realnumber^{\statespace\times \actionspace\times \statespace}$ and $r_{h}=r\in\realnumber^{\statespace\times \actionspace}$. Besides, the time horizon $H$ satisfies $H=\ceil{2(1-\gamma)^{-1}\log(\epsilon^{-1})}=\Theta((1-\gamma)^{-1}\log(\epsilon^{-1}))$. Besides, under any action $a\in\actionspace=\Tilde{\actionspace}$, there is a $(1-\gamma)$ probability for each state $s\in\Tilde{\statespace}$ to transition to $s_{0}$ and $\gamma$ probability to follow the original transitions in $\Tilde{\MDP}$. However, when the agent is in $s_{0}$, it can only transition to itself with probability $1$, no matter which action $a\in\actionspace$ it takes. Hence, $s_{0}$ is an absorbing state in $\MDP$. Overall, we have the following definitions for the transition probability kernel $P$ in $\MDP$.
    \begin{align}
        \forall s,s'\in\Tilde{\statespace}, a\in\actionspace, & P(s'|s,a)=\gamma\Tilde{P}(s'|s,a),P(s_{0}|s,a)=(1-\gamma),\\
        & P(s'|s_{0}, a)=0, P(s_{0}|s_{0},a)=1.
    \end{align}
    Besides, we define $r(s_{0},a)=0, r(s,a)=\Tilde{r}(s,a)\in[0,1]$ for all $s\in\Tilde{\statespace}$ and $a\in\actionspace$.

    Now, we proceed to prove that $\infiNorm{Q_{0}^{*}|_{\Tilde{\statespace}\times\actionspace}-\Tilde{Q}^{*}}\leq \epsilon$, i.e., $|Q_{0}^{*}(s,a)-\Tilde{Q}^{*}(s,a)|\leq \epsilon$ for all $s\in\Tilde{\statespace}$ and $a\in\Tilde{\actionspace}=\actionspace$. First, we note that $Q_{H-1}^{*}=r(s,a)\leq \Tilde{Q}^{*}$ by the definition of $\Tilde{Q}^{*}$. 
    In Lemma \ref{lem: quantum lower bound for epsilon optimal V value}, we see that it holds that $V_{h}^{*}(s)\leq \Tilde{V}^{*}(s)$ for all $h\in[H]$ and $s\in\Tilde{\statespace}$, and $V_{h+1}^{*}(s_{0})=0$ for all $h\in[H]$.
    Therefore, we have, 
    \begin{equation}
            \begin{aligned}
        Q_{h}^{*}(s,a)&=r(s,a)+\sum_{s'\in\statespace}P(s'|s,a)V_{h+1}^{*}(s')\\
        &=r(s,a)+\sum_{s'\in \Tilde{\statespace}}P(s'|s,a)V_{h+1}^{*}(s')+P(s_{0}|s,a)V_{h+1}^{*}(s_{0})\\
         &=r(s,a)+\sum_{s'\in \Tilde{\statespace}}P(s'|s,a)V_{h+1}^{*}(s')\\
     &=r(s,a)+\gamma\sum_{s'\in \Tilde{\statespace}}\Tilde{P}(s'|s,a)V_{h+1}^{*}(s')\\
     &\leq r(s,a)+\gamma\sum_{s'\in \Tilde{\statespace}}\Tilde{P}(s'|s,a)\Tilde{V}^{*}(s')\\
     &=\Tilde{Q}^{*}(s,a),
    \end{aligned}
    \end{equation}
    for all $h\in [H-1]$ and $(s,a) \in\Tilde{\statespace}\times \actionspace$. In particular, $Q_{0}^{*}(s,a)\leq \Tilde{Q}^{*}(s,a)$ for all $s\in\Tilde{\statespace}$ and $a\in\actionspace$.
    Let $\Tilde{\pi}^{*}\in\actionspace^{\statespace}$ be an optimal policy for the infinite-horizon MDP $\Tilde{\MDP}$. However, $\Tilde{\pi}\in\actionspace^{\statespace\times [H]}$, where $\Tilde{\pi}(\cdot, h)=\Tilde{\pi}^{*}$ for all $h\in[H]$, may not be an optimal policy for finite-horizon MDP $\MDP$. Then we have $Q_{0}^{\Tilde{\pi}}(s,a)\leq Q_{0}^{*}(s,a)$ for all $s\in\statespace$ and $a\in\actionspace$. In fact, for any $s\in\Tilde{\statespace}$, we have 
    \begin{equation}
            \begin{aligned}
        Q_{0}^{\Tilde{\pi}}(s,a)&=r(s, a)+ \sum_{s'\in\statespace}P(s'|s,a)r\bigl( s, \Tilde{\pi}^{*}(s) \bigr)+\cdots+\sum_{s'\in\statespace}P^{H}\bigl( s'|s,\Tilde{\pi}^{*}(s) \bigr)r\bigl( s, \Tilde{\pi}^{*}(s) \bigr)\\
        &=\Tilde{r}(s, a)+ \gamma\sum_{s'\in\Tilde{\statespace}}\Tilde{P}(s'|s,a)\Tilde{r}\bigl(s, \Tilde{\pi}^{*}(s) \bigr)+\cdots+\gamma^{H}\sum_{s'\in\Tilde{\statespace}}\Tilde{P}^{H}(s'|s,a)\Tilde{r}\bigl(s, \Tilde{\pi}^{*}(s) \bigr)\\
        &= \Tilde{Q}_{H}^{\Tilde{\pi}^{*}},
    \end{aligned}
    \end{equation}
    where $\Tilde{Q}_{H}^{\Tilde{\pi}^{*}}$ is the Q value of the infinite-horizon MDP $\Tilde{\MDP}$ induced by the policy $\Tilde{\pi}^{*}$ over $H$ iterations. Note that for any policy $\Tilde{\pi}$ for the infinite horizon MDP $\Tilde{\MDP}$, $\infiNorm{\Tilde{Q}_{k}^{\pi}- \Tilde{Q}^{\pi}}\leq \gamma^{k}\infiNorm{\Tilde{Q}_{0}^{\pi}-\Tilde{Q}^{\pi}}\leq 2\exp(-(1-\gamma)k)/(1-\gamma)$. Then, $\forall \epsilon$, it suffices to let $k\geq \log(2/((1-\gamma)\epsilon))/(1-\gamma)$ so that $\infiNorm{\Tilde{Q}_{k}^{\pi}- \Tilde{Q}^{\pi}}\leq \epsilon$. In fact, 
    \begin{equation}
            \begin{aligned}
        \frac{1}{1-\gamma}\log\left(\frac{2}{(1-\gamma)\epsilon}\right)&
        =\frac{1}{1-\gamma}\Biggl(\log\left(\frac{1}{1-\gamma}\right)+\log\left(\frac{2}{\epsilon} \right) \Biggr)\\
        &=\frac{1}{1-\gamma}\biggl(-\log(1-\gamma)+\log\left(\frac{2}{\epsilon} \right) \biggr)\\
        &\leq \frac{1}{1-\gamma}\biggl(\gamma+\log\left(\frac{2}{\epsilon} \right)
        \biggr)\\
        &\leq 2\frac{1}{1-\gamma}\log\left(\frac{2}{\epsilon} \right).
    \end{aligned}
    \end{equation}
    The third line comes from the fact that  $\log(1-\gamma)\leq -\gamma, \forall \gamma\in [0,1)$. The last line comes from the fact that $\log(2/\epsilon)>1, \forall \epsilon\in(0, 1/2)$.
    Since $H=\ceil{2\frac{1}{1-\gamma}\log\left(\frac{2}{\epsilon} \right)}$ and $\Tilde{\pi}^{*}$ is an optimal policy for $\Tilde{\MDP}$, then we must have $\Tilde{Q}_{H}^{\Tilde{\pi}^{*}}(s,a)\geq \Tilde{Q}^{\Tilde{\pi}^{*}}(s,a)-\epsilon=\Tilde{Q}^{*}(s,a)-\epsilon$.
    Therefore, we have $\Tilde{Q}^{*}(s,a)-\epsilon\leq Q_{0}^{*}(s,a)\leq \Tilde{Q}^{*}(s,a)$ for all $s\in\Tilde{\statespace}, a\in\actionspace$, which implies $\infiNorm{\hat{Q}_{0}^{*}|_{\Tilde{\statespace}\times \actionspace}-\Tilde{Q}^{*}}\leq \epsilon$. 
    Therefore, an $\epsilon$-approximation of $Q_{0}^{*}$ will give an $2\epsilon$-approximation to $\Tilde{Q}^{*}$. Specifically, if we let $\hat{Q}_{0}$ be an $\epsilon$-approximation of $Q_{0}^{*}$, then
    \begin{equation}
    \begin{aligned}
        \infiNorm{\hat{Q}_{0}|_{\Tilde{\statespace}\times \actionspace}-\Tilde{Q}^{*}}&\leq \infiNorm{\hat{Q}_{0}|_{\Tilde{\statespace}\times\actionspace}-Q_{0}^{*}|_{\Tilde{\statespace}\times \actionspace}}+\infiNorm{Q_{0}^{*}|_{\Tilde{\statespace}\times \actionspace}-\Tilde{Q}^{*}}\\
        &\leq \infiNorm{\hat{Q}_{0}-Q_{0}^{*}}+\infiNorm{Q_{0}^{*}|_{\Tilde{\statespace}\times \actionspace}-\Tilde{Q}^{*}}\\
        &\leq 2\epsilon.
    \end{aligned}        
    \end{equation}

    Therefore, obtaining $2\epsilon$-approximation $\Tilde{Q}^{*}$ for $\Tilde{\MDP}$ with a quantum generative oracle reduced to obtaining $\epsilon$-approximation value $\hat{Q}_{0}^{*}$ for $\MDP$ with a quantum generative oracle. Then, it implies that the algorithm $\mathcal{K}$ inherits the lower bound for obtaining $2\epsilon$-approximation $\Tilde{Q}^{*}$ for $\Tilde{\MDP}$ with a quantum generative oracle. Note that $\MDP$ is a time-independent MDP. Then the quantum generative oracle $\mathcal{G}$ is the same as $\Tilde{\mathcal{G}}$ defined in Theorem \ref{thm: quantum lower bound for infinite horizon MDP}. By Theorem \ref{thm: quantum lower bound for infinite horizon MDP}, we know that the lower bound for obtaining $2\epsilon$-approximation $\Tilde{Q}^{*}$ for $\Tilde{\MDP}$ with a quantum generative oracle is $\Omega(SA\Gamma^{1.5}/\epsilon)$. This implies the lower bound for obtaining $\epsilon$-optimal Q value function $\hat{Q}_{0}$ of finite horizon MDP $\MDP$ is $\Omega(SAH^{1.5}/(\epsilon\log^{1.5}(\epsilon^{-1})))$.

    Note that the above content also implies that obtaining $2\epsilon$-approximation $\Tilde{Q}^{*}$ for $\Tilde{\MDP}$ with a classical generative oracle reduced to obtaining $\epsilon$-approximation value $\hat{Q}_{0}^{*}$ for $\MDP$ with a classical generative oracle. By Theorem \ref{thm: quantum lower bound for infinite horizon MDP}, we know that the lower bound for obtaining $2\epsilon$-approximation $\Tilde{Q}^{*}$ for $\Tilde{\MDP}$ with a classical generative oracle is $\Omega(SA\Gamma^{3}/\epsilon)$. Therefore, the classical lower bound for finite horizon MDP is $\Omega(SAH^{3}/(\epsilon^{2}\log^{3}(\epsilon^{-1})))$.
\done

\subsubsection{Lower Bounds for Finite-horizon MDPs (Proof of Theorem \ref{thm: quantum lower bound for finite horizon MDP})}
\beginproof
    Since time-independent and finite-horizon MDP is a special case of time-dependent and finite-horizon MDP, we know that the lower bound of obtaining an $\epsilon$-approximation $\hat{V}_{0}$ of $V_{0}^{*}$ for time-dependent and finite-horizon MDP $\MDP$ with a classical or quantum generative oracle inherits the corresponding lower bound in Lemma \ref{lem: quantum lower bound for epsilon optimal V value}. Besides, obtaining $\epsilon$-approximations $\hat{V}_{0}$ of $V_{0}^{*}$ is a sub-task of obtaining $\epsilon$-approximations $\hat{V}_{h}$ of $V_{h}^{*}$ for all $h\in[H]$. Therefore, the lower bound of obtaining $\epsilon$-approximations $\hat{V}_{h}$ of $V_{h}^{*}$ for all $h\in[H]$ for time-dependent and finite-horizon MDP $\MDP$ with access to a classical or quantum generative oracle inherits the lower bound of obtaining $\epsilon$-approximations $\hat{V}_{0}$ of $V_{0}^{*}$ with a classical or quantum generative oracle. Therefore, algorithm $\mathcal{K}$ has the desired classical and quantum lower bounds for obtaining $\epsilon$-optimal V value functions $\{\hat{V}_{h}\}_{h=0}^{H-1}$. 
    With Lemma \ref{lem: quantum lower bound for epsilon optimal Q value}, similar idea also applies to obtain the classical and quantum lower bound of obtaining $\epsilon$-optimal Q value functions $\{\hat{Q}_{h}\}_{h=0}^{H-1}$. 
    
    Suppose $\mathcal{K}$ can output an $\epsilon$-optimal policy $\hat{\pi}$ for a finite horizon and time-dependent MDP $\MDP$, then the corresponding V-values $\{\hat{V}_{h}\}_{h=0}^{H-1}\define\{V^{\hat{\pi}}_{h}\}_{h=0}^{H-1}$ induced by $\hat{\pi}$ are $\epsilon$-optimal. Therefore, $\mathcal{K}$ has the desired classical and quantum lower bounds for obtaining the $\epsilon$-optimal policy $\hat{\pi}$ by inheriting the corresponding lower bound for obtaining $\epsilon$-optimal V-value functions $\{\hat{V}_{h}\}_{h=0}^{H-1}$.
\done

\end{document}